\documentclass[preprint]{aastex62}

\newcommand{\neswarrow}{%
  \mathrel{\text{\ooalign{$\swarrow$\cr$\nearrow$}}}%
}

\newcommand{\nwsearrow}{%
  \mathrel{\text{\ooalign{$\searrow$\cr$\nwarrow$}}}%
}

\usepackage{epstopdf}
\usepackage{graphicx}
\usepackage{amsmath}
\usepackage{hhline}
\usepackage{float}
\usepackage{longtable}
\usepackage{amssymb}
\usepackage{CJK}
\usepackage[whole]{bxcjkjatype} 
\usepackage{CJKutf8}
\usepackage[bookmarks=false]{hyperref}

\received{}
\revised{}
\accepted{}
\submitjournal{ApJ}

\shorttitle{SN\,2018gv}
\shortauthors{Yang et al.}

\begin{document}

\title{The Young and Nearby Normal Type Ia Supernova 2018gv: 
UV-Optical Observations and the Earliest Spectropolarimetry }

\correspondingauthor{Yi Yang}
\email{yi.yang@weizmann.ac.il}

\author[0000-0002-6535-8500]{Yi Yang\begin{CJK*}{UTF8}{gbsn}
(杨轶)
\end{CJK*}}
\affil{Department of Particle Physics and Astrophysics, 
Weizmann Institute of Science, Rehovot 76100, Israel}

\author{Peter Hoeflich}
\affiliation{Department of Physics, Florida State University, 
Tallahassee, Florida 32306-4350, USA}

\author{Dietrich Baade}
\affiliation{European Southern Observatory, Karl-Schwarzschild-Str. 2, 85748 Garching b.\ M{\"u}nchen, Germany}

\author[0000-0003-0733-7215]{Justyn R. Maund}
\affiliation{Department of Physics and Astronomy, University of Sheffield, Hicks Building, Hounsfield Road, \\
Sheffield S3 7RH, UK}

\author{Lifan Wang}
\affiliation{George P. and Cynthia Woods Mitchell Institute for Fundamental Physics $\&$ Astronomy, Texas A. $\&$ M. University, 4242 TAMU, College Station, TX 77843, USA}
\affiliation{Purple Mountain Observatory, Chinese Academy of Sciences, Nanjing 210008, China}

\author[0000-0001-6272-5507]{Peter.~J. Brown}
\affiliation{George P. and Cynthia Woods Mitchell Institute for Fundamental Physics $\&$ Astronomy, Texas A. $\&$ M. University, 4242 TAMU, College Station, TX 77843, USA}

\author[0000-0002-0504-4323]{Heloise F. Stevance}
\affiliation{Department of Physics and Astronomy, University of Sheffield, Hicks Building, Hounsfield Road, \\
Sheffield S3 7RH, UK}

\author[0000-0001-7090-4898]{Iair Arcavi}
\affiliation{The Raymond and Beverly Sackler School of Physics and Astronomy, Tel Aviv University, Tel Aviv 69978, Israel}

\author{Jamison Burke}
\affiliation{Las Cumbres Observatory, 6740 Cortona Drive, Suite 102, Goleta, CA 93117-5575, USA}
\affiliation{Department of Physics, University of California, Santa Barbara, CA 93106-9530, USA}

\author[0000-0001-7101-9831]{Aleksandar Cikota}
\affiliation{Physics Division, Lawrence Berkeley National Laboratory, 1 Cyclotron Road, Berkeley, CA 94720, USA}

\author{Alejandro Clocchiatti}
\affiliation{Instituto de Astrofisica, Pontificia Universidad Catolica (PUC), Casilla 306, Santiago 22, Chile}

\author[0000-0002-3653-5598]{Avishay Gal-Yam}
\affiliation{Department of Particle Physics and Astrophysics, 
Weizmann Institute of Science, Rehovot 76100, Israel}

\author[0000-0002-9154-3136]{Melissa.~L. Graham}
\affiliation{Department of Astronomy, University of Washington, Box 351580, U.W., Seattle, WA 98195, USA}

\author[0000-0002-1125-9187]{Daichi Hiramatsu}
\affiliation{Las Cumbres Observatory, 6740 Cortona Drive, Suite 102, Goleta, CA 93117-5575, USA}
\affiliation{Department of Physics, University of California, Santa Barbara, CA 93106-9530, USA}

\author[0000-0002-0832-2974]{Griffin Hosseinzadeh}
\affiliation{Harvard-Smithsonian Center for Astrophysics, 60 Garden Street, Cambridge, MA 02138, USA}
\affiliation{Las Cumbres Observatory, 6740 Cortona Drive, Suite 102, Goleta, CA 93117-5575, USA}
\affiliation{Department of Physics, University of California, Santa Barbara, CA 93106-9530, USA}

\author[0000-0003-4253-656X]{D. Andrew Howell}
\affiliation{Las Cumbres Observatory, 6740 Cortona Drive, Suite 102, Goleta, CA 93117-5575, USA}
\affiliation{Department of Physics, University of California, Santa Barbara, CA 93106-9530, USA}

\author[0000-0001-8738-6011]{Saurabh W. Jha}
\affiliation{Department of Physics and Astronomy, Rutgers, the State University of New Jersey, 136 Frelinghuysen Rd., Piscataway, NJ 08854, USA}
\affiliation{Center for Computational Astrophysics, Flatiron Institute, 162 5th Avenue, New York, NY 10010, USA}

\author[0000-0001-5807-7893]{Curtis McCully}
\affiliation{Las Cumbres Observatory, 6740 Cortona Drive, Suite 102, Goleta, CA 93117-5575, USA}

\author[0000-0002-0537-3573]{Ferdinando Patat}
\affiliation{European Southern Observatory, Karl-Schwarzschild-Str. 2, 85748 Garching b.\ M{\"u}nchen, Germany}

\author[0000-0003-4102-380X]{David.~J. Sand}
\affiliation{Steward Observatory, University of Arizona, 933 North Cherry Avenue, Rm. N204, Tucson, AZ 85721-0065, USA}

\author[0000-0001-6797-1889]{Steve Schulze}
\affil{Department of Particle Physics and Astrophysics, 
Weizmann Institute of Science, Rehovot 76100, Israel}

\author[0000-0001-6815-4055]{Jason Spyromilio}
\affiliation{European Southern Observatory, Karl-Schwarzschild-Str. 2, 85748 Garching b.\ M{\"u}nchen, Germany}

\author[0000-0001-8818-0795]{Stefano Valenti}
\affiliation{Department of Physics, University of California, Davis, CA 95616, USA}

\author[0000-0001-8764-7832]{J\'{o}zsef Vink\'{o}}
\affiliation{Konkoly Observatory, MTA CSFK, Konkoly Thege M. ut 15-17, Budapest, 1121, Hungary}
\affiliation{Department of Optics \& Quantum Electronics, University of Szeged, Dom ter 9, Szeged, 6720 Hungary}
\affiliation{Department of Astronomy, University of Texas at Austin, Austin, TX 78712, USA}

\author[0000-0002-7334-2357]{Xiaofeng Wang}
\affiliation{Physics Department and Tsinghua Center for Astrophysics (THCA), Tsinghua University, Beijing, 100084, People's Republic of China}

\author[0000-0003-1349-6538]{J. Craig Wheeler}
\affiliation{Department of Astronomy, University of Texas at Austin, Austin, TX 78712, USA}

\author[0000-0002-0301-8017]{Ofer Yaron}
\affil{Department of Particle Physics and Astrophysics, 
Weizmann Institute of Science, Rehovot 76100, Israel}

\author[0000-0002-8296-2590]{Jujia Zhang}
\affiliation{Yunnan Observatories (YNAO), Chinese Academy of Sciences, Kunming 650216, People's Republic of China}
\affiliation{Key Laboratory for the Structure and Evolution of Celestial Objects, Chinese Academy of Sciences, Kunming 650216, People's Republic of China}





\begin{abstract}
The non-detection of companion stars in Type Ia supernova (SN) 
progenitor systems lends support to the notion of double-degenerate 
(DD) systems and explosions triggered by the merging of two white dwarfs. 
This very asymmetric process should 
lead to a conspicuous polarimetric signature. By contrast, observations 
consistently find very low continuum polarization as the signatures from 
the explosion process largely dominate over the pre-explosion configuration 
within several days. Critical information about the interaction of the 
ejecta with a companion and any circumstellar matter is encoded in the 
early polarization spectra. In this study, we obtain spectropolarimetry of 
SN\,2018gv with the ESO Very Large Telescope at $-$13.6 days relative to the 
$B-$band maximum light, or $\sim$5 days after the estimated explosion --- the 
earliest spectropolarimetric observations to date of any Type Ia SN. These early 
observations still show a low continuum polarization ($\lesssim$0.2\%) and 
moderate line polarization (0.30$\pm$0.04\% for the prominent Si II 
$\lambda$6355 feature and 0.85$\pm$0.04\% for the high-velocity Ca component). 
The high degree of spherical symmetry implied by the low line and continuum 
polarization at this early epoch is consistent with explosion models of 
delayed detonations and is inconsistent with the merger-induced explosion 
scenario. The dense UV and optical photometry and optical spectroscopy within 
the first $\sim$100 days after the maximum light indicate that SN\,2018gv is 
a normal Type Ia SN with similar spectrophotometric behavior to SN\,2011fe. 
\end{abstract}



\section{Introduction \label{section_intro}}
Thermonuclear supernovae (Type Ia SNe) have been very well calibrated
at low redshift and are used as cosmic distance indicators out to redshift $z\sim$2 
\citep{Riess_etal_1998, Perlmutter_etal_1999, Riess_etal_2016, Riess_etal_2018_snia}. 
However, this calibration is purely empirical and parametric, lacking a detailed 
physical foundation. The general consensus is that Type Ia SNe are powered by the 
ignition of degenerate nuclear fuel from carbon/oxygen white dwarfs (CO WDs, 
\citealp{Hoyle_etal_1960}, see e.g. \citealp{Howell_2011, Hillebrandt_etal_2013, Maoz_etal_2014, Branch_Wheeler_2017, Hoeflich_2017} for recent reviews). 
However, it is still unknown how the thermonuclear runaway is triggered 
and propagates throughout the progenitor star \citep{Arnett_1969, 
Nomoto_etal_1976, Khokhlov_etal_1991, Niemeyer_etal_1996, Reinecke_etal_2002, 
Plewa_etal_2004, Ropke_2007, Pakmor_etal_2011, Pakmor_etal_2012, Seitenzahl_etal_2013}. 
A comprehensive summarize and comparison of these works is provided by 
\citet{Branch_Wheeler_2017}. 
The principal contenders are double-degenerate (DD) models, in which two WDs merge \citep{Iben_Tutukov_1984, Webbink_1984}, and single-degenerate (SD) models, in which 
a WD accretes matter from a companion \citep{Whelan_Iben_1973} until the critical 
Chandrasekhar mass of $M_{\rm Ch}\sim$1.4$M_{\odot}$ is reached. There is some 
evidence that the Type Ia SN population could consist of both the SD and DD 
progenitor systems (see e.g. \citealp{Maoz_etal_2014}). It also remains unclear what 
the fractions are for various processes contributing to the Type Ia SN population. 

Within this general picture for progenitors, three major classes of explosion 
have been proposed, which are distinguished by the triggering mechanism of 
the explosion, 
\begin{enumerate}
\item Explosion of a single CO WD with a mass close to $M_{\rm Ch}$. The explosion 
is triggered by compressional heating near the WD center as a {\sl deflagration} 
front. Because the compressional heat release increases rapidly towards $M_{\rm Ch}$, 
the exploding WD lands in a very narrow mass range \citep{Hoeflich_Khokhlov_1996} 
though actual triggers may differ as much as slow accretion, pulsational instabilities, 
or nova eruptions. The donor star may be a moderately evolved Roche-lobe-overflowing 
star in an SD system \citep{Whelan_Iben_1973} or a tidally disrupted WD in a 
DD system \cite{Whelan_Iben_1973, Piersanti_etal_2004}; 

\item Dynamical merging of two CO WDs in a binary system after losing or shedding 
angular momentum via gravitational radiation \citep{Iben_Tutukov_1984, Webbink_1984, 
Benz_etal_1990, Nugent_etal_1997, Pakmor_etal_2010}. The thermonuclear explosion 
is triggered by the heat of the merging process or as a consequence of the WD-WD 
collisions. It is unclear whether the dynamical merging process or a violent 
collision of two WD leads to a Type Ia SN, an `accretion induced collapse' (AIC), 
or a WD with high magnetic fields \citep{Rasio_Shapiro_1994, Hoeflich_Khokhlov_1996, Segretain_etal_1997, Yoon_etal_2007,Loren-Aguilar_etal_2009, 
Garcia-Berro_etal_2012}

\item Another class involves explosions of a sub-$M_{\rm Ch}$ CO WD triggered by 
detonating a thin surface He layer on the WD, which triggers a detonation front 
\citep{Woosley_etal_1980, Nomoto_1982_p1, Nomoto_1982_p2, Livne_1990, 
Woosley_Weaver_1994, Hoeflich_Khokhlov_1996, Kromer_etal_2010}.
\end{enumerate}

In the SD channel, the explosions of Type Ia SNe are triggered by a subsonic ignition 
of degenerate material. If the pure subsonic deflagration \citep{Reinecke_etal_2002} 
persists for the entire duration of the explosion, the front would result in a 
rather homogeneous angular distribution of the ejecta, in terms of both the material 
velocities and the chemical composition. The deflagration in $M_{\rm Ch}$ WDs, 
however, is not suitable to describe the class of normal Type Ia SNe because it 
fails to explain the rather high kinetic energy of the ejecta, the large amount of 
$^{56}$Ni production \citep{Hillebrandt_etal_2013}, together with the presence of 
unburnt carbon and oxygen. A pure detonation scenario was also ruled out given the 
fact that the whole star is not burnt to iron and nickel. The delayed detonation 
scenario initially requires a subsonic deflagration that, after a period of time, 
transitions to a supersonic detonation \citep{Khokhlov_etal_1991}. The detonation 
front burns the outer layers to intermediate mass elements (IME), i.e. from Si to Ca. 
The ejecta are predicted to be stratified in terms of density and chemical 
abundance, and hence exhibit significant homogeneity on large scales 
\citep{Gamezo_etal_2005,Seitenzahl_etal_2013, Sim_etal_2013}. 
Another mechanism suggests that the WD could explode at a sub-$M_{\rm Ch}$ 
\citep{Fink_etal_2010}. A CO WD accreting mass from a donor star, depending on 
the assumed mass transfer rates and the mass of the CO WD, could be able to 
accumulate a layer of helium which may develop to a degenerate-helium-shell flash 
under the right physical conditions \citep{Taam_1980}. Such unstable thermonuclear 
shell ignition could trigger a second detonation in the sub-$M_{\rm Ch}$ 
WD \citep{Shen_etal_2010}. 

In the DD channel, the explosion of the SN could be triggered through a dynamic 
merger of two CO WDs \citep{Pakmor_etal_2010}. Starting from two CO WDs with 
masses of 0.9$M_{\odot}$ and 1.1$M_{\odot}$, the so-called violent merger model 
is able to reproduce the peak brightness, the color, as well as spectral shape 
and the velocity profiles of most of the line features, although the predicted 
rise time of the $B-$band light curve is long compared to that of normal Type 
Ia SNe \citep{Hayden_etal_2010_rise}. Other configurations, i.e. with very low 
mass-transfer rates, could fail to produce an explosive phenomenon and burn the 
C-O mixture into an O-Ne-Mg WD \citep{Saio_etal_1998} or form a single neutron 
star \citep{Saio_etal_1985}. Thermonuclear explosions might also be triggered 
by direct, head-on collisions of WDs, and high-resolution numerical simulations 
are able to reproduce the primary observational signatures of Type Ia SNe 
\citep{Kushnir_etal_2013}. 

One difficulty in distinguishing models reliably by means of conventional 
photometry or spectroscopy results from the ambiguity of the shape of the SN 
explosion. The geometry of the explosion and the structure of the SN ejecta, 
which are too distant to be spatially resolved, can only be probed with 
polarimetry. The polarized emission from a SN arises from a departure from 
spherical symmetry \citep{Shapiro_etal_1982}. Electron scattering in 
asymmetric ejecta leads to the incomplete cancellation of electric field 
vectors (E-vectors), which produce nonzero degrees of observable polarization \citep{Hoeflich_1991, Hoeflich_etal_1995}. The continuum polarization tests 
whether the photosphere deviates from spherical symmetry, while line 
polarization traces mostly the distribution of elements in the SN ejecta. 
Material in the SN ejecta with considerable optical depth may unevenly block 
the photospheric light beneath, thereby producing a polarization variation 
and/or polarization position angle rotation in certain spectral features 
\citep{Kasen_etal_2003, Wang_etal_2006_04dt}. 

Recent studies, in at least one case, have firmly established the progenitor as a 
compact object consistent with a WD \citep{Nugent_etal_2011}. The same data also 
excluded any luminous red giant companion star \citep{Li_etal_2011}. The absence 
of luminous red giants in Type Ia SN progenitor systems is corroborated by studies 
of supernova remnants in the Milky Way \citep{Gonzalez-Hernandez_etal_2012} and in 
the LMC \citep{Edwards_etal_2012}. These results favor the DD channel for Type Ia 
SNe. This process could be sufficiently asymmetric and lead to a clear polarimetric 
signature. By contrast, observations consistently find a very low continuum 
polarization (i.e. $\lesssim0.2\%$, \citealp{Wang_Wheeler_2008}), with a diversity 
in subluminous events (i.e., SN\,1999by, \citealp{Howell_etal_2001} and SN\,2005ke, 
\citealp{Patat_etal_2012}). 
Spectropolarimetry has only been available from $\sim$seven days past explosion 
and cannot penetrate the opaque ejecta that have already expanded beyond the 
innermost interaction zone. Low continuum polarization at these intermediate 
epochs indicates the photosphere is remarkably spherical \citep{Wang_etal_1996, Wang_etal_2003_01el, Maund_etal_2013, Zheng_etal_2017, Hoeflich_etal_2017}. 
The approximate spherical symmetry is expected to be maintained in the SD models 
\citep{Khokhlov_etal_1991}. 

Recent high-cadence wide-field optical surveys and rapid follow-up observations 
of SNe, within hours of the explosion, are opening up a new phase in our 
understanding of SNe. This set of transients, discovered very early, will be 
extremely valuable for constraining the progenitor systems and explosion physics 
of SNe. The ejecta quickly sweep away almost all traces of the pre-explosion 
configuration within a few days. Such information is particularly valuable 
since it is directly connected to the final mass-loss history of the progenitor 
system right before the explosion. It is only accessible during the earliest 
phases. Early polarization measurements, before the pre-explosion configuration 
is left far behind by the rapidly advancing photosphere, can set constraints on 
the progenitor systems. Different degrees and types of asymmetry in the SN 
ejecta are produced by various multi-dimensional explosion models. 
Large departures in the global symmetry can be expected for dynamical processes. 
By contrast, $M_{\rm Ch}$ CO WD explosions triggered by a deflagration and 
sub-$M_{\rm Ch}$ CO WD explosions triggered by detonation of a He layer 
will mostly appear in the chemical distribution. Mergers can be expected to 
produce large, time-variant continuum polarization whereas the latter cases 
may produce abundance asymmetries manifesting themselves in line polarization. 
Early polarimetry also provides 
unrivalled clues for differentiating various progenitor scenarios: e.g., high 
continuum polarization is expected for DD mergers \citep{Pakmor_etal_2012, 
Pakmor_etal_2013, Moll_etal_2014, Raskin_etal_2014, Bulla_etal_2016a}, low 
continuum but significant line polarization is predicted for 
delayed-detonation explosions \citep{Khokhlov_etal_1991, Hoeflich_etal_1995, 
Bulla_etal_2016b}, and low line and continuum polarization should prevail in 
homogeneously mixed structures of deflagrations \citep{Gamezo_etal_2004}. 

Polarimetry can also provide diagnostics for other Type Ia SN hypotheses. 
Theoretical models by \citet{Kasen_2010} show that the ejecta-companion 
interaction may be detected a few days past explosion. These predictions may 
be supported by a UV light curve excess (iPTF14atg, \citealp{Cao_etal_2015}) 
and a clearly-resolved blue bump in the light curve of SN\,2017cbv 
\citep{Hosseinzadeh_etal_2017}. The early blue excess could, however, also 
be explained by vigorous mixing of radioactive $^{56}$Ni in the SN ejecta 
\citep{Hosseinzadeh_etal_2017, Miller_etal_2018}. If such an early blue bump 
was caused by ejecta-companion interaction, the brightest bump would be 
observed from looking straight down the companion, from which the view is 
symmetric so that there is no net polarization. Large polarimetric 
signatures would come from an off-axis viewing angle in which the bump is 
fainter or invisible. Early polarimetry and its correlation with the light 
curve morphology would provide an important diagnostic of the 
ejecta-companion interaction case. 

We present extensive UV-optical photometry, optical and near-infrared (NIR) 
spectroscopy, and optical spectropolarimetry of the nearby Type Ia SN\,2018gv 
in the host galaxy NGC 2525. The organization of this paper is as follows: 
Observations and data reductions are detailed in Section~\ref{section_obs}. 
Section~\ref{section_lc} describes the photometric evolution and estimates 
derived for the extinction arising in the host galaxy. 
Section~\ref{section_spec} presents the spectral evolution and 
Section~\ref{section_specpol} investigates the spectropolarimetric properties 
of the SN. Discussions and a brief summary of the study are given in 
Section~\ref{section_discussion} and ~\ref{section_summary}, respectively. 

\begin{figure}[!h]
\epsscale{1.0}
\plotone{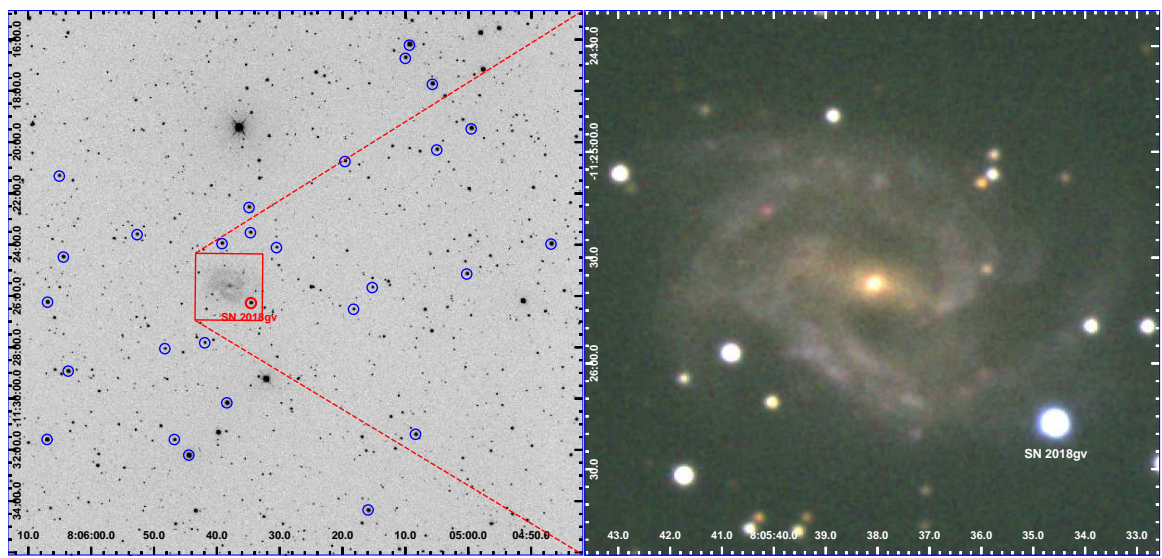}
\caption{Left panel: LCO $V-$band image showing the location of SN\,2018gv. 
Reference stars are marked with blue circles. The red square outlines the 
right panel. Right panel: color image of SN\,2018gv in NGC 2525 from LCO $B$, 
$V$, and $r'-$band exposures. North is up, east is left. 
\label{Fig_image}
}
\end{figure}

\section{Observations and Data Reduction~\label{section_obs}}
SN\,2018gv was discovered at UT 2018-01-15 16:21:06 with an 0.5-m/f6.8 telescope 
at an unfiltered magnitude of 16.5 mag \citep{Itagaki_etal_2018}. 
Follow-up spectroscopy on UT 2018-01-16 12:41:15 \citep{Siebert_etal_2018} 
reveals that SN\,2018gv was a very young, normal Type Ia SN, at 
$\sim11-13$ days before the maximum luminosity according to the 
classification with SNID \citep{Blondin_etal_2007}. 
We measure its J2000.0 coordinates on the images obtained by the Sinistro 
cameras on Las Cumbres Observatory (LCO) Global Network of 1 m telescopes to be 
$\rm \alpha = 08^{h}05^{m}34\farcs{58}$, $\rm \delta = -11^{\circ}26'16\farcs{77}$. 
SN\,2018gv exploded $3\farcs{46}$ W and $39\farcs{15}$ S of the nucleus of the 
host spriral galaxy NGC 2525 (see Fig.~\ref{Fig_image}). We queried the 
SIMBAD\footnote{http://simbad.u-strasbg.fr/simbad/} 
database to retrieve reported heliocentric radial velocity measurements of 
the host galaxy. Among six reported measurements, we adopt a median value 
of 1582 km s$^{-1}$ with a standard deviation of 12.6 km s$^{-1}$, which 
falls into the value obtained by the 2MASS catalog \citep{Tully_2015}. 
This implies a redshift of $z=0.00527 \pm 0.00004$ for SN\,2018gv, which is 
used throughout the paper together with the adopted Hubble constant of 
$H_0 = 73.24 \pm 1.74$ km s$^{-1}$ Mpc$^{-1}$ \citep{Riess_etal_2016}. 

\subsection{Optical Photometry}
\subsubsection{Ground-Based Photometry}
Extensive $UBg'Vr'i'$ photometry was obtained with the Sinistro cameras on 
the LCO network of 1 m telescopes. The images were pre-processed including 
bias subtraction and flat-field correction using the BANZAI automatic pipeline 
\citep{McCully_etal_2018}. Figure~\ref{Fig_image} shows the field around 
SN\,2018gv. Because the SN is bright and in the outskirts of the galaxy, 
template subtraction and PSF-fitting are not necessary. Therefore, the flux 
of the SN and the local reference stars were measured with a circular aperture 
of 3$\farcs{0}$ in radius. The background was estimated by the median pixel 
value of an annulus around the SN with an inner radius of 9\farcs{0} and an 
outer radius of 12\farcs{0}. We calibrated the instrumental $BV$ and $g'r'i'$ 
magnitudes of SN\,2018gv to the standard Johnson $BV$ system 
\citep{Johnson_1966} in Vega Magnitude and the SDSS photometric system 
\citep{Fukugita_etal_1996} in AB magnitude, respectively, based on the 
magnitude of local comparison stars from the AAVSO Photometric All Sky Survey 
(APASS) DR9 Catalogue (\citealp{Henden_etal_2016}, see 
Table~\ref{Table_phot_standard}). $U-$band magnitudes are only available for 
five of the comparison stars from the All-sky spectrally matched Tycho2 stars 
catalogue \citep{Pickles_etal_2010}. The final $Bg'Vr'i'$ and $U-$band 
calibrations were carried out based on the median of the difference between 
catalogue magnitude and the instrumental magnitude of the 27 comparison stars 
and the five comparison stars, respectively. The field containing the 
comparison stars is also shown in the left panel of Fig.~\ref{Fig_image}. 

\subsubsection{SWIFT UVOT Photometry} 
Ultraviolet and optical photometry was obtained using the Ultra-Violet/Optical 
Telescope (UVOT; \citealp{Roming_etal_2005}) on the Neil Gehrels Swift 
Observatory \citep{Gehrels_etal_2004}. Photometry was reduced using the 
pipeline of the Swift Optical Ultraviolet Supernova Archive (SOUSA; 
\citealp{Brown_etal_2014_SOUSA}) using the zero-points of 
\citet{Breeveld_etal_2011}.  

\subsection{Optical/Near-Infrared (NIR) Spectroscopy}
A journal of spectroscopic observations of SN\,2018gv is provided in 
Table~\ref{Table_log_spec}. The spectral sequence of SN\,2018gv 
spans $t=-15.2$ to $+83.6$ days. All phases are given relative to the 
$B-$band maximum at MJD 58,149.698 or UT2018-01-31.698 (see 
Section~\ref{section_lc}) throughout the paper. 

\subsubsection{LCO Optical Spectroscopy}
LCO optical spectra were taken with the FLOYDS spectrographs mounted on 
the 2m Faulkes Telescope North and South at Haleakala, USA and Siding 
Spring, Australia, respectively, through the Global Supernova Project. 
A 2$\arcsec$ slit was placed on the target at the parallactic angle. 
One-dimensional spectra were extracted, reduced, and calibrated following 
standard procedures using the FLOYDS pipeline 
\footnote{\url{https://github.com/svalenti/FLOYDS_pipeline}} 
\citep{Valenti_etal_2014}. 

\subsubsection{Gemini Optical Spectroscopy}
We obtained an optical spectrum of SN\,2018gv using the Gemini Multi-Object 
Spectrographs (GMOS, \citealp{Davies_etal_1997}) on Gemini North telescope 
on 2018-01-17 UT with an airmass of 1.22 (GN-2017B-Q-12; PI Howell). 
For this spectrum, we obtained 2$\times$300 s exposures in both the B600 and 
R400 gratings with central wavelengths of 450 nm and 750 nm respectively. 
We covered the chip gap by moving the central wavelength for the second 
exposure by 5 nm and 7.5 nm for the blue and red setups respectively. 

The Gemini data were reduced using the standard techniques using a combination 
of the Gemini-IRAF\footnote{\url{https://www.gemini.edu/node/11823}} 
and custom procedures written in 
Python\footnote{\url{https://github.com/cmccully/LCOgemini}}. 
We obtained observations of the {\it HST} spectrophotometric standard star, 
HZ44, as part of the same program. The sensitivity function was derived using 
the {\it HST} spectrophotometric data. The telluric correction was then 
derived from the standard star observation. 

\subsubsection{Hobby-Eberly Telescope (HET) Spectroscopy}
Additional spectra covering the optical regime between 3700 and 10500 \AA\
were taken with the Low Resolution Spectrograph 2 (LRS2, 
\citealp{Chonis_etal_2016}) on the 10-meter {\sl Hobby-Eberly Telescope} 
\citep{Ramsey_etal_1998}. LRS2 is composed of two dual-arm spectrographs, 
LRS2-B and LRS2-R,  each having two spectral regions with a $\sim 100$ \AA\ 
overlap. The LRS2-B UV-arm extends from 3700 \AA\ to 4700 \AA, while the 
Orange arm covers the 4600$-$7000 \AA\ interval, with a resolving power of 
1900 and 1100, respectively. The Red arm of LRS2-R records spectra from 
6500 to 8420 \AA\ while the range 8180 to 10500 \AA\ is covered by the 
Far-Red arm, both having $R \sim 1800$ spectral resolving power. Both arms 
are fed by their own $12\arcsec \times 6\arcsec$ Integral Field Unit (IFU) 
that contains 280 densely packed fibers with lenslet coupling. The diameter 
of a single fiber/lenslet is $\sim 0\farcs{6}$ on sky. The fill factor of 
both IFUs is $\sim 98$\% which provides very good spatial sampling on the 
sky without the need for dithering (see \citealt{Chonis_etal_2016} for 
additional details). 

The reduction of the LRS2 IFU data was done with self-developed 
{IRAF}\footnote{{IRAF} is distributed by the National Optical Astronomy 
Observatories, which are operated by the Association of Universities for 
Research in Astronomy, Inc., under cooperative agreement with the National 
Science Foundation.} 
and Python scripts. Fiber-to-fiber transmission variations were corrected by 
observing at least one frame of blank sky during twilight and requiring a 
homogeneous, flat, output signal on the reconstructed image within the 
field of view. Wavelength calibration was performed by a combination of Hg and 
Cd spectral lamp exposures for LRS2-B and an FeAr spectral lamp for LRS2-R. 
For sky subtraction, the mean sky spectrum was constructed by 3$\sigma$-clipping 
the fibers having signal exceeding the median of all fibers, then computing the 
median combination of all remaining fibers. Flux calibration was completed 
based on nightly observations of spectrophotometric standard stars taken at 
approximately similar airmasses to the SN. Telluric lines were removed from the 
final spectra by using a mean telluric spectrum (constructed from multiple 
observations of telluric standard stars) scaled to the flux level of the actual 
SN spectrum. 

\subsubsection{SALT Optical Spectroscopy}
We observed SN\,2018gv with the Southern African Large Telescope (SALT) using 
the Robert Stobie Spectrograph (RSS; \citealp{Smith_etal_2006}) on 
2018-01-25.8 UT under Rutgers University program 2017-1-MLT-002 (PI: SWJ). 
We used the PG0900 grating and 1$\farcs$5 wide longslit with a typical 
spectral resolution $R =\lambda/\Delta \lambda \approx 1000$. 
Exposures were taken in four grating tiltpositions to cover the optical 
spectrum from 350 to 930 nm. The data werereduced using a custom pipeline 
based on standard Pyraf spectral reductionroutines and the PySALT package 
\citep{Crawford_etal_2010}. 

\subsubsection{ARC Optical Spectroscopy}
On 2018-03-22 UT we obtained one low-resolution spectrum with the Dual 
Imaging Spectrograph 
(DIS\footnote{\url{https://www.apo.nmsu.edu/arc35m/Instruments/DIS/}}, 
mounted on the 3.5 m Astrophysics Research Consortium (ARC) telescope at 
the Apache Point Observatory. The B400 and R300 gratings were used with 
central wavelengths of 4500 and 7500 \AA\ , respectively. The instrument 
was rotated to the parallactic angle and 3$\times$300 second exposures 
were obtained. The data were reduced using standard procedures and 
calibrated to a standard star obtained the same night using the PyDIS 
package \citep{pydis}. 

The $UBg'Vr'i'$ lightcurves and FLOYDS/LCO spectra were obtained as part of 
the Global Supernova Project. All photometry 
and spectroscopy will become available via WISeREP 
\footnote{https://wiserep2.weizmann.ac.il/} \citep{Yaron_Gal-Yam_2012}. 

\subsection{VLT Spectropolarimetry}
Spectropolarimetry of SN\,2018gv was conducted using the Focal Reducer and 
low dispersion Spectrograph (FORS2) on UT1 (Antu) of the ESO Very Large 
Telescope (VLT). Observations were carried out in the FORS2 Polarimetry 
with Multi-Object Spectroscopy (PMOS) mode \citep{Appenzeller_etal_1998} 
on January 18 2018 (epoch 1) and January 31 2018 (epoch 2), corresponding 
to $t=-13.6$ day and $t=-0.5$ day, respectively. 
Details of the VLT spectropolarimetry are available in Appendix~\ref{App_VLT}. 
The intensity-normalized Stokes parameters ($I,Q,U$) are binned 
in $\sim$25 \AA\ wide bins ($\sim$7.5 pixels) to further increase the 
signal-to-noise ratio (S/N). The observed degree of linear polarization 
($p_{\rm obs}$) and its position angle ($PA_{\rm obs}$) are given by: 
\begin{equation}
\begin{aligned}
p_{\rm obs} = \sqrt{Q^2 + U^2}, \\
PA_{\rm obs} = \frac{1}{2} {\rm arctan} \bigg{(} \frac{U}{Q} \bigg{)}. 
\end{aligned}
\label{Eqn_stokes0}
\end{equation}
The calculated $p_{\rm obs}$ is by definition a positive number and 
is therefore biased toward larger values than the true degree of 
polarization $p$. We correct the polarization bias following the 
equation given in \citet{Wang_etal_1997}: 
\begin{equation}
p = (p_{\rm obs} - \sigma_{p}^2 / p_{\rm obs}) \times h(p_{\rm obs} - \sigma_p), 
{\rm \ and \ } PA = PA_{\rm obs},  
\label{Eqn_pol_biascorr}
\end{equation}
where $\sigma_p$ gives the 1-$\sigma$ uncertainty in $p_{\rm obs}$, $h$ 
is the Heaviside step function. Calculation and bias correction of the 
polarization, as well as the estimation of associated uncertainties 
were performed by our own specially written software, following the 
prescriptions described by \citet{Patat_etal_2006_polerr} and the 
scheme presented by \citet{Maund_etal_2007_05bf}. 

\begin{figure}[!h]
\epsscale{0.8}
\plotone{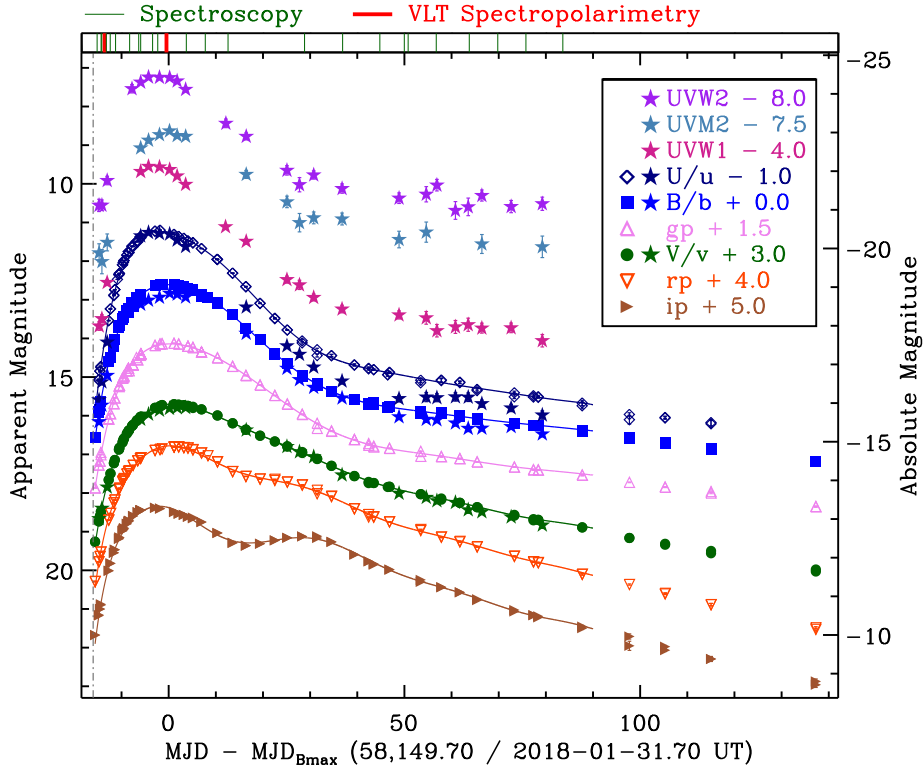}
\caption{The optical light curves of SN\,2018gv. The vertical dot-dashed 
line shows the time of discovery. Short vertical lines at the top of the 
panel mark the spectroscopic and spectropolarimetric observations. 
Solid curves present polynomial fits to the LCO photometry. 
\label{Fig_lc}
}
\end{figure}

\section{Light Curves of SN\,2018gv \label{section_lc}}
In Fig.~\ref{Fig_lc}, we show the $UBg'Vr'i'-$band light curves after 
correction for interstellar extinction. The $uvw2$, $uvm2$, $uvw1$, $u$, $b$, 
and $v$ photometry without the extinction corrections are also presented. 
The $UBg'Vr'i'-$band light curves were sampled during the period $t\approx-$15 
to $+$137 days relative to the $B-$band maximum. We conduct high-order 
polynomial fits to the light curves before day $+$110 and plot the fits 
between the first observation and day $+$90. Short lines on the top 
axis indicate the epochs of spectroscopy and spectropolarimetry.

We list the calibrated LCO $UBg'Vr'i'$ Photometry in Table~\ref{Table_phot}. 
All tabulated photometry was not corrected for the extinctions from the 
host and the Milky Way. Photometric parameters including the times of the 
light curve peaks, maximum brightness, the decline in the magnitude from 
the maximum light to the magnitude 15 days after, i.e. $\Delta m(15)$, 
which characterizes the width of the light curve, are reported in 
Table~\ref{Table_phot_para}. K-corrections were applied to our photometry 
when calculating $\Delta m(15)$. The presented values of the peak 
brightness have been corrected for the extinction from the Milky but not 
the host galaxy because we estimate little or no extinction from the host 
galaxy towards SN\,2018gv. A detailed analysis of the extinction will be 
provided in the following subsection. 

\subsection{Extinction\label{Section_extinction}}
The Galactic reddening towards the SN\,2018gv line of sight has been 
estimated as $E(B-V)^{MW}_{\rm 18gv} = 0.051$ mag using the NASA/IPAC 
NED Galactic Extinction Calculator adopting the $R_V = 3.1$ extinction 
law \citep{Cardelli_etal_1989} and the extinction map given by 
\citet{Schlafly_etal_2011}. We consider the host galaxy reddening to 
be low for the following reasons: 
1) the SN's position in the outskirts of the host galaxy NGC 2525; 
2) the evolution of the $B-V$ color of the SN is consistent with the 
Lira-Phillips relation at 30$-$90 days after the $B-$band maximum 
light (\citealp{Phillips_etal_1999}). 
Although an empirical relation between the 
amount of dust extinction and the strength of the absorption doublet 
of Na {\sc ID} 5890 and 5896 \AA\ has been proposed by 
\citet{Munari_etal_1997} and widely used, the validity of the 
application of the methodology has been questioned for use with 
low-resolution spectra \citep{Poznanski_etal_2011}. All spectroscopic 
observations discussed in this study were carried out in a 
low-resolution regime, and thus we have not considered extinction 
estimation based on Na {\sc ID} absorption features. 

A linear relationship between the $B-$band magnitude and the $B-V$ 
color evolution of Type Ia SNe within a few days to approximately a 
month relative to the $B-$band peak brightness has been found by 
\citet{Wang_etal_2003}. This color-magnitude (CMAG) relation provides 
a robust way (denoted as the `CMAGIC' method hereafter) to deduce the 
SN distance as well as the dust extinction from the host galaxy when 
the magnitude at the SN maximum light was also measured. Applying the 
`CMAGIC' method, we deduce the color excess of SN\,2018gv as 
$E^{\rm 18gv}_{BV} = 0.028 \pm 0.027$ mag, see Appendix~\ref{App_CMAGIC}. 
A sanity test of the method based on the photometry of SN\,2011fe 
\citep{Munari_etal_2013} is also provided therein. 
Note that the color-magnitude relation has also been applied to the 
entire set of sample I of Carnegie-Supernovae Program (CSP I)
\citep{Krisciunas_etal_2017}. For normal-bright Type Ia SNe like 
SN\,2018gv, the color-magnitude diagram is stable but there is a 
systematic shift for transitional and subluminous Type Ia SNe as 
predicted by models \citep{Hoeflich_etal_2017}. 

We also compare the $B-V$ color of SN\,2018gv at 30$-$90 days after 
the $B-$band maximum light to the color evolution described by the 
Lira-Phillips relation (\citealp{Phillips_etal_1999, Folatelli_etal_2010}, 
see Equations~\ref{Eqn_lira12}a and \ref{Eqn_lira12}b, respectively). 
\begin{equation}
\begin{aligned}
(B-V)_{0} = 0.725 - 0.0118 (t_V - 60) \  {\rm (a)}, \\
(B-V)_{0} = 0.732(0.006) - 0.0095(0.0005)(t_V - 55) \  {\rm (b)}; \\ 
\end{aligned}
\label{Eqn_lira12}
\end{equation}
The mean difference between the $B-V$ color and the Lira law given by 
Equation~\ref{Eqn_lira12}(a) is $E(B-V) = 0.036 \pm 0.018$ mag, which 
has a dispersion of 0.06 mag (see the lower left panel of 
Fig.~\ref{Fig_color}). This is in good agreement with the 
estimation based on the CMAGIC method. The 
mean difference between the SN\,2018gv observations and the Lira law 
fitted by \citet{Folatelli_etal_2010} gives $E(B-V) = 0.092 \pm 0.036$ mag. 
Considering a dispersion of 0.077 mag in the Lira law suggested by 
Equation~\ref{Eqn_lira12}(b), we conclude that the extinction estimated 
by the CMAGIC method agrees with that interpreted from the Lira law. 

\begin{figure}[!h]
\epsscale{0.6}
\plotone{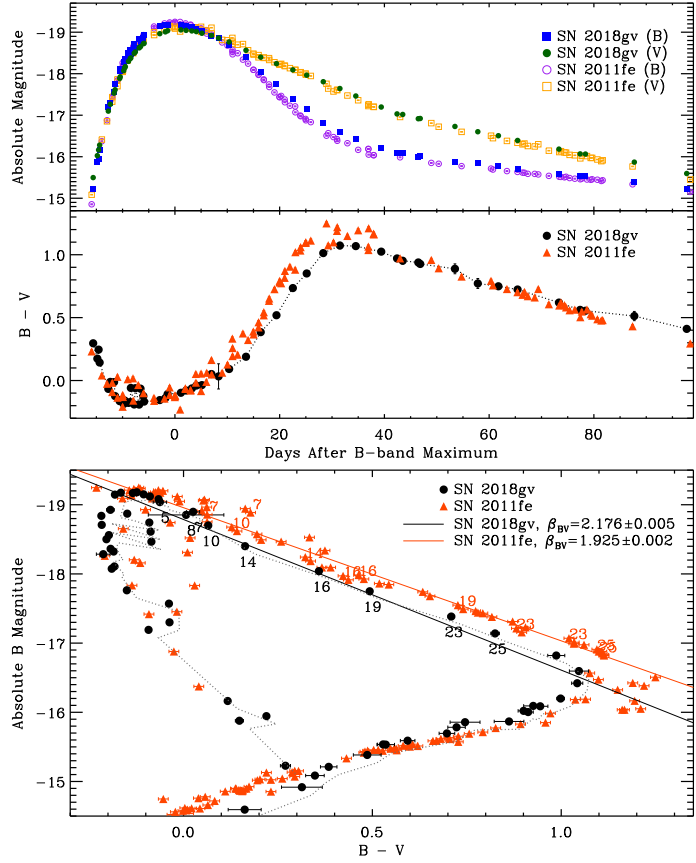}
\caption{The upper panel presents the $B$ and $V-$band light curves of 
SN\,2018gv and SN\,2011fe. The middle panel shows the color vs. rest-frame 
day plot. The dotted black line shows the $B-V$ color without the extinction 
correction of the host galaxy. In the bottom panel, the black circles and 
the orange triangles show the color-magnitude diagrams of SN\,2018gv and 
SN\,2011fe, respectively. The dotted black line is shown for comparison 
which represents the color-magnitude plot of SN\,2018gv before correcting 
for the host extinction. The epochs within the linear regions are labeled by 
corresponding rest-frame days. The black and the orange solid lines are the 
linear fit to the linear regions of SN\,2018gv and SN\,2011fe, respectively. 
\label{Fig_cmagic}
}
\end{figure}

The final estimated values of the Galactic and the host galaxy extinction in 
different bandpasses are listed in Table~\ref{Table_phot_para}. Here we have 
assumed that the dust in the host of SN\,2018gv has similar properties as 
Galactic dust with $R_V = 3.1$. In this study, we only apply the 
extinction corrections for both the Galactic and the host component to 
SN\,2018gv photometry when estimating the UV-optical pseudo-bolometric 
luminosities. This will be discussed in Section~\ref{section_bolo}. 
Figure~\ref{Fig_cmagic} illustrates the CMAGIC method applied to the $B-$band with 
$B-V$ color of SN\,2018gv after correcting for the Galactic extinction. Similar 
diagrams derived from the well-sampled light curves of SN\,2011fe are presented. 

\subsection{Color Curves}
\begin{figure}[!h]
\epsscale{1.0}
\plotone{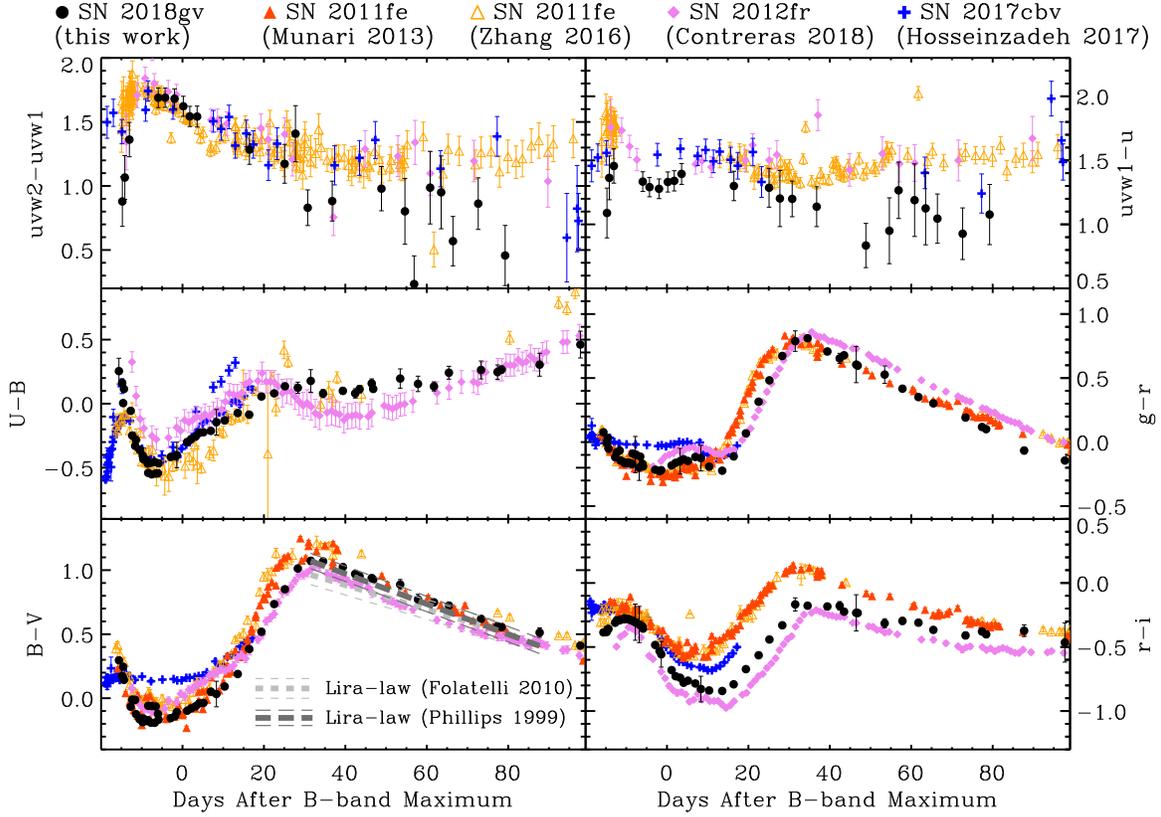}
\caption{The Galactic and host reddening-corrected $uvw2-uvw1$, $uvw1-u$, 
$U-B$, $B-V$, $g'-r'$, and $r'-i'$ color curves of SN\,2018gv compared with 
the Galactic reddening corrected color curves of SN\,2011fe, SN\,2012fr, 
and SN\,2017cbv. 
\label{Fig_color}
}
\end{figure}

Figure~\ref{Fig_color} shows the color evolution of SN\,2018gv ($uvw2-uvw1$, 
$uvw1-u$, $U-B$, $g'-r'$, $B-V$, and $r'-i'$), corrected for the 
Galactic and host reddening derived in Section~\ref{Section_extinction}. 
The color curves of SN\,2018gv are overplotted with those of the Type Ia 
SNe 2011fe \citep{Munari_etal_2013, Zhang_etal_2016}, 2012fr 
\citep{Contreras_etal_2018}, and 2017cbv \citep{Hosseinzadeh_etal_2017}, 
corrected for reddening in both the Milky Way and the host galaxies. 

The $uvw2-uvw1$ and $uvw1-u$ colors of SN\,2018gv appear to have similar 
trends to those of the selected SNe, except that SN\,2018gv has bluer 
colors than the comparison SNe before $t\approx -10$ days, i.e. by 
$\sim$0.3 index. 
The early $U-B$ color of SN\,2018gv shows a rapid decline and reached a 
minimum of $\sim-0.5$ mag at $t\approx-7$ days and then became redder 
in a linear fashion in magnitude space until t$\sim25$ days. The color 
curve hereafter entered a plateau phase and slowly turns red over time. 
SN\,2012fr, exhibited a similar behavior, whereas SN\,2017cbv and 
SN\,2011fe became redder until $t \approx -14$ days, and then started 
to become bluer and reached the turning point at $t \approx -7$ days. 
The $B-V$ and the $g'-r'$ colors of the selected SNe show a similar 
evolution, except for SN\,2017cbv, which displayed constant colors from 
the earliest epoch to t$\approx +7$ days. The $r'-i'$ color curves of 
SN\,2018gv and other selected SNe have very similar shapes but the 
rising phases to the secondary maximum display offsets of up to 10 days. 

\subsection{Early Light Curves \label{Section_earlylc}}
If SN 2018gv exploded as an ideal, expanding fireball with constant 
temperature and velocity, its luminosity should scale as the surface area 
of the expanding fireball, and therefore the SN flux ($f$) should increase 
quadratically with time \citep{Riess_etal_1999}. This $f \propto t^2$ 
relation reasonably describes the composite light curves collected from 
surveys (i.e. \citealp{Riess_etal_1999, Goldhaber_etal_2001, 
Conley_etal_2006, Garg_etal_2007, Hayden_etal_2010_rise, 
Ganeshalingam_etal_2011, Firth_etal_2015}) and some of the very well 
sampled individual Type Ia SNe (i.e. SN\,2011fe, \citealp{Nugent_etal_2011}). 
A more fundamental explanation of the $f \propto t^2$ relation is provided 
by \citet{Arnett_1982} which considers radioactive heating and photon 
diffusion. Nonetheless, such an oversimplified expanding fireball may not 
be sufficient to explain the non-uniform and time-variant rising of Type Ia 
SNe in the first $\sim$3 days (for instance, SN\,2013dy, 
\citealp{Zheng_etal_2013} and iPTF16abc, \citealp{Miller_etal_2018}). 

We model the early $g'-$band flux as a function of time from SN\,2018gv as 
a power law: 
\begin{equation}
f(t) \propto (t-t_0)^{n}, 
  \label{Eqn_expmodel}
\end{equation}
where $t_0$ gives the time of first light, and $n$ denotes the index 
of the power law. The best fit to the first 6.5 days after the estimated 
explosion gives $n=2.43\pm0.48$ and a rise time 
$t_{\rm rise} = t_{Bmax} - t_0 = 18.51\pm0.92$ day, this is consistent 
with the fit using the first 6.0 days light curve, i.e. $n=2.59\pm0.56$ 
and $t_{\rm rise} = 18.79\pm1.07$ day. The fit adopting the photometry 
within the first 7.0 days gives $n=1.89\pm0.18$ and 
$t_{\rm rise} = 17.53\pm0.38$ days. One can see that the power-law index 
and the estimated rising time are broadly consistent with the mean values 
deduced by previous studies based on relatively arbitrary choices of the 
time range. Stringent constraints on the early photometric evolution of 
SN\,2018gv, however, cannot be placed without a complete light curve 
coverage of the early phases. 

For comparison, the model was also fitted to the early $g-$band flux 
of SN\,2011fe, which yields one of the most comprehensive observed 
normal Type Ia SNe with photometry at $\sim$1 day after the explosion. 
In addition to adopting the $g'-$band photometry from 
\citet{Nugent_etal_2011}, we also apply the transformations of 
\citet{Jordi_etal_2006} to convert the photometry by 
\citet{Zhang_etal_2016} from the $UBVRI$ magnitude system to the Sloan 
$ugriz$ system. A better sampled early $g-$band flux curve of 
SN\,2011fe was then obtained by combining the two sources. The 
$f\propto t^n$ power law was fitted to the early (i.e. within 
4 days of the estimated time of explosion) $g-$band flux. We obtain 
$n=2.20\pm0.09$, which is marginally consistent with $n=2.01\pm0.01$ 
derived by \citet{Nugent_etal_2011}. The same conclusion holds for 
the data within the first five days, e.g. $n=2.03\pm0.14$. 

\begin{figure}[!h]
\epsscale{1.0}
\plotone{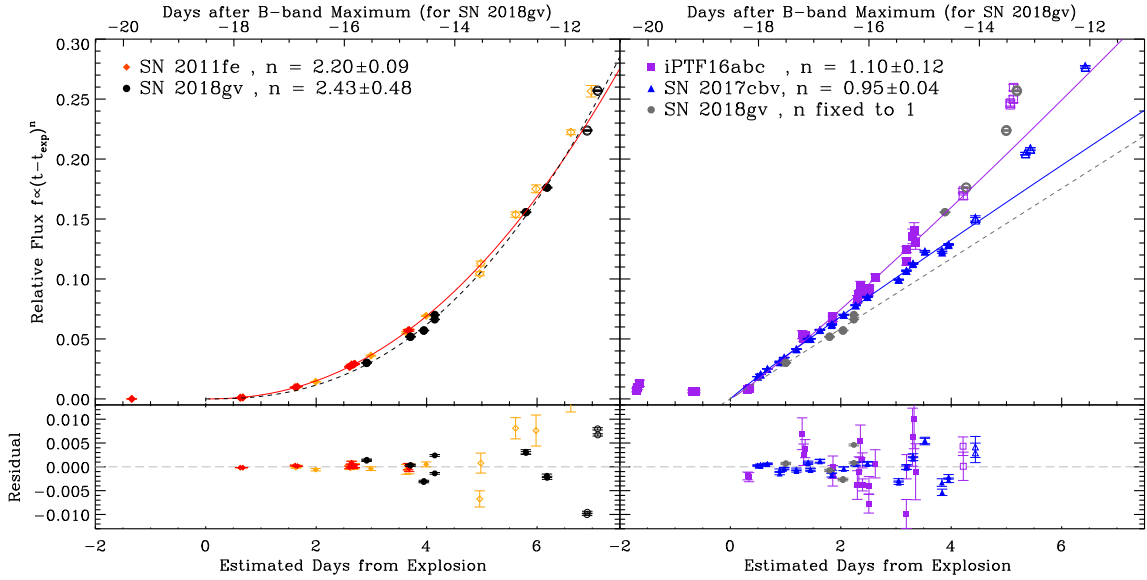}
\caption{The best-fit $f\propto t^{n}$ model to describe the early 
$g'-$band flux from SN\,2018gv compared to the early flux evolution of other 
SNe that have well-sampled early light curves. In the upper-left panel, 
the black-dashed line shows the fitting of SN\,2018gv which allows $n$ 
to vary. The red-solid line fits the early $g-$band flux of SN\,2011fe. 
The orange diamonds indicate the $B-$band flux of SN\,2011fe. In the 
upper-right panel, the gray-dashed line indicates the linear fit assuming 
the early flux evolution of SN\,2018gv is similar to SN\,2017cbv and 
SN\,2018gv, a blue bump has been observed within 5 days after their 
estimated time of explosion. The residuals are shown in the bottom panels. 
\label{Fig_earlylc}
}
\end{figure}

In a few cases, the early flux from the SNe exhibits significant 
deviations from the $f\propto t^2$ model. For example, the blue bump 
in the $U$, $B$, and $g'-$band light curves of SN\,2017cbv during the 
first five days can be explained by interactions between the ejecta 
and the companion or immediate circumstellar material 
\citep{Hosseinzadeh_etal_2017}. The flux evolution of iPTF16abc at the 
first $\sim$four days can be modeled by the power law with $n=0.98$ 
\citep{Miller_etal_2018}. The photometric and spectroscopic behaviors 
of iPTF16abc can be best explained by strong $^{56}$Ni mixing in the 
SN ejecta. In these two cases, the early flux can be well characterized 
by $n\sim1$. 

To test whether SN\,2018gv has such a nature, we fit the $f\propto t^n$ 
law to its early flux by fixing $n=1$. The fluxes during the first four 
days after the estimated time of the explosion were adopted. After day 4, 
the fluxes of SN\,2018gv display large deviations from the linear model. 
For comparison, the $f\propto t^n$ modeling to the fluxes of iPTF16abc 
and SN\,2017cbv in similar phase spanning gives $n=1.10\pm0.12$ and 
$n=0.95\pm0.04$, respectively, both coincide with the linear model. We 
conclude that the early flux of SN\,2018gv does not favor an $n\sim$1 
index due to the large deviations from the linear model after day 4. 
Figure~\ref{Fig_earlylc} summarizes our investigation of the early flux 
evolution of SN\,2018gv and several other SNe discussed above. In all 
cases the fits were were forced to go through the origin, which 
determines the time of explosion. Filled symbols represent the data 
point used in each case. Data points shown by the corresponding open 
symbols were not used in the fitting. The left panel presents the 
fitting to the early flux curve of SN\,2018gv which allows $n$ to vary. 
The right panel tests the case of $n$ fixed to 1. The best fits for 
SN\,2018gv and other SNe are shown by dashed lines and solid lines, 
respectively. Estimates based on different assumptions for the 
light-curve shape at early phases yield different times of explosion. 
We concede that the relatively sparsely sampled early light curve is 
not sufficient to constrain the first light of SN\,2018gv. Overall, 
based on the test and comparisons above, we consider that the early 
flux evolution of SN\,2018gv more resembles the average behavior of 
normal Type Ia SNe. 

\section{Spectroscopy\label{section_spec}}
\begin{figure}[!h]
\epsscale{0.73}
\plotone{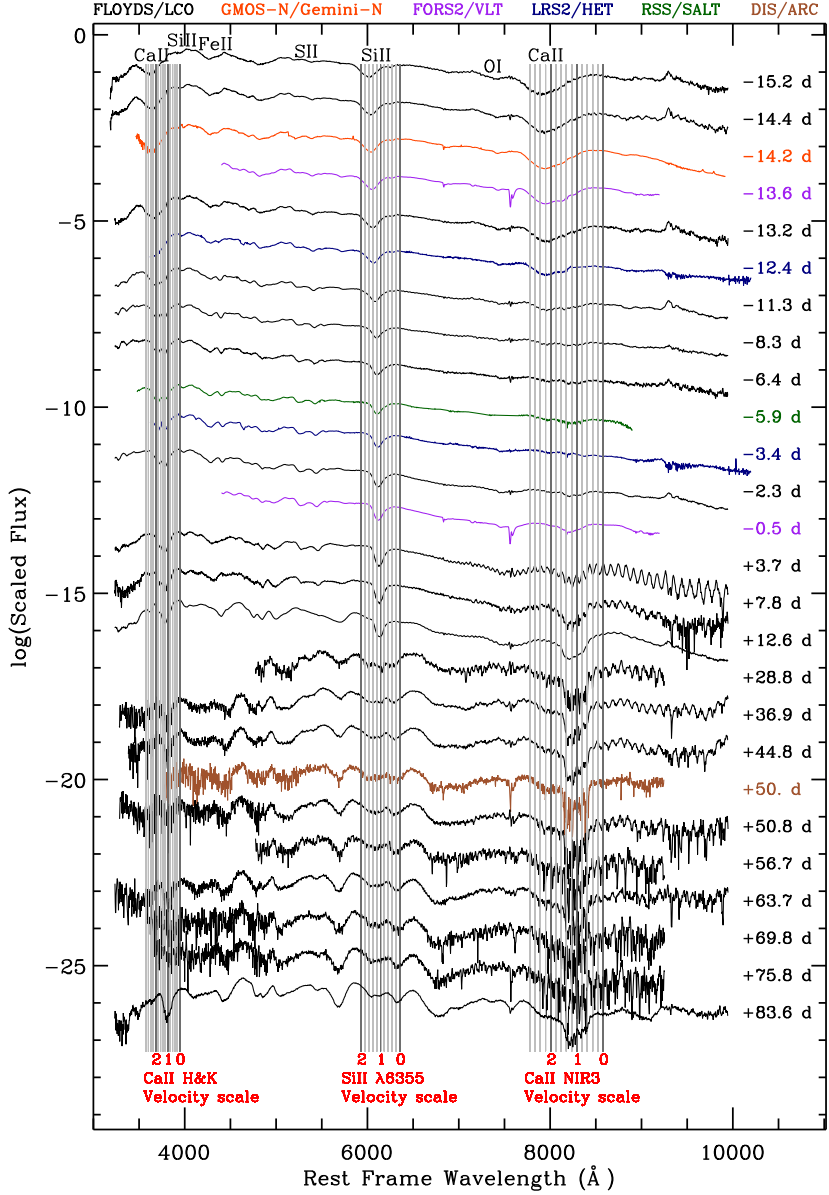}
\caption{Spectral time series of SN\,2018gv 
(solid curves, phase is label on right). The wavelengths corresponding 
to 0 km s$^{-1}$, $-$10,000 km s$^{-1}$ and $-$20,000 km s$^{-1}$ around 
the features indicated by the text are shown as thick gray lines, with 
2000 km s$^{-1}$ intervals denoted by thin gray lines. For the purpose 
of presentation, all spectra were binned to 3 \AA .
\label{Fig_spec}
}
\end{figure}

Figure~\ref{Fig_spec} presents the spectral sequence of SN\,2018gv. 
A total number of 26 optical low-resolution spectra spanning from 
$t\sim-15.2$ to $+83.6$ days relative to the $B-$band maximum light 
were obtained. All spectra were corrected for the redshift of the host 
galaxy and smoothed by rebinning the data to 3 \AA. The wavelength 
scale was corrected to the rest-frame using the host galaxy recessional 
velocity (1582 km s$^{-1}$) as described above. 

\begin{figure}[!h]
\epsscale{1.0}
\plotone{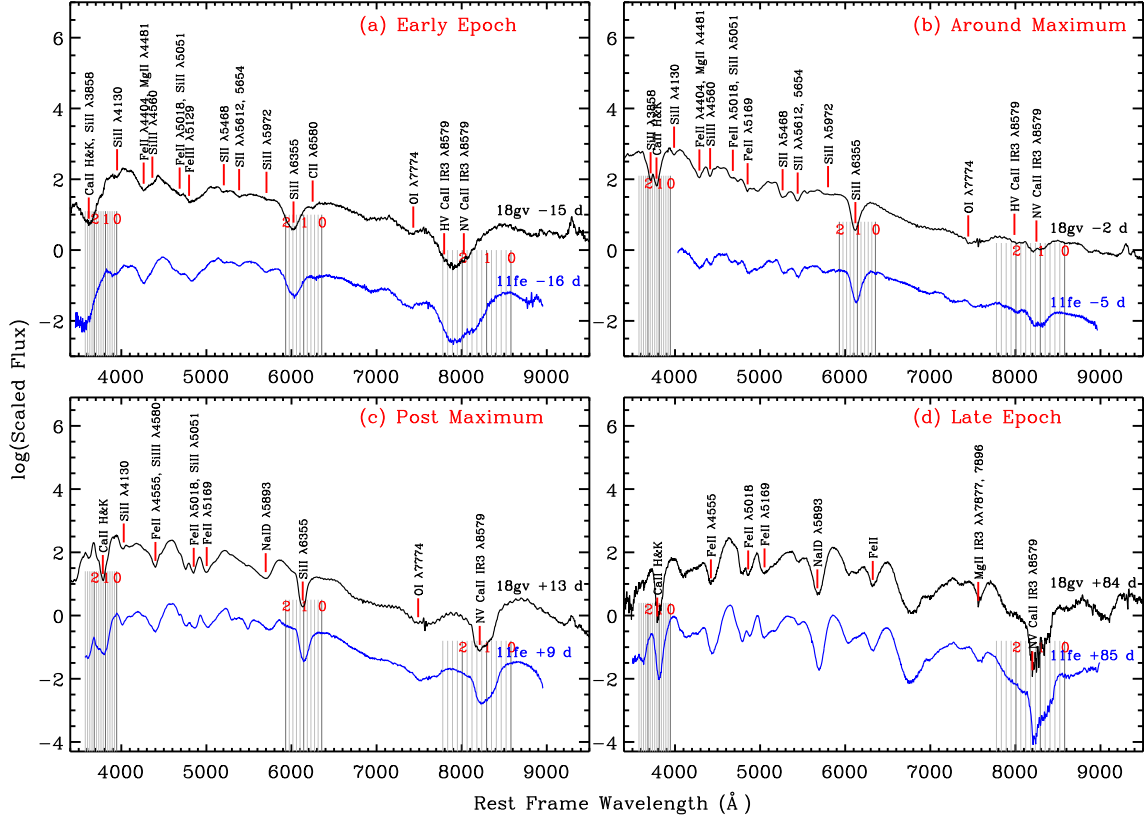}
\caption{Spectra of SN\,2018gv (black) at different epochs 
compared with the spectra of SN\,2011fe (blue) at similar phases 
\citep{Zhang_etal_2016}. The wavelengths corresponding 
to 0 km s$^{-1}$, $-$10,000 km s$^{-1}$ and $-$20,000 km s$^{-1}$ around 
the features indicated by the text are shown as thick gray lines, with 
2000 km s$^{-1}$ intervals denoted by thin gray lines. All spectra 
have been corrected for the redshift of the host galaxy. 
Several prominent lines at different epochs are labeled. The line 
identifications adopted here are obtained from \citet{Branch_etal_2005, 
Branch_etal_2006}. For the purpose of presentation, all spectra were 
binned to 3 \AA, shifted arbitrarily, and presented in logarithmic scale. 
\label{Fig_spec_compare}
}
\end{figure}

Figure~\ref{Fig_spec_compare}a compares the earliest spectrum of 
SN\,2018gv at day $-$15.2 with the spectrum of SN\,2011fe at day $-$16. 
The earliest spectrum of SN\,2018gv exhibits several prominent broad and 
blue-shifted absorption features near 3630 \AA\ due to \ion{Ca}{2} H\&K, 
(rest-frame wavelength $\lambda_0 \sim$3969, 3934 \AA); the 
characteristic `W'-shaped \ion{S}{2} lines of the Type Ia SN spectra 
before and around the peak ($\lambda_0 \sim$5454, 5640 \AA); the 
distinctive strong absorption features around 6020 \AA\ due to \ion{Si}{2} 
($\lambda_0 \sim 6355$ \AA, denoted as \ion{Si}{2} 
$\lambda$6355 hereafter) and around 7890 \AA\ due to the \ion{Ca}{2} NIR 
triplet ($\lambda_0 \sim 8579$ \AA, denoted as \ion{Ca}{2} NIR3 hereafter). 
The \ion{C}{2} $\lambda$6580 and the \ion{O}{1} $\lambda$7774 are both 
identified in the earliest spectra. Oxygen in the ejecta of Type Ia SNe 
can be unburned fuel or a product of carbon burning. 
Apart from an absence of a shoulder on the red wing of the \ion{Ca}{2} 
NIR3 in the SN\,2018gv spectrum, we suggest that the spectral 
features and their strength in the earliest spectrum of 
SN\,2018gv exhibit considerable similarities to those of the SN\,2011fe. 

By $t\sim-$2 day (shown in Fig.~\ref{Fig_spec_compare}b), the width of 
the broad absorption feature associated with \ion{Si}{2} $\lambda$6355 
develops into a narrower profile, while the adjacent \ion{C}{2} 
$\lambda$6580 feature is no longer distinguishable in our observations 
of SN\,2018gv. \ion{C}{2} $\lambda$6580 is marginally detected in the 
comparison spectrum of SN\,2011fe at an earlier epoch of $-$5 day. An 
asymmetric and broad absorption profile is identified in the wavelength 
range of \ion{Ca}{2} NIR3, indicating the existence of a high-velocity 
(HV) component and a normal-velocity (NV) component of the \ion{Ca}{2} 
feature. These two components were blended in the earliest spectra and 
become shallower. The decomposition and a detailed study of the 
\ion{Ca}{2} NIR3 features of SN\,2018gv will be presented in the 
following subsection. The HV and NV components of the \ion{Ca}{2} NIR3 
become shallower than those observed at $-15$ day. 
At $+$13 (Fig.~\ref{Fig_spec_compare}c) and $+$84 day 
(Fig.~\ref{Fig_spec_compare}d), significant similarities between the 
spectral evolution of SN\,2018gv and SN\,2011fe can be identified. 
In the following subsections, we detail the spectral evolution of 
SN\,2018gv within the first $\sim$100 days. 

\subsection{Evolution of the Features Around \ion{Si}{2} $\lambda$6355 and \ion{Ca}{2} NIR Triplet}
The prominent \ion{Si}{2} $\lambda$6355 absorption feature in the early 
spectra suggests the spectral evolution of SN\,2018gv resembles other 
normal Type Ia SNe. The highest velocity measured across the \ion{Si}{2}
$\lambda$6355 line is $\sim-$16,000 km s$^{-1}$ and a blue wing out to about 22,000 km s$^{-1}$ at t$\sim-$15.2 day. The 
\ion{Si}{2} $\lambda$6355 velocity for SN\,2018gv measured from the spectra 
at $-$0.5 day is found to be $\sim-$11,000 km s$^{-1}$. This is consistent 
with the interpolation of the exponential fitting to the time-evolution 
of the velocity profile, i.e. $-$10,870 km s$^{-1}$. These measurements 
suggest that SN\,2018gv should be classified as a normal-velocity (NV) 
Type Ia SN in the scheme of \citet{Wang_etal_2009_HV}. The expansion 
velocity gradient calculated directly from the \ion{Si}{2} $\lambda$6355 
at day $-0.5$ gives $\dot{v} = 36.6 \pm 6.4$ km s$^{-1}$ day$^{-1}$. The 
calculation is also consistent with the velocity gradient measured by 
interpolating the exponential fitting to the same velocities during the 
period from $t = 0$ to $+10$ days, i.e. $\dot{v} = 33.3$ km s$^{-1}$ 
day$^{-1}$. This is below the threshold of the high-velocity gradient 
subclass in the classification scheme of \citet{Benetti_etal_2005}, i.e. 
$\dot{v} \sim 70$ km s$^{-1}$ day$^{-1}$. Therefore, we suggest that 
SN\,2018gv belongs to the low-velocity gradient group. 

\begin{figure}[!h]
\epsscale{1.0}
\plotone{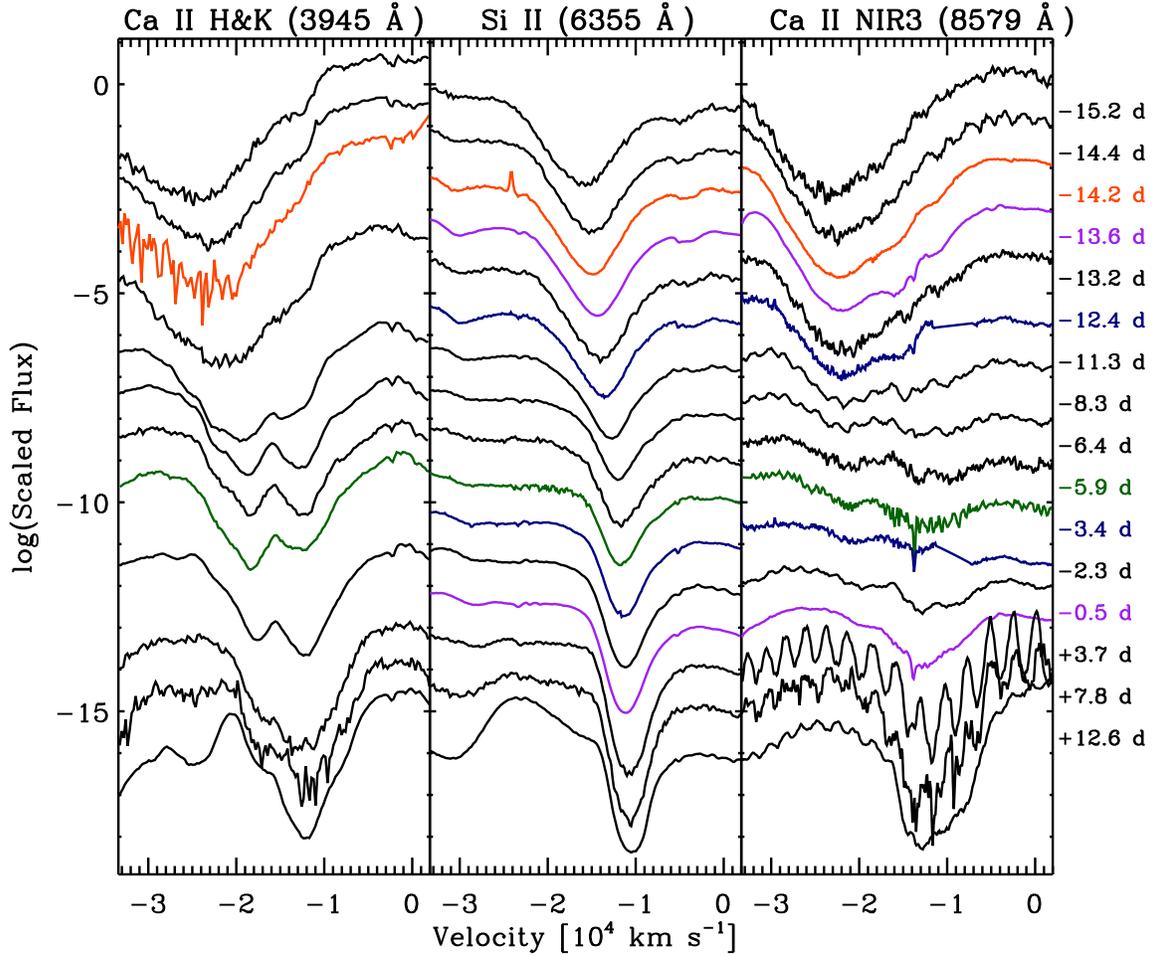}
\caption{Evolution of \ion{Ca}{2} H\&K (left panel), \ion{Si}{2} $\lambda$6355 
(middle panel), and \ion{Ca}{2} NIR3 (right panel) of SN\,2018gv in velocity 
space. The semi-regular fluctuations in some of the \ion{Ca}{2} NIR3 profiles 
are caused by fringing in the detectors. 
\label{Fig_line_region}
}
\end{figure}

Figure~\ref{Fig_line_region} provides the detailed evolution of the \ion{Ca}{2} 
H\&K, \ion{Si}{2} $\lambda$6355, and \ion{Ca}{2} NIR3. HV features (HVF) are 
ubiquitous in the \ion{Ca}{2} NIR3 in early phases of Type Ia SNe 
\citep{Mazzali_etal_2005_hv}. We find no sign of an HV component of \ion{Si}{2} 
$\lambda$6355 in the spectral time-sequence of SN\,2018gv. The broad and 
asymmetric \ion{Ca}{2} NIR3 profiles indicate multiple velocity components 
associated with \ion{Ca}{2}. The \ion{Ca}{2} H\&K line may have similar HV 
components, but they appear to overlap with the \ion{Si}{2} $\lambda$3858 
line at early phases. 

In addition to displaying an absorption profile around the `photospheric' 
velocity, some elements show a HVF which would require noticeably higher 
velocities (typically a few thousand km s$^{-1}$ above the photospheric 
velocity). These absorption lines can be attributed to the materials above 
the SN photosphere, and has been interpreted as the exponentially declining 
abundances of Si and other intermediate mass elements, a signature of 
detonations in C/O rich material \citep{Quimby_etal_2007}. 
The Doppler-broadened profiles at different velocities 
often overlap with each other as well as the continuum profile. In order 
to mitigate the line blending and identify different line profiles, we 
adopt multiple-component Gaussian functions to fit the \ion{Si}{2} 
$\lambda$6355 and the \ion{Ca}{2} NIR3, separately. 
Note that the shape of the absorption feature is dependent on the structure 
and the velocity distribution of the absorbing material, therefore the 
profile is intrinsically non-Gaussian and can be complex 
\citep{Mulligan_Wheeler_2018, Mulligan_etal_2019}. 
One should be cautious about the inferred line properties based on 
multiple-component Gaussian fitting. 

First, following \citet{Childress_etal_2014_hv}, we assign the blue and red 
ends of the feature of interest by visually inspecting the data; second, we 
identify a line segment as the `pseudo-continuum' by connecting the blue 
and the red end; third, we subtract the flux of the segment profile from 
the feature of interest; finally, we fit the `pseudo-continuum' removed 
regions with the multiple-component Gaussian function: 
\begin{equation}
f_c(\lambda) = \sum_{i=1}^n A_{i} \  {\rm exp} 
\bigg{(} - \frac{(\lambda_{i} - \lambda_{i}^{\rm rest})^2}{2 \sigma_{i}^2} \bigg{)}. 
\label{Eqn_gaussian}
\end{equation}

\begin{figure}[!h]
\epsscale{1.17}
\plotone{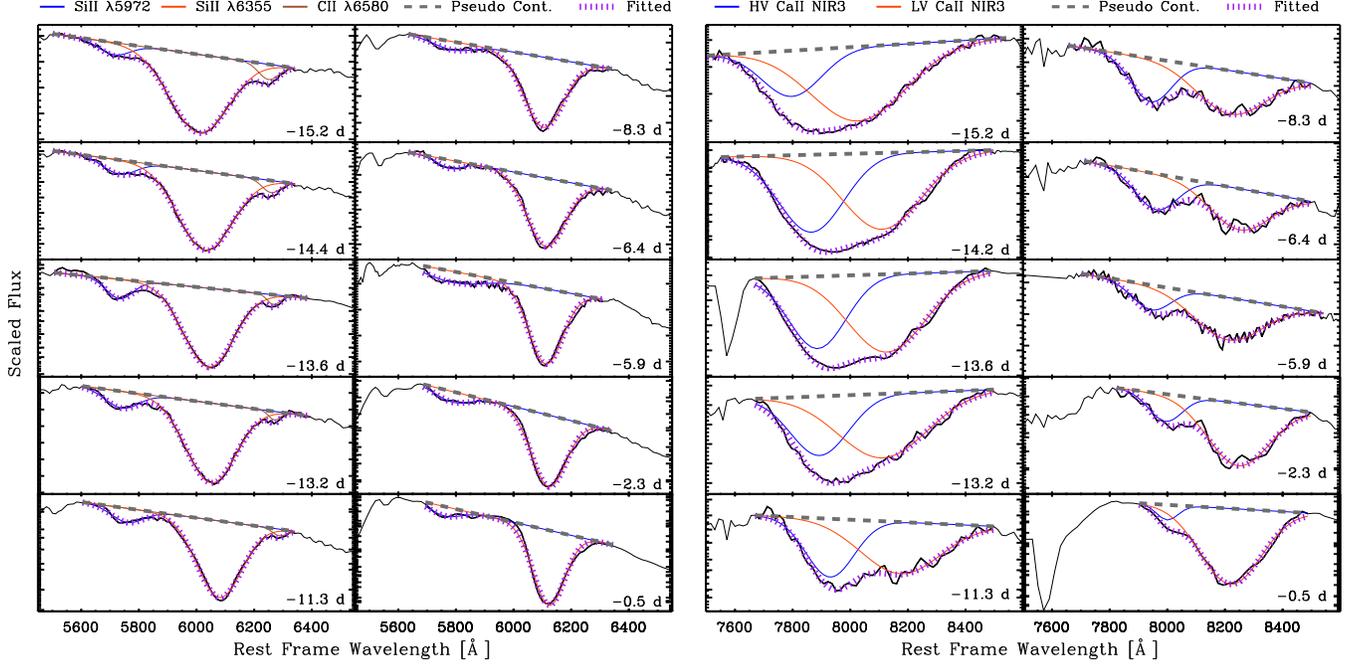}
\caption{Gaussian fit to the early-time spectra of SN\,2018gv. The left panel 
shows the fit to the region dominated by the \ion{Si}{2} $\lambda$6355 feature. 
The fitting shown in the leftmost subpanels has taken into account the 
\ion{Si}{2} $\lambda$5972, \ion{Si}{2} $\lambda$6355, and \ion{C}{2} 
$\lambda$6580 features, and the remaining five subpanels show the fitting that does
not include \ion{C}{2} $\lambda$6580. The right panel presents the fit for the 
region covering both the HVF and photospheric velocity \ion{Ca}{2} NIR3 features. 
\label{Fig_fit2gauss}
}
\end{figure}

\begin{figure}[!h]
\epsscale{0.8}
\plotone{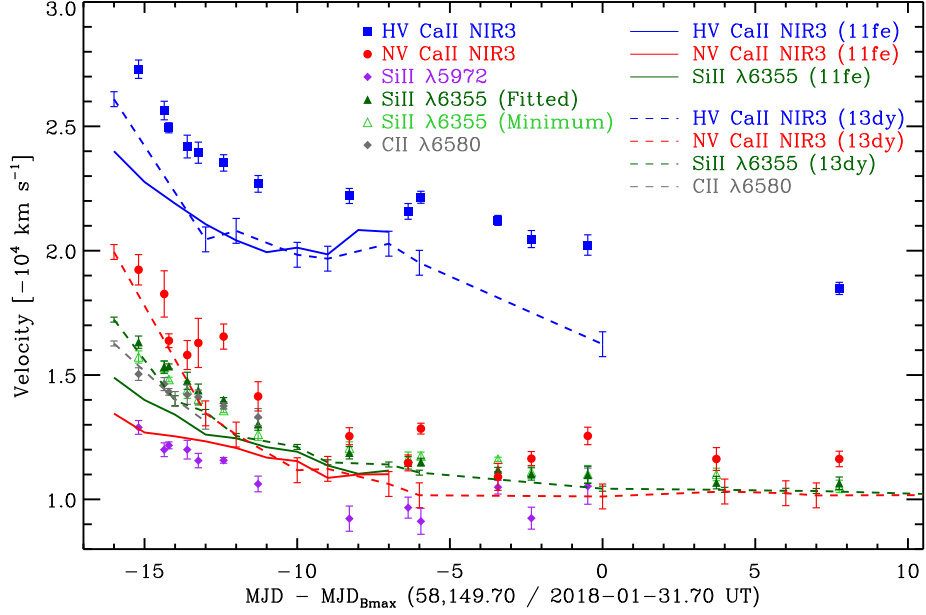}
\caption{Evolution of the expansion velocity of SN\,2018gv in early 
phases measured from different spectral features including \ion{Si}{2} 
$\lambda$5972/6355, \ion{C}{2} $\lambda$6580, and the HVF together with 
the photospheric velocity features of the \ion{Ca}{2} NIR3. The central 
wavelength of each absorption component was determined through the 
multiple-component Gaussian fitting procedure. The local minima of 
\ion{Si}{2} $\lambda$6355 were also overplotted and examined to be 
consistent with the central wavelength of the same feature obtained by 
the fitting process. The velocity evolution of the line absorptions for 
SN\,2011fe \citep{Zhao_etal_2015} and SN\,2013dy at similar phases 
\citep{Zhai_etal_2016} is shown for comparison. 
\label{Fig_velo}
}
\end{figure}

The fitted parameters are the center wavelength in the rest-frame ($\lambda_i$), 
pseudo-equivalent width ($\sigma_i$), and absorption depth ($A_i$) of each 
component. A three-component Gaussian function is used to fit the absorption 
features of the \ion{Si}{2} $\lambda$5972, \ion{Si}{2} $\lambda$6355, and 
\ion{C}{2} $\lambda$6580 line complex before $t\sim-11$ day. 
A fourth-component Gaussian function did not find any signature of 
HV \ion{Si}{2} $\lambda$6355. The unburned-carbon 
lines are typically singly ionized \citep{Tanaka_etal_2008} and manifest 
themselves at the earliest spectra of Type Ia SNe as weak and time-evolving 
\ion{C}{2} $\lambda$6580 absorption lines (for example, see recent studies 
of \citealp{Parrent_etal_2011, Thomas_etal_2011, Folatelli_etal_2012, 
Silverman_etal_2012_IV}). The unburned carbon and oxygen would be accelerated 
with the expanding layers and trace the ejecta. The \ion{C}{2} $\lambda$6580 
line in SN\,2018gv was as strong as the \ion{Si}{2} $\lambda$5972 in the 
earliest spectra around two weeks before the $B-$band light curve peak. It 
faded rapidly and became indiscernible in the spectra obtained after 
approximately one week before the $B-$band maximum light, suggesting it is 
mostly concentrated in the outer part of the ejecta. Therefore, the 
three-component Gaussian function fitting was only applied before $t\sim-11$ 
day. The two-component Gaussian fitting procedure has been carried out 
since $t\sim-11$ day to fit the \ion{Si}{2} $\lambda$5972 and \ion{Si}{2} 
$\lambda$6355 simultaneously. The HV and the photospheric velocity component 
of the \ion{Ca}{2} NIR3 features around the rest-frame wavelength 
$\lambda$8579 \AA\ are also fitted with a two-component Gaussian function. 
The HV feature was prominent at early phases and faded to almost unseen 
after the $B-$maximum. 
The fitting procedure together with the temporal evolution of these two 
line regions is illustrated in Fig.~\ref{Fig_fit2gauss}. The velocity 
evolution of SN\,2018gv in early phases is presented in Fig.~\ref{Fig_velo}. 

Starting from about two weeks after the $B-$maximum, as the photosphere 
recedes more deeply into the ejecta, lines of intermediate-mass elements 
(9$\leqslant$Z$\leqslant$20) become significantly weaker while the iron-group 
elements (21$\leqslant$Z$\leqslant$30) like \ion{Fe}{2} start to dominate the 
spectra. The \ion{Si}{2} $\lambda$6355 absorption feature remains distinct at 
$v=$10,700 km s$^{-1}$ on day $+$12.6 and has become difficult to identify in 
the spectra on $+$28.8 day and $+$36.9 day. 

\begin{figure}[!h]
\epsscale{1.0}
\plotone{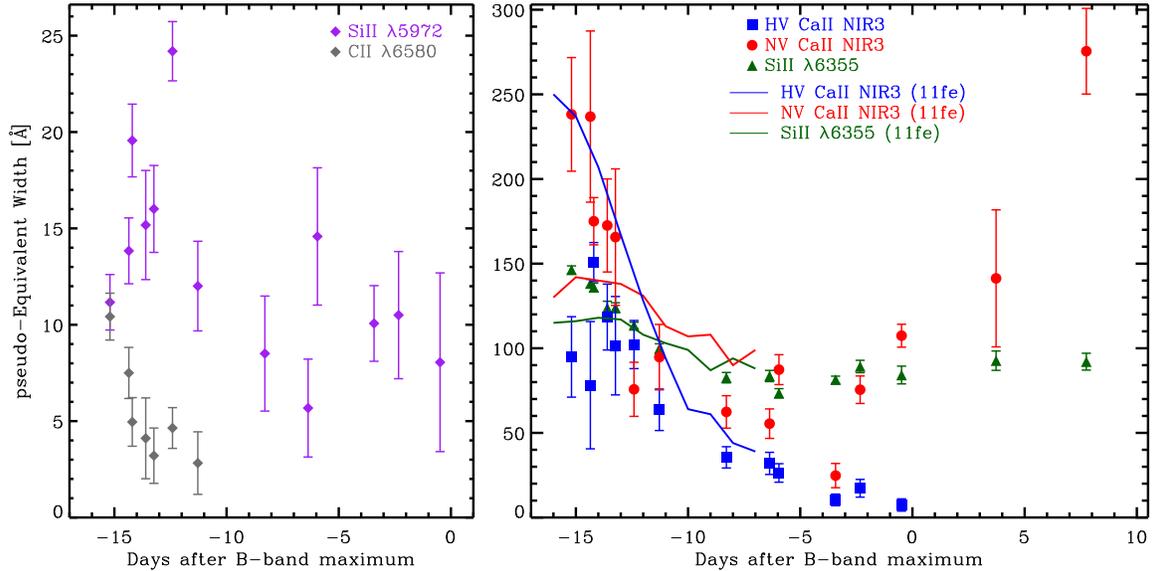}
\caption{Evolution of the pseudo-equivalent width of SN\,2018gv measured 
for the absorption lines \ion{Si}{2} $\lambda$5972, \ion{C}{2} $\lambda$6580 
(left panel), \ion{Si}{2} $\lambda$6355, HVF and photospheric velocity 
\ion{Ca}{2} NIR3 (right panel). In the right panel we also compare the 
measurement of SN\,2018gv with SN\,2011fe. 
\label{Fig_pew}
}
\end{figure}

We measure the pesudo-equivalent width ($pEW$) of the absorptions of the 
\ion{Si}{2} $\lambda$5972, \ion{Si}{2} $\lambda$6355, \ion{C}{2} $\lambda$6580, 
the HV and the NV \ion{Ca}{2} NIR3 features and analyze their time-evolutions. 
We also compare the results with those from SN\,2011fe as shown in 
Fig.~\ref{Fig_pew}. According to Fig.~\ref{Fig_fit2gauss}, these features 
are all embedded in a well-determined pseudo-continuum. We calculate the 
$pEW$ (e.g., \citealp{Garavini_etal_2007, Silverman_etal_2012_II}) defined as 
\begin{equation}
pEW = 
\sum_{i=1}^n \Delta \lambda_{i} \bigg{(} \frac{f_c(\lambda_i) - f(\lambda_i)}{f_c(\lambda_i)} \bigg{)}, 
\label{Eqn_pew}
\end{equation}
where $\lambda_i$ denote the wavelengths of each one of the total $N$ 
resolution elements in the pseudo-spectrum ranging from the blue endpoint to 
the red endpoint, $\Delta \lambda_i$ gives the width of the $i$th resolution 
element, $f(\lambda_i)$ represents the spectral flux at $\lambda_i$ and 
$f_c(\lambda_i)$ is the flux of the pseudo-continuum at $\lambda_i$. The 
$1\sigma$ uncertainty of the $pEW$ was derived by error propagation of the 
uncertainty in the measured flux at each resolution element. 

In Fig.~\ref{Fig_pew}, the $pEW$ of the \ion{Si}{2} $\lambda$6355 reached 
a minimum around 73 \AA\ at $\sim-$6 day. The $pEW$ of the NV component 
of \ion{Ca}{2} NIR3 reached a minimum of $\sim$25 \AA\ at a similar epoch 
around $-$6 day to $-$2day and became larger until approximately one week 
after the $B-$band maximum light. The evolution of the $pEW$ of SN\,2011fe 
adopted from \citet{Zhao_etal_2015} is shown for comparison. The 
absorption strength of the HV \ion{Ca}{2} NIR3 displays a slower decline 
in SN\,2018gv compared to that of SN\,2011fe. Other major absorption 
features exhibit similar trends as seen in SN\,2011fe, as well as several 
other Type Ia SNe, i.e. see Fig.~15 of \citet{Zhang_etal_2016}. 

The fit results of the different velocity components are presented in 
Table~\ref{Table_fitv}. The uncertainty in the measurement of the 
absorption velocity was converted from the associated error in wavelength, 
which has been estimated by adding the propagated error of the fitted center 
wavelength (see Equation~\ref{Eqn_gaussian}), the rms value in wavelength 
calibration (typically less than 0.3 \AA), and half of the size of the 
smallest resolution element ($\Delta \lambda/2 = \lambda/2R$),  where $R$ 
gives the spectral resolution in quadrature. The estimated uncertainty 
mainly depends on the spectral resolution and the S/N of the spectrum. 

%

\subsection{Pseudo-bolometric Luminosity \label{section_bolo}}
As discussed by \citet{Brown_etal_2016}, estimating the bolometric luminosity 
of Type Ia SNe could be better carried out by employing flux-calibrated 
spectrophotometry. The diversity in the UV flux distributions, as well as 
the lack of NIR information could introduce erratic systematical 
uncertainties. To better quantify the bolometric characteristics of SN\,2018gv, 
we compute the pseudo-bolometric luminosity of SN\,2018gv over a wavelength 
range from $\sim1660-8180$ \AA\ based on the Swift $uvw2$, $uvm2$, $uvw1$, and 
the LCO $UBg'Vr'i'$-band photometry. 
We also conduct a similar calculation for SN\,2011fe and compare the result to 
SN\,2018gv. The construction of the spectral energy distribution (SED) of 
SN\,2018gv is illustrated by Fig.~\ref{Fig_construct_swift_bolo} and the 
steps are detailed in Appendix~\ref{App_SED}. 

\begin{figure}[!h]
\epsscale{0.6}
\plotone{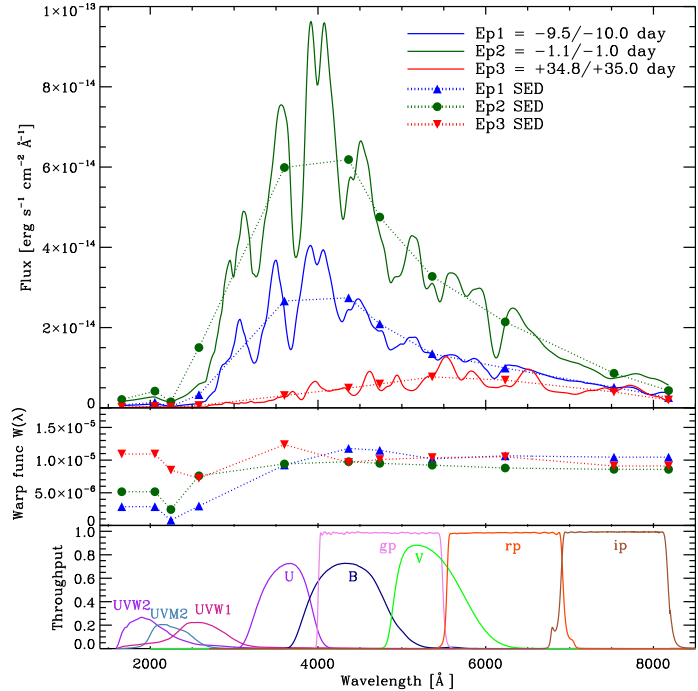}
\caption{The constructed SED for SN\,2018gv. In the upper panel, dots in three 
different colors show the bandpass monochromatic flux from each of three 
observations at their effective wavelengths. Dotted lines show the SED 
constructed with the warping procedure, and solid lines present the warped 
spectra from Hsiao's template \citep{Hsiao_etal_2007} for the nearest integer 
epoch (see legend). The middle panel gives the value of the warp function. 
The lower panel presents the associated bandpasses throughput curves for the 
SN\,2018gv observations. 
\label{Fig_construct_swift_bolo}
}
\end{figure}

\begin{figure}[!h]
\epsscale{0.8}
\plotone{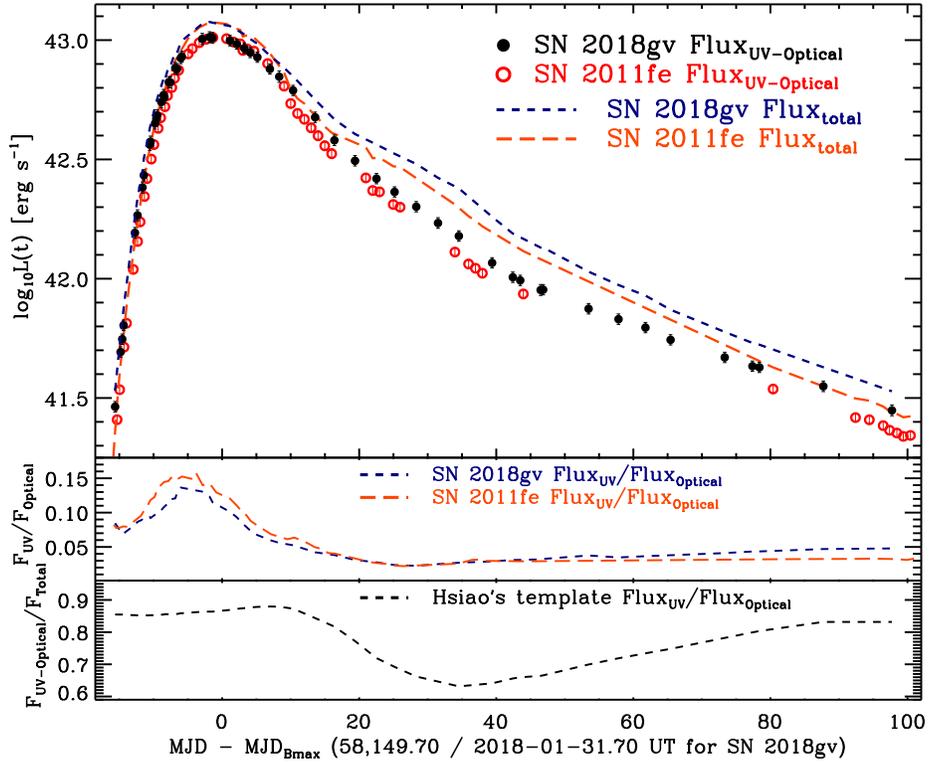}
\caption{The constructed quasi-bolometric light curves of SN\,2018gv 
(upper panel). The filled black circles represent the bolometric 
luminosity integrated within a wavelength range of 1660$-$8180 \AA, 
the open red circles estimate the total bolometric luminosity based on 
the fraction of optical to total luminosity computed based on SN\,2012fr, 
which is shown by the gray line in the lower panel. {The middle panel 
presents the ratio of the UV (1660$-$3200 \AA) to optical 
(3270$-$8180 \AA) flux of SN\,2018gv compared to that of SN\,2011fe. 
The fluxes were derived by integrating the composite spectral template 
\citep{Hsiao_etal_2007} as described in Section~\ref{section_bolo}. 
}
\label{Fig_pseudo_swift_bolo}
}
\end{figure}

The UV-optical pseudo-bolometric light curve of SN\,2018gv is shown 
in Fig.~\ref{Fig_pseudo_swift_bolo}. For comparison, we also apply 
the same procedure to the Swift $uvw2$, $uvm2$, $uvw1$ and the $UBVRI$ 
photometry \citep{Zhang_etal_2016} of SN\,2011fe. We adopt a Cepheid 
distance modulus of 6.4 Mpc for SN\,2011fe \citep{Shappee_etal_2011}, 
which is the same as the distance applied in the bolometric luminosity 
calculation conducted in \citet{Zhang_etal_2016}. The calculated 
pseudo-bolometric light curve of SN\,2011fe is also shown in 
Fig.~\ref{Fig_pseudo_swift_bolo}. The integration of the SN\,2011fe 
SED was performed over the same wavelength range as SN\,2018gv. The 
middle panel presents the ratio of the UV (1660$-$3200 \AA) to optical 
(3270$-$8180 \AA). The UV/optical flux ratio ($F_{UV}/F_{\rm Optical}$) 
of SN\,2018gv is comparable to that of SN\,2011fe, both of which reached 
their peak about six days earlier than the bolometric luminosity. 
This suggests a relatively short diffusion time for their higher-energy 
photons. We also calculated the fraction of UV to optical flux relative 
to the total bolometric luminosity over the wavelength range of 
1660$-$24000 \AA\ ($F_{UV-{\rm Optical}}/F_{\rm Total}$) by integrating 
the composite spectral template created by \citet{Hsiao_etal_2007}. This 
fraction was divided from the pseudo-bolometric luminosity for both SNe 
to provide a raw estimation of their bolometric luminosity, which is 
shown by the dashed lines in the first panel. 

We tabulate the pseudo-bolometric luminosity of SN\,2018gv and SN\,2011fe 
calculated within the UV to optical wavelength ranges within the first 100 
days after the $B-$band peak in Table~\ref{Table_bolo}. 
For each SN, the estimated UVOIR luminosity is also presented, which has 
similar error as the UV-Optical pseudo-bolometric luminosity. 
The maximum value of the UV-optical pseudo-bolometric luminosities of SN\,2018gv 
is consistent with the peak value of SN\,2011fe. Assuming SN\,2018gv has the 
same NIR/optical flux ratio ($F_{\rm NIR} / F_{\rm Optical}$) as SN\,2011fe, 
we estimate the peak bolometric luminosity of SN\,2018gv to be 
log$L$ = (43.074$\pm$0.023$\pm$0.008) erg s$^{-1}$. The first and the second 
uncertainties represent the statistical error and the error due to the 
distance, respectively. Following \citet{Stritzinger_etal_2005}, the peak 
bolometric luminosity produced by the radioactive $^{56}$Ni can be written as: 
\begin{equation}
L_{\rm max} = (6.45 {\rm e}^{-\frac{t_r}{8.8 \rm \ day}} + 1.45 {\rm e}^{-\frac{t_r}{111.3 \rm \ day}}) 
\bigg{(} \frac{M_{\rm Ni}}{M_{\odot}} \bigg{)} \times 10^{43} {\rm erg \ s^{-1}}, 
\label{Eqn_ni_mass}
\end{equation}
where $t_r$ is the rise time of the bolometric light curve, and $M_{\rm Ni}$ 
denotes the synthesized nickel mass in the SN ejecta. A high-order polynomial 
fit to the bolometric light curve from $t=-15$ to $+15$ days suggests that the 
bolometric luminosity peaked around $t = -1.1\pm2.4$ days relative to the 
$B-$band maximum. This is consistent with the mean of the distribution of the 
time difference between the bolometric light curve peak and the $B-$band 
maximum (see, e.g. \citealp{Scalzo_etal_2014_bolo}). Adopting a $B-$band rise 
time of 18.51$\pm0.92$ days estimated in Section~\ref{Section_earlylc} and peak 
luminosity of $L = (1.186\pm0.064\pm0.022) \times 10^{43}$erg s$^{-1}$, we 
derive a nickel mass of 0.56$\pm$0.08$M_{\odot}$ for SN\,2018gv. This is also 
consistent to the estimation assuming a typical bolometric rise time of 
19$\pm$3 days, i.e. $M_{\rm Ni}\approx0.59\pm0.09 M_{\odot}$ according to the 
Arnett law (e.g. see \citealp{Arnett_1982}, and Equation 7 in 
\citealp{Stritzinger_etal_2005}). These results are consistent with the nickel 
mass derived based on the peak of the UVOIR luminosity of 
$L = 1.191 \times 10^{43}$erg s$^{-1}$.

Regarding the sanity test simultaneously performed on SN\,2011fe, we noticed 
that the peak pseudo-bolometric luminosity of SN\,2011fe derived over a 
wavelength range of $1660-8180$ \AA\ (log\,$L$ = 43.011$\pm$0.016 erg s$^{-1}$) 
is systematically lower than the UVOIR bolometric luminosity calculated 
considering the SED over $1600-24000$ \AA\ (log\,$L$ = 43.05 erg s$^{-1}$, with 
a typical uncertainty of 0.07 dex, dominated by the uncertainty in the distance 
\citealp{Zhang_etal_2016}). This discrepancy of $\sim$8.6\% in the peak 
bolometric luminosity is due to the construction of the pseudo-bolometric 
light curves because we do not account for the NIR fluxes. After correcting 
for the $\sim$9\% $F_{\rm NIR} / F_{\rm Optical}$ ratio around the peak, the 
bolometric luminosity of SN\,2011fe log\,$L$ = 43.07 erg s$^{-1}$ is consistent 
with \citet{Zhang_etal_2016} and confirms the sanity of the method. 

\section{Spectropolarimetry\label{section_specpol}}
\begin{figure}[!h]
\epsscale{0.7}
\plotone{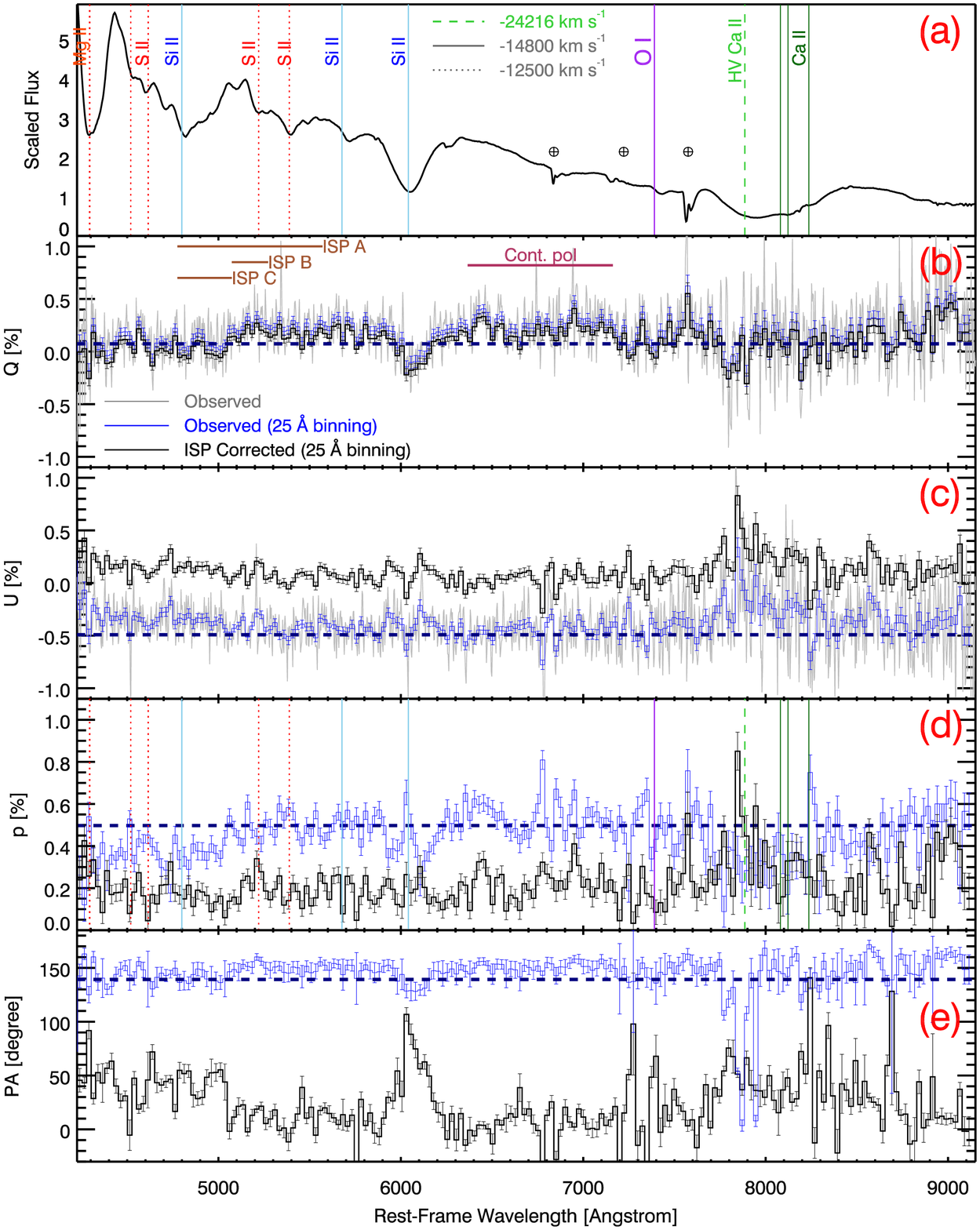}
\caption{Spectropolarimetry of SN\,2018gv at $-$13.6 day (epoch 1) relative 
to the $B-$band maximum light at MJD 58149.698. The five panels (from top to 
bottom) give (a) the scaled observed flux spectrum with \ion{Si}{2}, 
\ion{Ca}{2}, \ion{O}{1} and \ion{Mg}{2} lines labeled for different velocities; 
(b) the normalized Stokes parameter $Q$; (c) the normalized Stokes parameter 
$U$; (d) the polarization spectrum ($p$); and (e) the polarization position 
angle $PA$. Line identifications  provided in the top panel. The diagrams 
in panels (b)-(e) represent the polarimetry before (blue) and after (black) 
the ISP correction. The $Q$ and $U$ components of the ISP and the 
corresponding $p$ and $PA$ are shown by the horizontal gray lines in panels 
(b)-(e), respectively. The data have been rebinned to 25 \AA\ for clarity. 
\label{Fig_iqu_ep1}
}
\end{figure}

\begin{figure}[!h]
\epsscale{0.7}
\plotone{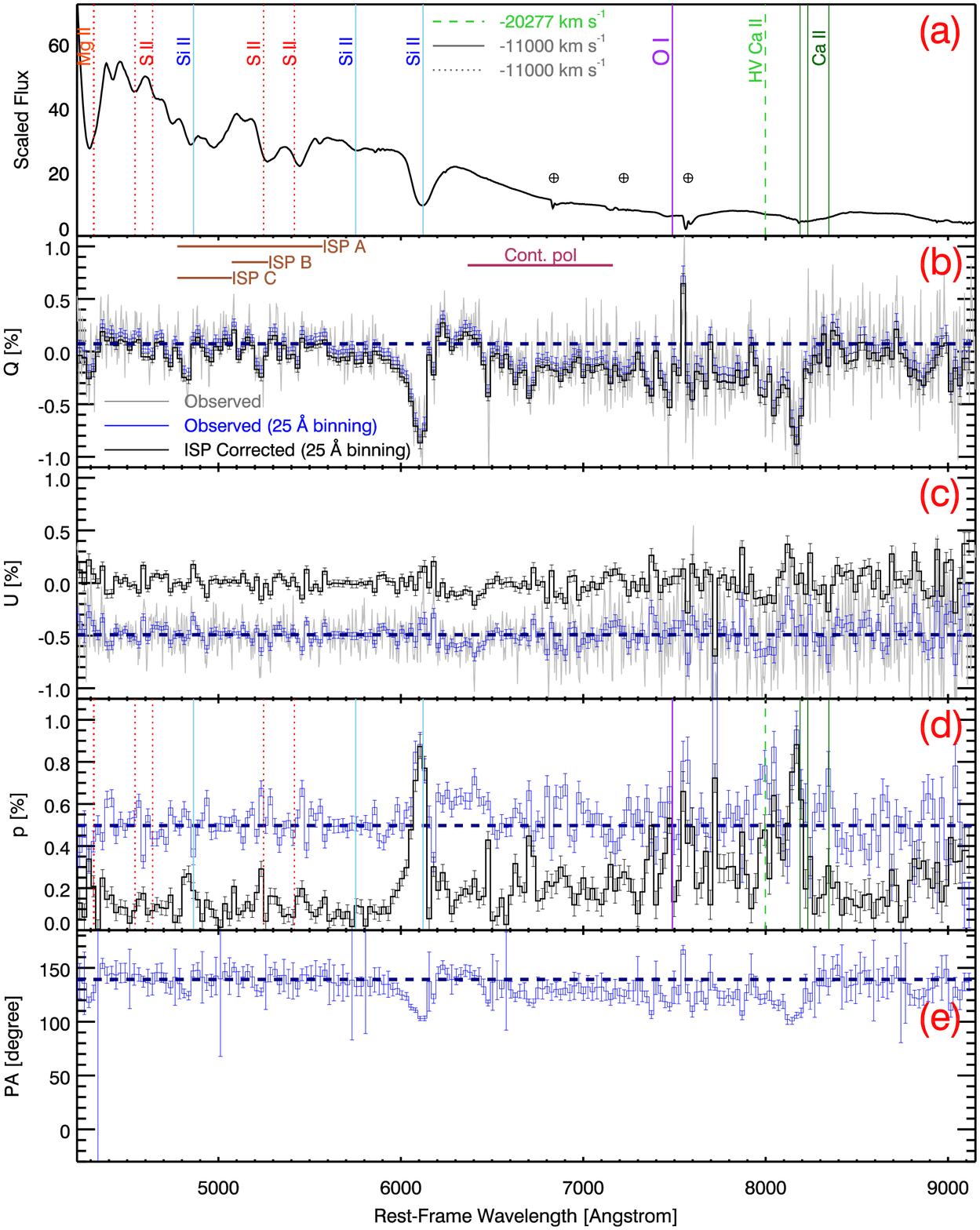}
\caption{Same as Fig.~\ref{Fig_iqu_ep1}, but for spectropolarimetry 
of SN\,2018gv at $-$0.5 day (epoch 2). Because the intrinsic continuum 
polarization is very low around maximum brightness, the ISP-subtracted 
$PA$ exhibits quasi-random values. Therefore, in panel (e), we only 
present the $PA$ before the ISP-correction (blue). 
\label{Fig_iqu_ep2}
}
\end{figure}

Spectropolarimetry of SN\,2018gv obtained at day $-$13.6 (epoch 1) and day 
$-$0.5 (epoch 2) relative to the $B-$band maximum light, together with the 
associated flux spectra in the rest frame, is shown in 
Figs.~\ref{Fig_iqu_ep1} and \ref{Fig_iqu_ep2}, respectively. Because of the 
low and decreasing level of the continuum polarization, the Stokes parameters 
measured at different wavelengths at the second epoch are overall close to 
zero, which leads the deduced position angle ($PA$) measured for 
interstellar-polarization (ISP)-subtracted data to display random 
orientations; therefore, we only show the $PA$ calculated before ISP removal. 

\subsection{Interstellar Polarization~\label{section_isp}}
Light from SNe always suffers from the extinction caused by the interstellar 
dust grains along the line of sight, in both the Milky Way and their host 
galaxies. Dichroic extinction by partially aligned non-spherical paramagnetic 
interstellar dust grains will polarize the traversed photons, which causes 
the observed ISP. The removal of the ISP is essential to determine the 
intrinsic polarization of SNe. An upper limit on the dichroic 
extinction-induced polarization by Milky Way-like dust grains yields 
$p_{\rm ISP} \textless 9\times E(B-V)$ \citep{Serkowski_etal_1975}. Assuming 
both the Galactic and the SN\,2018gv host dust follow a similar $R_V=3.1$ 
extinction law \citep{Cardelli_etal_1989}, the upper limits on the ISP 
derived from the Milky Way and the host galaxy NGC 2525 reddenings (derived 
from the values listed in Table~\ref{Table_phot_para}) yield 
$p_{\rm ISP}^{MW} \textless 0.46\%$ 
and $p_{\rm ISP}^{Host} \textless 0.20\%$, respectively. 

Following a similar procedure to the spectropolarimetric analysis of 
SN\,2012fr \citep{Maund_etal_2013}, we estimate the ISP towards SN\,2018gv 
as $Q_{\rm ISP} =  0.07\% \pm0.16$\%, $U_{\rm ISP} = -0.49\% \pm0.09$\%. 
The corresponding degree and position angle are 
$p_{\rm ISP} = 0.50\% \pm0.09$\% and 
${PA_{\rm ISP}} = 139\fdg{0} \pm 5\fdg{2}$, respectively. Detailed 
information of the ISP estimation is reported in Appendix~\ref{App_ISP}. 
Notice that due to the relatively low ISP suggested by the low extinction 
towards SN\,2018gv, we adopted a wavelength-independent ISP correction to 
the observations. Before the maximum luminosity, the presence of Fe 
absorption wings and its line-blanketing depolarization over the wavelength 
ranges $4800-5600$ \AA\ may not be sufficient to characterize 
the true ISP. Therefore, we would like to stress that the ISP determination 
described here is only tentative. 
However, the ISP only resets the origin of the polarization on the $Q-U$ 
diagram, and should not alter the polarized spectral features. 
The consistent ISP estimates derived for the two epochs also suggest 
that the uncertainties in the ISP do not have a strong impact on the 
interpretation of the intrinsic polarization of SN\,2018gv. 

\subsection{The $Q-U$ Plane And The Dominant Axis~\label{section_quplane}}
Plotting the observed spectropolarimetry on the Stokes $Q-U$ plane 
provides an intuitive diagnostic for investigating the relative behavior 
of the Stokes parameters of the continuum and for different spectral 
features \citep{Wang_etal_2001}. In most of the cases, the observations 
indicate the roles of both axisymmetric and nonaxisymmetric components. 
In the latter case, axisymmetry may be broken by the presence of clumps 
of different composition, optical depth, and even different velocities. 
For an axially symmetric structure, the resulting polarization can be 
described by a single, straight line on the Stokes $Q-U$ plane, namely 
the `dominant axis' \citep{Wang_etal_2003_01el}, i.e. 
\begin{equation}
U = \alpha + \beta Q. 
\label{Eqn_daxis}
\end{equation}
Recalling Equation~\ref{Eqn_stokes0}, the slope of the line characterizes 
the $PA$, which is only related to 
the position angle of the symmetry axis on the sky. The distance to the 
origin gives the degree of polarization, which is determined by the 
scattering opacity of different elements. 

The polarimetry can therefore be decomposed into two components relative 
to the dominant axis. Linear least-squares fitting to the polarimetry on 
the $Q-U$ plane can be used to determine the dominant axis (axis $d$). 
Such a dominant direction can sometimes be recognized through 
the distribution of the points in the $Q-U$ plane. Deviations from the 
dominant axis, in the perpendicular direction along the orthogonal axis 
(axis $o$), indicate departures from axial symmetry. This projection is 
equivalent to finding the first two Principal Components ($P_{d}$, $P_{o}$) 
of the data or applying a rotation to the $Q-U$ plane, i.e. see Equation 
1$-$2 of \citet{Wang_etal_2003_01el} and Equation 2$-$3 of 
\citet{Stevance_etal_2017}. 

\begin{figure}[!h]
\epsscale{1.0}
\plotone{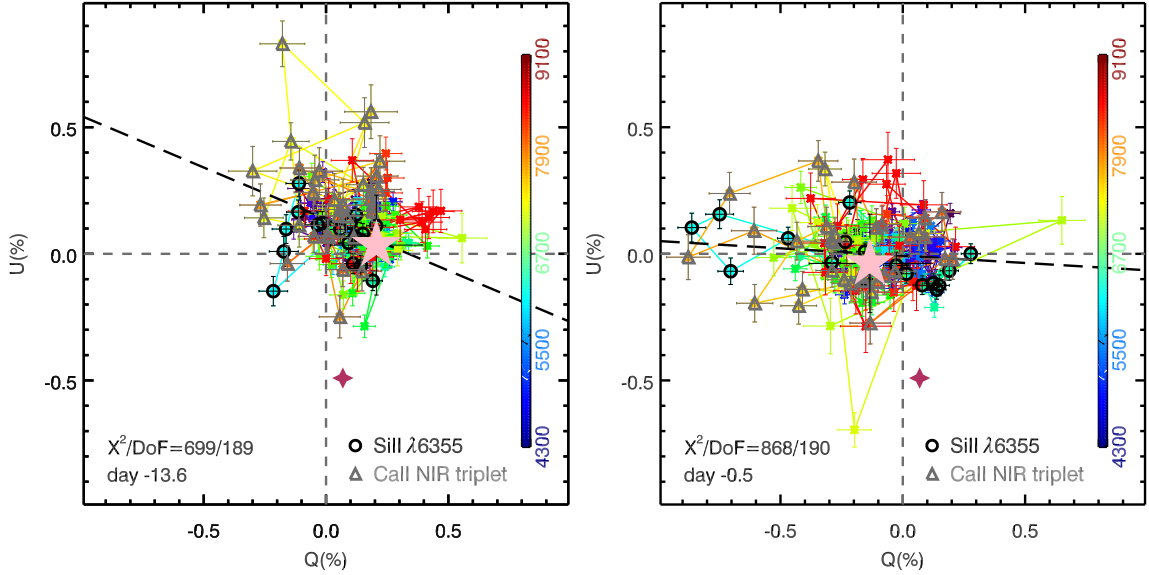}
\caption{The ISP-corrected Stokes parameters of SN\,2018gv, on the Stokes 
$Q-U$ plane. The data have been rebinned to 25 \AA. In each panel, the color 
bar indicates the wavelength and the filled maroon star shows the position 
of the estimated ISP. The black line traces the dominant axis computed using 
the data in a wavelength range $4300 \leq \lambda \leq 9100$ \AA. The solid 
pink $\star$ in each panel indicates the deduced continuum polarization over 
the wavelength range $6400 - 7200$ \AA. The open black circles and the open 
gray triangles mark the spectral regions covering the \ion{Si}{2} 
$\lambda$6355 and the \ion{Ca}{2} NIR3 features, respectively. 
\label{Fig_qu_plot}
}
\end{figure}

Figure~\ref{Fig_qu_plot} shows the ISP-corrected Stokes parameters on the 
Stokes $Q-U$ plane for both epochs. The determination of the dominant axis 
of SN\,2018gv was conducted for both epochs, by performing an error-weighted 
linear least squares fitting to the data. In Fig.~\ref{Fig_qu_plot}, the 
black long-dashed lines fit the dominant axis to the observed polarization 
in the wavelength range $4300 \leq \lambda \leq 9100$ \AA , representing 
the direction of axial symmetry. The direction of the dominant axis changed 
from $PA = 168\fdg{9}\pm1\fdg{2}$ to $178\fdg{3}\pm0\fdg{7}$. 
We suggest that a dominant axis seems to be present at both epochs, and it 
is almost parallel to the $U$ axis. Though the observations complied with a 
dominant axis, the large values of $\chi^2$ as labeled on the lower-left 
corner of each panel indicate that the data are poorly described by just 
a linear relation. Large deviations on the $Q-U$ plane from the dominant 
axis indicate a significant departure from axial symmetry. 
In the classification scheme of \cite{Wang_Wheeler_2008}, the spectropolarimetry of 
SN~2018gv is an example of spectropolarimetric (SP) Type D1. 
This type is characterized by data that show an elongated ellipsein a $Q-U$ diagram 
so that a dominant axis can be identified, but for which a straight line does not 
provide a satisfactory fit. Significant deviations are found orthogonal to the dominant axis.

\subsection{Continuum Polarization~\label{section_contpol}}
Linear polarization in the continuous spectrum is due to the Thomson 
scattering of free electrons, and it is independent of wavelength. 
After correcting for the ISP, the continuum polarization of SN\,2018gv 
at the two epochs was estimated based on the Stokes parameters over the 
wavelength range 6400$-$7200 \AA, which is known to be free of strongly 
polarized lines \citep{Patat_etal_2009}. 
Scattering in the degrees of polarization can still be seen within the 
selected polarized lines-free wavelength ranges which used to estimate 
the continuum polarization. Such fluctuation arises from the bound-bound 
transitions, primarily of iron-peak elements, which modify the emergent 
radiation by depolarizing the continuum flux and produce some line polarization \citep{Hoeflich_etal_1996_93J,Hoeflich_etal_2006}. 
The error-weighted mean across 
this region at the two epochs gives $Q_{\rm Cont1} =  0.20\% \pm 0.08\%$, 
$U_{\rm Cont1} =  0.04\% \pm 0.09\%$, and 
$Q_{\rm Cont2} = -0.13\% \pm 0.13\%$, 
$U_{\rm Cont2} = -0.04\% \pm 0.09\%$, respectively. 
The error has been estimated by adding the statistical uncertainties and 
the standard deviation calculated from the 25\AA\ binned spectra within 
the continuum wavelength range in quadrature. At epoch 1, the level 
of continuum polarization of SN\,2018gv is consistent with the low levels 
of continuum polarization measured from Type Ia SNe at early epochs, i.e. 
$0.06\% \pm 0.12 \%$ at day $-$11 for SN\,2012fr \citep{Maund_etal_2013}, 
and $\sim0.3\%$ at $-$9.1 day for SN\,2016coj without the removal of ISP 
\citep{Zheng_etal_2017}. At epoch 2, a difference of 
$\Delta Q_{\rm Cont} = -0.33\% \pm 0.15$\% and 
$\Delta U_{\rm Cont} = -0.08\% \pm 0.13$\% can be identified. This marks 
the time-evolution of the degree and the position angle of the continuum 
polarization, i.e. from $p_{\rm Cont1} = 0.20\% \pm 0.08$\%, and 
$PA_{\rm Cont1} = 4\fdg{8} \pm 11\fdg{3}$, to 
$p_{\rm Cont2} = 0.14\% \pm 0.13$\% and 
$PA_{\rm Cont2} = -82\fdg{4} \pm 27\fdg{2}$, respectively. At both 
epochs, the measured degrees of continuum polarization are consistent with 
the low levels typically measured for Type Ia SNe \citep{Wang_Wheeler_2008}, 
indicating an approximately spherical symmetry. 

\subsection{Line Polarization~\label{section_linepol}}
In addition to the wavelength independent continuum polarization arising 
from a globally axisymmetric ejecta, we observe wavelength dependent 
polarization features associated with spectral lines at both epochs in 
association with spectral lines identified in the Stokes $I$ flux spectrum. 
This indicates a departure from axisymmetry. 
The most likely interpretation of line polarization in the context of Type 
Ia SNe gives that the underlying electron-scattering photosphere is covered by 
certain element-rich clumps with high optical depth. These indicate additional 
asymmetries, that are dependent on the chemical composition, ionization, and 
velocity structure of the ejecta, exterior to the photosphere.  Different 
elements may have different electron scattering opacities as well as different 
geometric distributions. Polarimetry measures the degree of incomplete 
cancellation of photon `E-vectors' as the photons interact with 
line-transitions in, possibly, asymmetrically-distributed material. 

For prominent, broad lines, the polarization profile over wavelength probes 
the structure of the ejecta in velocity space. For complex structures, which 
depart from a simple axial symmetry, the resulting polarization across the 
wavelengths associated with particular spectral features, may deviate from 
the straight line which is the tracer of axial deviation from spherical 
symmetry \citep{Wang_Wheeler_2008}. The presence of asymmetric structures in 
the ejecta can be inferred from the observation of loops in the $Q-U$ plane. 
Such loops are defined as a gradual rotation of the $PA$ as a function of 
observed wavelength across a spectral line. The SN with loops across spectral lines on the $Q-U$ diagram were assigned SP Type L in \cite{Wang_Wheeler_2008}. The presence of one or more loops 
across a specific spectral feature indicates a significant variation in the 
degree of polarization and the polarization position angle at different 
velocities (corresponding to different depths into the ejecta; 
\citealp{Wang_Wheeler_2008}). 


\subsubsection{\rm \ion{Si}{2} Lines}
At $-$13.6 day, the peak around 5900$-$6200 \AA\ in $Q$ and the 
troughs at similar wavelengths in the $U$ spectrum can be attributed 
to the \ion{Si}{2} $\lambda$6355 line at $v\sim-$14,000 km s$^{-1}$ (see 
Fig.~\ref{Fig_iqu_ep1}). After correcting for the 
ISP, the degree of polarization across this line did not show any 
distinguishable features in $p$ (see the black line in 
Fig.~\ref{Fig_iqu_ep1}d), but exhibited a difference of $\sim90$ degrees 
in $PA$ compared to the continuum (see Fig.~\ref{Fig_iqu_ep1}e). Such a 
pattern in the $PA$ over the \ion{Si}{2} $\lambda$6355 feature indicates that 
the Si-rich portion of the ejecta is likely to be oriented perpendicular to
the principal axis of symmetry of the total ejecta. The line profile at 
$\sim$4810 \AA\ can be identified as the blue-shifted \ion{Si}{2} 
$\lambda\lambda$5041, 5056 doublet which has a similar velocity to \ion{Si}{2} 
$\lambda$6355; however, the feature is also blended with a series of iron 
lines, i.e. \ion{Fe}{2} $\lambda$4913, 5018, 5169. At $-$0.5 day, the 
polarization peak around \ion{Si}{2} $\lambda$6355 occurs at $\sim$6100 \AA. 
The central velocity of the feature is measured as $\sim-$11,000 km s$^{-1}$. 
The polarization is the highest at the absorption minimum of this feature. The 
polarization profile is asymmetric with a sharp drop towards lower velocities.  

\subsubsection{\rm \ion{Ca}{2} NIR Triplet}
At $-$13.6 day troughs in q and peaks in u are also seen between 7800 and 
8300 \AA\ (see Fig.~\ref{Fig_iqu_ep1}). 
Though the observed line profiles of \ion{Ca}{2} NIR3 
at early phases can be well fit by a double Gaussian function describing 
a normal and HV components with separate central wavelengths, the 
polarization spectra (with low levels of S/N) do not exhibit separate 
features for the HV and photospheric velocity components. One can still 
identify a `notch' around 7800$-$7900 \AA\ in $Q$ accompanied by a `peak' 
at a similar wavelength in $U$. These features are associated with the 
HV \ion{Ca}{2} NIR3 at $\sim-$24,200 km s$^{-1}$. Less obvious features 
can be seen within the wavelength region of 8000$-$8300 \AA. These 
features have velocities similar to that inferred for the photosphere 
from the \ion{Si}{2} $\lambda$6355 at $\sim-$14,000 km s$^{-1}$ and 
also deviate from the overall spectral ranges in $p$ and $PA$. After 
correcting for the ISP, the $PA$ across the \ion{Ca}{2} NIR3 shows 
different values compared to either that of the continuum or the 
\ion{Si}{2} $\lambda$6355, indicating the configuration of the Ca-rich 
ejecta is not identical to the \ion{Si}{2} $\lambda$6355 line forming 
regions and distinct from the photosphere as well. At $-0.5$ days, the 
\ion{Ca}{2} NIR3 is dominated by the photospheric component at 
$\sim-$11,500 km s$^{-1}$. The HV component is still discernible and 
located at $\sim-$20,200 km s$^{-1}$. The polarization behaviors over 
the HV regions and the photosphere do not show distinct variation. 

\subsubsection{\rm \ion{Mg}{2} $\lambda$4481, \ion{Si}{2} $\lambda$5041, \ion{Si}{2} $\lambda$5454, 5640 and \ion{O}{1} $\lambda$7774 Lines\label{section_other_lines}}
Despite approaching the blue end of spectral coverage, a $\sim$0.3\% 
polarization can be marginally identified with the line of 
\ion{Mg}{2} $\lambda$4481 after ISP-correction (see 
Fig.~\ref{Fig_iqu_ep1}d). \ion{Si}{2} $\lambda$5041 does not exhibit 
clear polarization, but the $PA$ across \ion{Si}{2} $\lambda$5041 and 
\ion{Mg}{2} $\lambda$4481 are both larger compared to the $PA$ from 
5100$-$7200 \AA, except for the \ion{Si}{2} $\lambda$6355 and telluric 
features (see Fig.~\ref{Fig_iqu_ep1}e). This suggests some commonality 
in their geometry. Same interpretation also holds for epoch 2, in which 
\ion{Mg}{2} $\lambda$4481 and \ion{Si}{2} $\lambda$5041 polarized at 
$\sim$0.3$-$0.4\%. At all epochs, the distinctive `W'-shaped 
\ion{S}{2} ($\lambda$5454, 5640) absorption features appear at lower 
velocities compared to the velocity derived from the 
\ion{Si}{2} $\lambda$6355 line. For instance, from epoch 1 to epoch 2, 
the \ion{S}{2} $\lambda$5454, 5640 velocities evolve from $-$13,800 km 
s$^{-1}$ and $-$13,500 km s$^{-1}$ to $-$10,100 km s$^{-1}$ and 
$-$10,400 km s$^{-1}$, respectively. The slower velocities measured for 
\ion{S}{2} $\lambda$5454, 5640 are due to this feature being optically 
thin at larger radii compared to \ion{Si}{2} and \ion{Ca}{2}. This is 
corroborated by the consistent low polarization measured across this 
feature, which suggests that sulfur is more concentrated in 
lower-velocity regions than calcium and silicon. Additionally, we find 
that the \ion{O}{1} $\lambda$7774 line appears to be weak in the flux 
spectrum, and no sign of that feature can be identified in the 
polarization spectrum. 
The primordial oxygen from the WD maintains its initial spherically 
symmetric distribution despite the asymmetric explosion. This gives an
important constraint on the explosion models \citep{Hoeflich_etal_2006}. 

\subsection{The Inferred Ejecta Geometry}
\begin{figure}[!h]
\epsscale{1.0}
\plotone{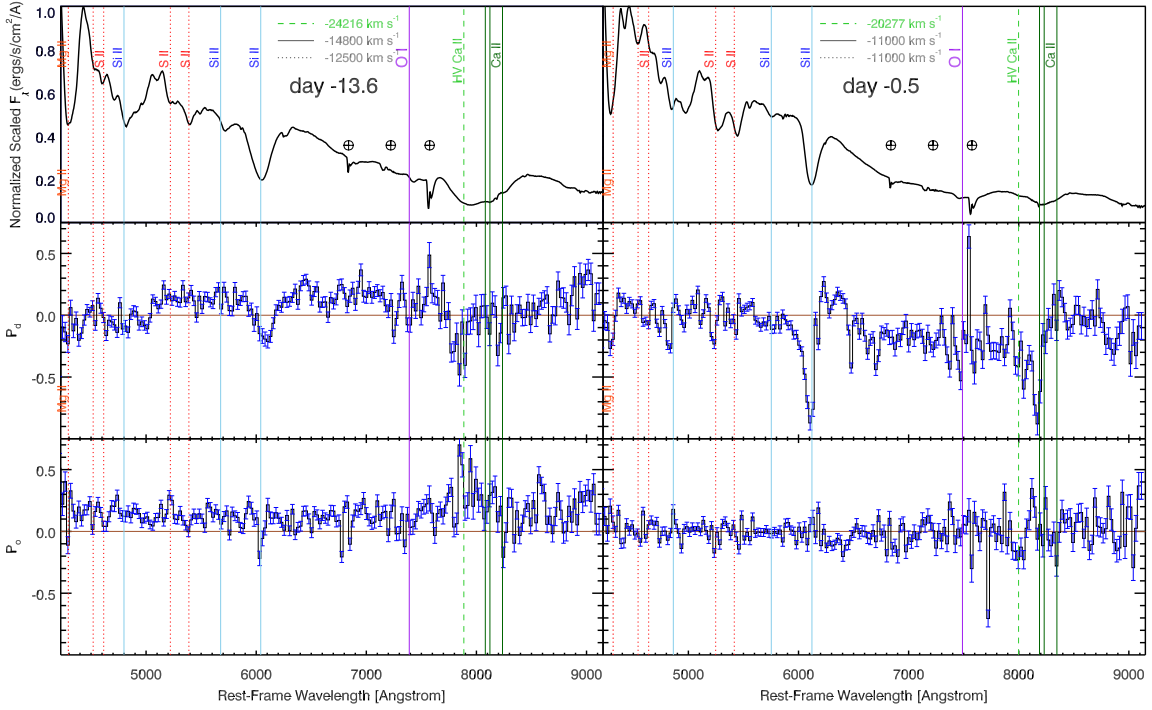}
\caption{The normalized flux spectra together with the PCA of SN\,2018gv 
spectropolarimetry at $-$13.6 day (epoch 1, left panels) and $-$0.5 day 
(epoch 2, right panels). The top row gives the flux spectra normalized 
to the maximum value within the range. The middle and the bottom rows 
illustrate the polarization spectra projected onto the dominant axis and 
the orthogonal axis, respectively. The vertical solid lines mark the 
positions of certain spectral features at different velocities as labeled 
in the figure. Some major tellurics are labeled by $\Earth$. At $-$13.6 
day, the \ion{Si}{2} $\lambda$6355 and the \ion{Ca}{2} NIR3 features 
exhibit similar but not identical deviations from the dominant axis, 
suggesting that the geometrical structures of \ion{Si}{2}, \ion{Ca}{2} 
and the photosphere are different. No polarization is detected across 
other weaker lines. At $-$0.5 day as shown by the right panels, the 
\ion{Si}{2} $\lambda$6355 and the \ion{Ca}{2} NIR3 features appear to 
have the same dominant axis. Over the wavelength range 6400$-$7200 \AA 
characterizing the continuum polarization, the error-weighted mean of 
$P_{O}$ has evolved from 0.109\%$\pm$0.095\% to $-$0.044\%$\pm$0.089\% 
from $-$13.6 to $-$0.5 days, indicating a more spherically symmetric 
geometrical structure of the inner ejecta. Notice that the two epochs 
are characterized by different position angles of the dominant axis 
and therefore do not share the same orientation of the photosphere.
\label{Fig_stokes_pca}
}
\end{figure}

\begin{figure}[!h]
\epsscale{0.7}
\plotone{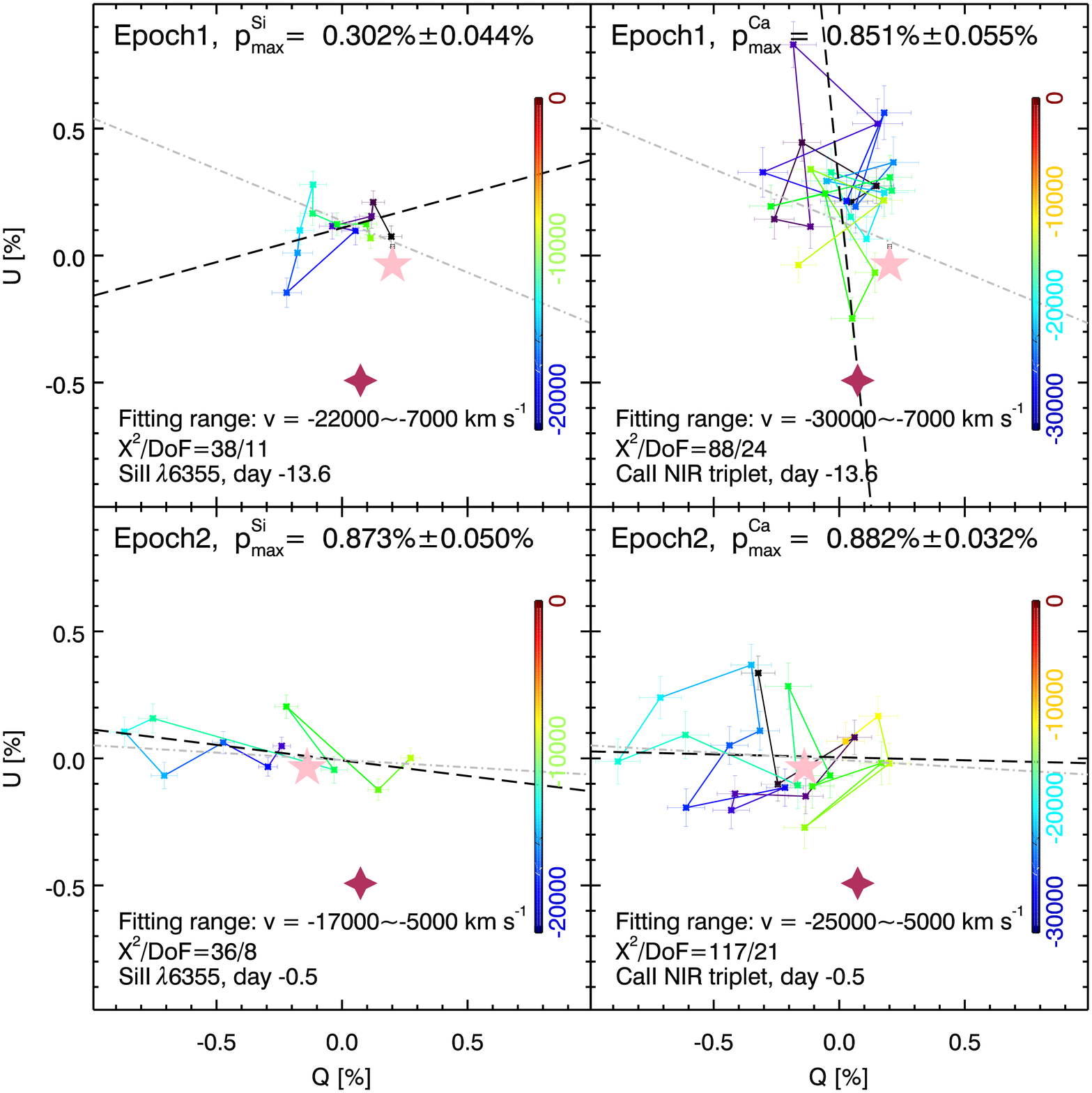}
\caption{The polarization for the \ion{Si}{2} $\lambda$6355 (left panels) and the 
\ion{Ca}{2} NIR3 (right panels) features on the Stokes $Q-U$ plane at $-$13.6 day 
and $-$0.5 day. The data have been rebinned to 25 \AA. In each panel, the points 
are color coded according to the velocity relative to the respective wavelengths 
measured in the rest frame for the two features. The data were corrected for the 
ISP (the location of which is indicated by the filled maroon star) and the 
continuum polarization is shown by the solid pink $\star$. The maximum 
polarization across the feature after ISP-correction is also presented. The dashed 
black line fits the displayed data points and the dotted gray line traces the 
dominant axis computed in the wavelength range $4300 \leq \lambda \leq 9100$ \AA. 
\label{Fig_qu_si_ca}
}
\end{figure}

Following from Section~\ref{section_quplane}, we present the dominant 
and the orthogonal polarization components after correcting for ISP in 
Fig.~\ref{Fig_stokes_pca}. The dominant component, that is, the 
polarization projected onto the dominant axis, represents global 
geometric deviations from spherical symmetry. Any polarization signal 
orthogonal to the dominant axis carries information about deviations 
from axial symmetry. In Fig.~\ref{Fig_qu_si_ca}, we show the loops 
across the \ion{Si}{2} $\lambda$ 6355 and \ion{Ca}{2} NIR3 features. 
At epoch 1, the polarization modulations associated with \ion{Si}{2} 
$\lambda$6355 are principally observed in the dominant polarization 
component $P_d$ (i.e. the middle-left panel in 
Fig.~\ref{Fig_stokes_pca}). In the orthogonal polarization component 
(see bottom left panel in Fig.~\ref{Fig_stokes_pca}) a narrow 
polarization profile can be identified at the wavelength coincident 
with the position of the absorption minima of \ion{Si}{2} $\lambda$6355. 
Such a residual indicates that the bulk orientation of Si-rich ejecta is 
different from that of the dominant axis of the photosphere (as inferred 
from the continuum polarization). A line complex can be observed in the 
polarization modulations associated with \ion{Ca}{2} NIR3. The HV 
component shows modulation in both the dominant and the orthogonal 
polarization components. The NV component, however, exhibits 
significantly less modulation in both the dominant and orthogonal components. 

At epoch 2, the polarization modulations associated with \ion{Si}{2} 
$\lambda$6355 and \ion{Ca}{2} NIR3 both fall predominantly along the 
$P_d$ axis. The signal in the orthogonal polarization component, over 
the observed wavelength range is consistent with the expected level of 
uncertainty. The rotation of the $PA$ over \ion{Si}{2} $\lambda$6355 
(i.e see Fig.~\ref{Fig_qu_si_ca}) can be interpreted as the rotation 
of the principal axis of symmetry with depth into the ejecta. The HV 
component of \ion{Ca}{2} NIR3 has become significantly shallower by 
epoch 2. Due to insufficient S/N, it is unclear if the HV \ion{Ca}{2} 
NIR3 component carries the same axial asymmetry as other lines. The 
polarization signal associated with \ion{Ca}{2} NIR3 is dominated by 
the NV component. Moreover, the $PA$ of the \ion{Ca}{2} NIR3 are 
almost aligned with the \ion{Si}{2} $\lambda$6355 at epoch 2, 
suggesting that the line-forming regions of silicon and calcium have 
already settled to a relatively similar geometric configuration. 

In epochs 1 and 2 (see Fig.~\ref{Fig_qu_si_ca}), the \ion{Si}{2} 
$\lambda$6355 and the \ion{Ca}{2} NIR3 both exhibit loops in the $Q-U$ 
space. In epoch 1, linear fittings of Stokes $Q-U$ data points across 
each line suggest that the Si-rich and the Ca-rich regions reside in 
different positions of the ejecta, which are also incompatible with the 
global geometry indicated by fitting all the data points across the 
entire observed wavelength. In epoch 2, the data points across the 
Si-rich and the Ca-rich ejecta tend to fall along a single locus. 
The orientation of these ejecta is also consistent with the overall 
shape of the ejecta. 

\begin{figure}[!h]
\epsscale{1.0}
\plotone{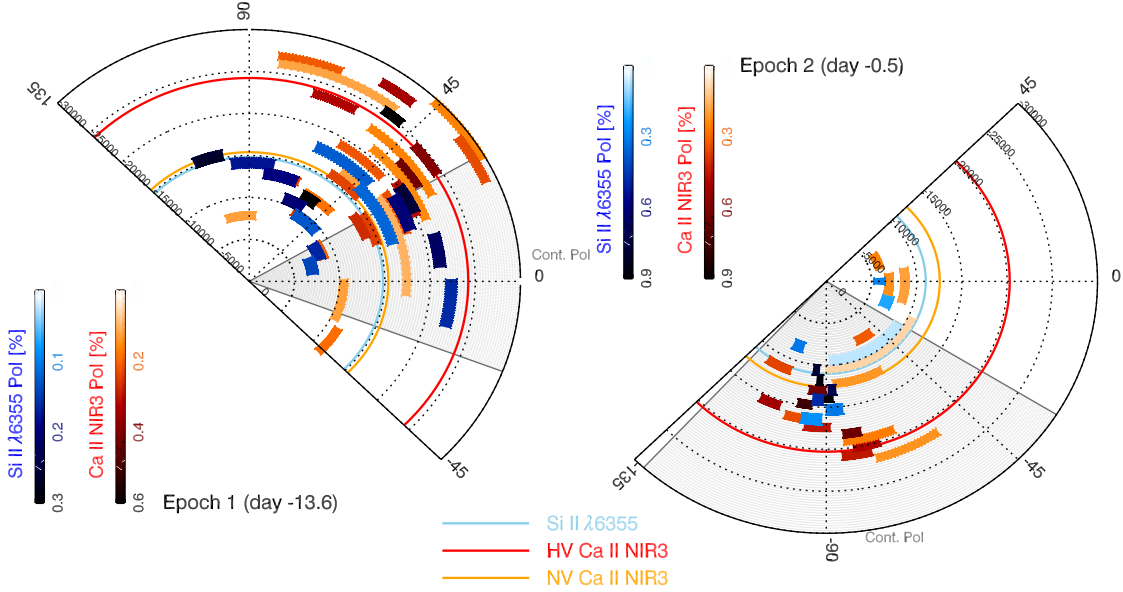}
\caption{
Polar plots of SN\,2018gv showing the velocity and polarization angle across 
the profiles for \ion{Si}{2} $\lambda$6355 and \ion{Ca}{2} NIR3 as a function 
of radial velocity. In each panel, the color bars indicate the ISP-corrected 
polarization degrees with the continuum polarization not subtracted. The 
numbers along the radial and position angle are labeled in km s$^{-1}$ and 
in degrees, respectively. The data have been rebinned to 25 \AA\ for better 
visualization. The center and the angular extent of each colored-bin 
represent the average $PA$ and the associated error, respectively. 
Note that larger angular extent suggests greater uncertainty on the $PA$. 
The continuum polarization angle is shown in grey, and its angular extent 
indicates the 1-$\sigma$ uncertainty. 
Typical uncertainties in the data points are smaller in epoch 2 compared with 
epoch 1 due to higher S/N. 
Velocities measured at the minima of the flux spectra of each species are 
indicated by colored semicircles as labeled. 
\label{Fig_polar_plot}
}
\end{figure}

The geometry of the Si-rich and the Ca-rich ejecta can also be visualized in 
polar plots (e.g. see 
\citealp{Maund_etal_2009, Reilly_etal_2016, Hoeflich_etal_2017, Stevance_etal_2019}). 
Such plots present the position angles measured for selected spectral features at 
different radial velocities in polar coordinates. In Fig.~\ref{Fig_polar_plot} 
we show polar plots for the \ion{Si}{2} $\lambda$6355 and the \ion{Ca}{2} NIR3 
profiles at both epochs. At epoch 1 ($-$13.6 day, left panel), the large offset 
in the radial profile of \ion{Si}{2} $\lambda$6355 (shown in blue) compared to 
the total ejecta (the gray sector) confirms the large deviations from the 
dominant axis of the continuum polarization. The even more erratic radial 
profile of the Ca-rich component also indicates a complex structure of the 
line forming regions. At epoch 2 ($-$0.5 day, right panel), the change in the 
position angle of the continuum polarization (see Section~\ref{section_contpol}) 
can be seen as the rotation of the gray-shaded sector. The radial profiles of 
both the Si-rich and the Ca-rich components have settled to a similar 
orientation as the total ejecta indicated by the gray sector. 

\section{Discussion of The Spectropolarimetry Observations~\label{section_discussion}} 
\subsection{The \ion{Si}{2} $\lambda$6355 Polarization Compared with a Larger Sample}
First, we notice that the polarization of \ion{Si}{2} on the $Q-U$ diagram 
can be well fitted by a straight line (e.g. see the right panel of 
Fig.~\ref{Fig_qu_plot} and the lower-left panel of Fig.~\ref{Fig_qu_si_ca}). 
With a clearly defined dominate axis, SN\,2018gv would thus be classified as 
SP Type D0 \cite{Wang_Wheeler_2008}. Such significantly polarized \ion{Si}{2} 
feature is suggestive of a global asymmetry of silicon layer. 

The peak polarization across the \ion{Si}{2} $\lambda$6355 line evolves 
with time. In order to compare with different SNe at similar phases, 
\citet{Wang_etal_2007} fitted the data with a second order polynomial to 
describe the time dependence of the degree of the \ion{Si}{2} polarization: 
$p_{\rm Si{\sc II}}(t) = 0.65 - 0.041(t-5)-0.013(t+5)^2$. Here $t$ denotes 
the time (in days) after the $B-$band maximum light and $p_{\rm Si{\sc II}}(t)$ 
gives the measured polarization (in percent) of \ion{Si}{2} $\lambda$6355. 
\citet{Wang_etal_2007} derived a correlation between the maximum line 
polarization of \ion{Si}{2} $\lambda$6355 and $\Delta m_{15} (B)$:  
\begin{equation}
p_{\rm Si{\sc II}}^{corr-5} = 0.48(03) + 1.33(15) (\Delta m_{15}-1.1), 
\label{Eqn_p_dm15}
\end{equation}
where $p_{\rm Si{\sc II}}^{corr-5}$ is defined as the polarization of the 
\ion{Si}{2} $\lambda$6355 corrected to $-$5 day in a unit of percent. This 
correlation indicates that at a given epoch, dimmer SNe exhibit higher Si 
line polarization, hence higher chemical nonuniformity. This can be 
understood if the dimmer the SN, the less material was burnt. Such a 
incomplete burning is not sufficient to erase chemically-lumpy 
configurations. The quantity $p_{\rm Si{\sc II}}^{corr-5}$ can be obtained 
by applying the correction given by \citet{Wang_etal_2007}: 
\begin{equation}
P_{\rm Si{\sc II}}(t) = P_{\rm Si{\sc II}}^{corr-5} - 0.041(t-5) - 0.013(t+5)^2;
\label{Eqn_psi_correction}
\end{equation}
For SN\,2018gv, adopting $p_{\rm Si{\sc II}}(t) = 0.87\% \pm 0.05\%$ at 
$t = -0.5$ day gives $p_{\rm Si{\sc II}}^{corr-5} = 0.91\% \pm 0.05\%$. 
This is significantly larger than the typical 1-$\sigma$ upper range of 
polarization shown as Fig.~2 of \cite{Wang_etal_2007}. 


Such a big discrepancy does not necessarily mean that SN\,2018gv has a 
peculiar \ion{Si}{2} $\lambda$6355 polarization because the empirical 
relationship proposed by \citet{Wang_etal_2007} does not account for the 
considerable intrinsic polarization variations from SN to SN. The high 
degree of polarization observed for SN\,2018gv suggests that the presence 
of large scale departure from spherical symmetry. Cikota et al. (2019, 
ApJS accepted) re-analyzed this relationship using a larger sample of 35 
Type Ia SNe and found a significantly larger scatter. The observed degree 
of silicon polarization of SN\,2018gv at $-$0.5 days falls in the 
$\sim$0.1\%$-$1.7\% range determined from a larger sample. Future 
high-precision spectropolarimetry of more Type Ia SNe with complete 
coverage of the rising phase is essential to determine the commonality 
and probe the intrinsic diversity in the geometric structure of the ejecta. 

According to \citet{Maund_etal_2010}, the peak polarization degree of the 
\ion{Si}{2} $\lambda$6355 line at $-$5 day is correlated with the average 
daily decline rate of the expansion velocity measured from the same line:  
\begin{equation}
P_{\rm Si{\sc II}}^{corr-5} = 0.267 + 0.006 \times \dot{v}_{\rm Si{\sc II}}.  
\label{Eqn_p_vgrad}
\end{equation}
The velocity gradients were derived based on the measurements taken  
between maximum and approximately two weeks after, when the \ion{Si}{2} 
feature disappears \citep{Benetti_etal_2005}, while the corrected day 
$-5$ \ion{Si}{2} polarizations were obtained based on 
Equation~\ref{Eqn_psi_correction}. Adopting the velocity gradient of 
SN\,2018gv around the peak or between days 0 to $+$10, i.e. 
$\dot{v_{\rm SiII}} = 36.6 \pm 6.4$ km s$^{-1}$ or 34.3 km s$^{-1}$, the 
estimated polarization gives $p_{\rm Si{\sc II}} = 0.49\% \pm 0.04\%$ 
or 0.47\%, respectively. These values are lower compared to the 
determined $p_{\rm Si{\sc II}}^{corr-5}$. 

Cikota et al. (2019) explored the possible correlations among the 
polarimetric and other observational properties of Type Ia SNe. Based 
on an archival data sample of 23 SNe that have at least one observation 
between day $-$11.0 and day 1.0, they found a strong linear correlation 
between the \ion{Si}{2} $\lambda$6355 polarization and the expansion 
velocity traced by the same line: 
\begin{equation}
P_{\rm Si{\sc II}} = (6.40 \pm 1.28) \times 10^{-5} \times v_{\rm Si{\sc II}@-5d} -(0.484 \pm 0.147), 
\label{Eqn_p_vsi}
\end{equation}
where $p_{\rm Si{\sc II}}$ is the maximum polarization of the \ion{Si}{2} 
$\lambda$6355 line between day $-$11 and $+$1, and $v_{\rm Si{\sc II}@-5d}$ 
gives the velocity measured from the \ion{Si}{2} $\lambda$6355 profile at 
$-$5 days relative to the $B-$band peak brightness. Adopting an 
interpolated $v_{\rm Si{\sc II}@-5d}=-$11,300 km s$^{-1}$ at day $-$5 and 
a typical uncertainty of 27 km s$^{-1}$, Equation~\ref{Eqn_p_vsi} yields 
a maximum \ion{Si}{2} $\lambda$6355 polarization of 0.24\%$\pm$0.21\%. 
Therefore, we conclude that SN\,2018gv is consistent with the \ion{Si}{2} 
velocity-polarization relationship at day $-$13.6 
($p_{\rm SiII} = 0.30\% \pm 0.04\%$), but exhibits a significantly 
higher \ion{Si}{2} $\lambda$6355 polarization at day $-$0.5 
($p_{\rm SiII} = 0.87\% \pm 0.05\%$). 

\subsection{The Asymmetric Polarization Profile of \ion{Si}{2} $\lambda$6355}
\begin{figure}[!h]
\epsscale{0.7}
\plotone{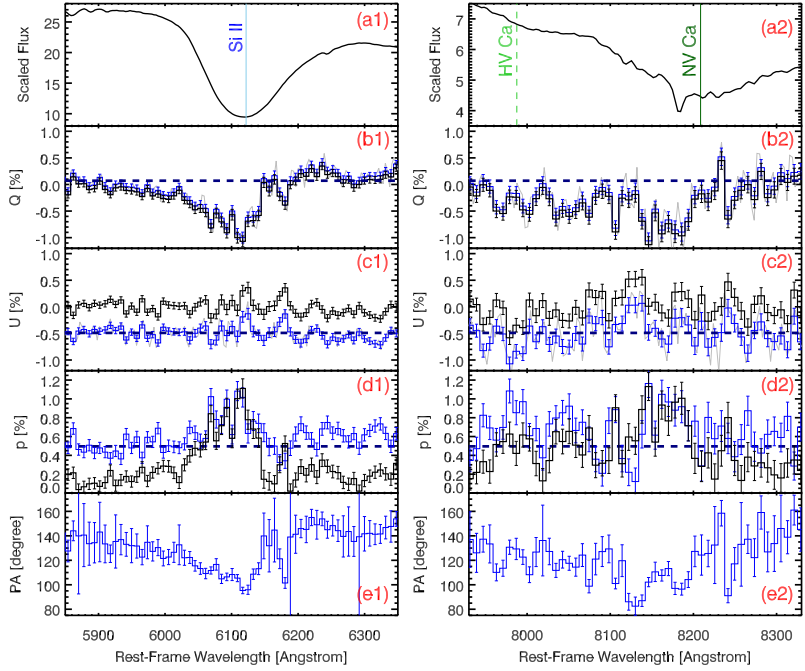}
\caption{
Same as Fig.~\ref{Fig_iqu_ep2}, but on the spectroplarimetry of SN\,2018gv 
across \ion{Si}{2} $\lambda$6355 (the left panels) and the \ion{Ca}{2} NIR3 
(the right panels) at day $-$0.5 (epoch 2). The data have been rebinned to 
8 \AA\ . In the left panels, i.e. d1, the \ion{Si}{2} polarization declines 
much faster from its peak around $\sim$6200 \AA\ ($v\sim-$11,000 km s$^{-1}$) 
towards longer wavelengths compared to shorter wavelengths. Multiple peaks 
around $\sim$6100 \AA\ may indicate the presence of various Si-rich 
components, which were not identified in the flux spectrum. 
\label{Fig_iqu_ep2_bin8}
}
\end{figure}

We notice that the most prominent \ion{Si}{2} $\lambda$6355 feature in the 
polarization spectrum around the $B-$maximum exhibits an asymmetric profile. 
Fig.~\ref{Fig_iqu_ep2_bin8} portrys the polarization profiles of 
\ion{Si}{2} $\lambda$6355 and \ion{Ca}{2} NIR3 at day $-$0.5 with an 8\AA\ bin 
size. In Fig.~\ref{Fig_iqu_ep2_bin8}-d1, one can see that the polarization 
signal across the \ion{Si}{2} $\lambda$6355 feature declines gradually from 
the absorption minimum in the flux spectrum towards shorter wavelengths. 
In contrast, the polarization profile exhibits a sharp drop from its peak 
towards longer wavelengths. We roughly estimated the blue side of the wing 
ranges from 5950$-$6120 \AA, and the red side covers $6120-6170$ \AA, 
corresponding to $-$19,100 to $-$11,100 km s$^{-1}$ and $-$11,100 to $-$8700 
km s$^{-1}$ in velocity space, respectively. Considering the \ion{Si}{2} 
$\lambda$6355 velocity measured from the same spectrum as $-$10,984$\pm$345 
km s$^{-1}$, the velocity range measured from the polarized spectrum gives 
$-$8100 to $+$2400 km s$^{-1}$ relative to the SN photosphere velocity 
traced by the same line. Such an asymmetric profile was not identified at 
epoch 1. The \ion{Ca}{2} NIR3 displays a complex polarization profile. 
At epoch 2, a weaker separate peak over $\sim$8020$-$8080 \AA\ (see 
Fig.~\ref{Fig_iqu_ep2_bin8}-d2) can be identified between the HV and NV 
Ca components, corresponding to a radial velocity of $\sim-$18,000 km s$^{-1}$. 
Without a finer temporal sampling of the spectropolarimetric evolution, 
it is not clear whether this component 
could be the analog of the asymmetry observed in \ion{Si}{2} $\lambda$6355. 

Around maximum light, the broad and symmetric flux profile across the 
\ion{Si}{2} $\lambda$6355 feature suggests the absence of a distinct 
chemical `boundary layer' at any drastically different velocities. 
Therefore, the sharp cut-off in the corresponding polarization profile 
at the low-velocity end can be attributed to the presence of a separate 
Si-rich component at a slightly higher velocity. It may be indicative of 
Si-rich matter at lower velocity and with similar geometry as the 
Thomson-scattering photosphere. The apparent asymmetry of the HV 
components of the Si-rich ejecta is consistent with an off-center 
delayed detonation. Such an asymmetric explosion would compress the core 
more significantly towards the direction away from the center, therefore 
pushing the intermediate mass elements and developing the asymmetric HV 
components (i.e. see Section 4.3 of \citealp{Fink_etal_2010} and Fig.~1 
of \citealp{Bulla_etal_2016b}). The absence of a similar structure at the 
first epoch would imply that the asymmetry becomes visible as more inner 
parts of the WD are exposed. This seems consistent with an off-center 
delayed detonation. 

We suggest that the width of the wing towards the lower velocities 
characterizes the scale height of the Si-rich ejecta above the photosphere. 
Such an asymmetric polarization profile may not necessarily be a ubiquitous 
phenomenon since ignition of detonation at small off-center distances 
would lead to a relatively uniform distribution of silicon (see 
\citealp{Hoeflich_etal_2002, Hoeflich_etal_2006}). A slightly off-centered 
detonation leads to a relatively uniform distribution of the elements 
synthesized near the central region. The velocity spread is, therefore, 
smaller so that there would be no splitting into NV and HV components of 
\ion{Si}{2} $\lambda$6355. This is in agreement with the study based on 
larger samples \citep{Silverman_etal_2015}. For example, HV features of 
\ion{Si}{2} $\lambda$6355 are seen in only $\sim$one-third of the Type Ia 
SNe even at early times ($t\textless-5$ days). Such HV features tend to 
appear at the earliest phases and in objects with large photospheric 
velocities \citep{Silverman_etal_2015}. By contrast, HV features of 
\ion{Ca}{2} NIR3 are observed in $\sim$91\% of the Type Ia SNe with 
spectra obtained earlier than day $-4$ \citep{Silverman_etal_2015}. 
Spectropolarimetry of a solid sample would deliver an alternative method 
of identifying and resolving the different components in the ejecta, 
which often are ambiguous in flux spectra. At earlier phases, the absence 
of such asymmetry would suggest that the Si-rich ejecta are well above the 
photosphere. A gradually-changing polarization profile towards the higher 
velocities may imply that the outer part of the Si-rich ejecta has already 
developed a diffuse structure above the photosphere. If indeed the 
polarized Ca-rich component at $\sim$8020$-$8080 \AA\ ($\sim-$18,000 km 
s$^{-1}$) is related to the asymmetric polarization profile of \ion{Si}{2} 
$\lambda$6355, it could be interperated as a HV Ca-rich layer in the 
ejecta formed further out relative to the Si-rich layers. This is 
corroborated by the higher velocities measured across the NV \ion{Ca}{2} 
NIR3 feature. 

\subsection{Implications of the Polarization and Different Models}
We compare the polarization of SN\,2018gv to simulations by 
\citet{Bulla_etal_2016a, Bulla_etal_2016b}, who calculated polarization 
spectra for different Type Ia SN explosion models at various equatorial 
viewing angles, and off-center delayed-detonation models with various
off-center points for a delayed detonation model for a normal-bright SNe \citep{Hoeflich_etal_2006}. The low continuum and line polarizations as 
early as day $-$13.6 contradicts the high polarization levels of 
violent merger models \citep{Pakmor_etal_2012, Bulla_etal_2016a}. 
The observed low continuum, the moderate levels of \ion{Si}{2} and 
\ion{Ca}{2} NIR3 line polarizations, and the low polarization of 
\ion{O}{1} $\lambda$7774 lead us to consider that SN\,2018gv is broadly 
consistent with the overall geometric properties predicted by 
off-centered delayed-detonation models (i.e. 
\citealp{Seitenzahl_etal_2013,  Bulla_etal_2016b}) and sub-$M_{\rm Ch}$ 
double-detonation models (i.e. \citealp{Fink_etal_2010, Bulla_etal_2016b}). 

Regardless of the presence of HV \ion{Si}{2} wing which is unresolved 
in the flux spectra, the observed pre- and near-maximum light 
polarization properties of SN\,2018gv are similar to those of other 
normal Type Ia SNe, i.e. SNe\, 1996X \citep{Wang_etal_1997}, 2001el 
\citep{Wang_etal_2003_01el},  2004dt \citep{Wang_etal_2006_04dt}, 
2006X \citep{Patat_etal_2009}, 2011fe \citep{Milne_etal_2017}, 
2012fr \citep{Maund_etal_2013}, 2014J \citep{Porter_etal_2016}, and 
2016coj \citep{Zheng_etal_2017}. 
\citet{Kasen_etal_2003} presented models of SN\,2001el and provided 
a comprehensive parametrization for several possible configurations 
of the HV \ion{Ca}{2} loop. The complex \ion{Ca}{2} loop profile 
observed from SN\,2018gv could be interpreted as multiple clumped 
shells, or a superposition of multiple clumped shells with a 
characteristic ellipsoidal shell. More work is needed to account 
for the formation of such a complex HV polarization feature and its 
different orientation from the photosphere. 

Additionally, comparing SN\,2018gv to the simulations of 
\citet{Bulla_etal_2016b}, we infer that the relatively high line 
polarization of SN\,2018gv around the peak (i.e. $\sim$0.87$\pm$0.05\%) 
does not favor a large number of ignition kernels in delayed-detonation 
models \citep{Seitenzahl_etal_2013}. A large number of ignition kernels 
(i.e. N$\textgreater$100) would lead to a more symmetric distribution 
of intermediate-mass elements and thus produce a generally lower degree 
of line polarization. More theoretical modeling is needed to account 
for the HV features observed in \ion{Si}{2} $\lambda$6355 and \ion{Ca}{2} 
NIR3 lines and to investigate the shape and time-evolution of the flux 
and polarized spectra \citep{Mulligan_Wheeler_2018}. 

In the following and guided by the flux spectra, we discuss the 
polarization characteristics of SN\,2018gv in terms of off-center 
delayed detonations \citep{Hoeflich_etal_2006} which are based on 
spherical delayed detonation models for the deflagration phase 
\citep{Hoeflich_etal_2017}. In addition, we perturb the underlying 
structure to use the observed polarization to identify and quantify 
the underlying aspherical components, namely asymmetry in density 
and abundances. The treatment of the asymmetries as perturbation is 
justified because, overall, they are small. We use spherical models 
as baseline because they suppress deflagration mixing, a requirement 
from the discussion above and many observations from the 
\citep{Hoeflich_Stein_2002, Fesen_etal_2007, Diamond_etal_2015}, and 
references therein. Possible reasons may include the effect of 
high-magnetic fields as indicated by late-time light curves, near 
mid-IR spectra 
\citep{Penney_etal_2014, Remming_Khokhlov_2014, Hristov_etal_2018}. 
We choose deflagration to detonation transition (DDT) models for 
normal-bright Type Ia SNe because the photometric characteristics of 
SN\,2018gv are very similar to the observations of the light curves, 
color curves and color-magnitude diagrams as shown and as discussed 
in Section~\ref{section_lc}. As baseline-model we use model 27 with 
the velocity, density and abundance distribution given in Figures 1 
and 2 of \citet{Hoeflich_etal_2017}. Model 27 has a rise time  of 
$18.8$ d and $\Delta m_{15}(B/V)= 0.62/0.98$ mag compared to 
$18.51 \pm 0.92$ day and $0.63/0.96$ mag of SN\,2018gv. 

\begin{figure}[!h]
\epsscale{0.8}
\plotone{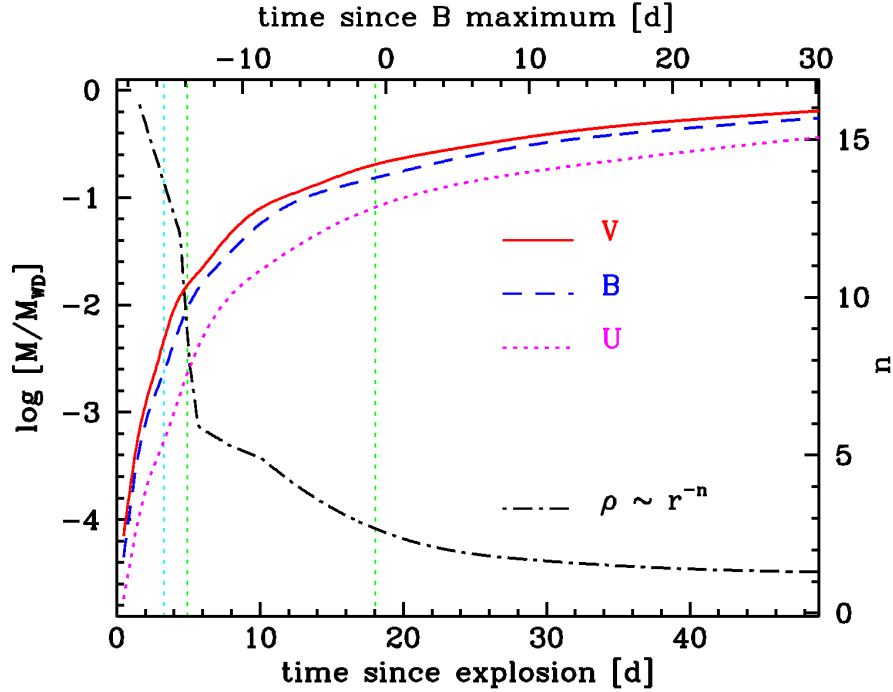}
\caption{Mass above the photosphere as a function of time for the 
normal-bright delayed-detonation model 27 \citep{Hoeflich_etal_2017}. 
In addition, we give exponent $n$ which approximates the density 
structure at the photosphere in $B$ and $V$. The vertical lines at 
3.3, 4.9 and 18.0 days since explosion mark the times with the first 
flux (cyan) and the polarization spectra (green), respectively. The 
uncertainty in the phase are $\approx 0.92$ days corresponding. In 
the $V-$band range, we probe the outer $(0.3-1)\times10^{-2} M_{\odot}$, 
$(1-3) \times 10^{-2} M_{\odot}$ and $0.2-0.3 M_{\odot}$, respectively.
} 
\label{model_struc}
\end{figure}

In Fig.~\ref{model_struc}, we show the mass above the photosphere 
as a function of time, and the corresponding density slope $n$ for 
$\rho \propto r^{-n}$ of the delayed detonation model cited above.
Note that the overall structures are rather similar for a wide range 
of models including sub-$M_{\rm Ch}$ but the mass fraction scales 
with the total mass. 

As discussed in previous sections, SN\,2018gv shows many characteristics 
of model 27: a) We see little \ion{C}{2} and \ion{O}{1} early on 
corresponding to the outer layers at $\approx 5 \times 10^{-3} M_\odot$; 
b) line wings of \ion{Si}{2} $\lambda$6355 extends to the very outer 
layers, i.e. $\textgreater$23,000 km s$^{-1}$, and evolves smoothly with 
time which starts at about $3 \times10^{-3} M_{\odot}$ in the wings, 
and the \ion{Si}{2} line extends down to about 0.3 to 0.4 $M_{\odot}$ 
from the outside. We note that this early start and such a large, 
continuous range for Si is barely compatible with double-detonations of 
a sub-$M_{\rm Ch}$ mass WD which would require at least 
$2 \times 10^{-2}$ and $3\times 10^{-3} M_\odot$ for a WD with 0.6 and 
1.2$M_\odot$, respectively \citep{Shen_Moore_2014}. 
The long, logarithmic decline of the Si abundances till close to the 
surface is inherent to detonations going through C/O matter in $M_{\rm Ch}$ 
mass models. It also leads to the inverted triangular shaped Si 
line-wings ($\triangledown$) visible till about $-$13.2 day in 
Fig.~\ref{Fig_line_region} corresponding to a photospheric velocity 
of $\approx$15,000 km s$^{-1}$ which, in the model, marks the 
transition to the Si/S dominated layers in the envelope. A linear line 
wing is a consequence of the logarithmic decline of the Si abundance in 
combination of low optical depth \citep{Quimby_etal_2007}.

Based on the delayed-detonation model, we want to discuss the 
polarization of SN\,2018gv in terms of parameterized asphericity. 
Polarization can be formed by i) aspherical density distributions 
\citep{Van_de_Hulst_1957, Hoeflich_1991}; ii) partial obscuration 
of the underlying Thomson-scattering dominated photosphere by 
line-absorption \citep{Kasen_etal_2006,Hoeflich_etal_2006}, and 
iii) off-center energy sources \citep{Chugai_1992, Hoeflich_etal_1995}. 
SN\,2018gv does not show a flip in the polarization angle and, thus, 
this case is not applicable. 

We first want to consider asymmetric chemical distributions produced by 
off-center DDT models in which the axis of chemical asymmetry is given 
by the center of the density distribution and the point of the DDT. The 
polarization of SN\,2018gv can be understood in very similar terms as 
SN\,2004dt, a normal-bright Type Ia SN but with a slightly steeper 
decline rate (i.e., $\Delta m_{B} (15) = 1.13\pm0.04$ mag, 
\citealp{Wang_etal_2012_UV}). At about one week before maximum, the 
polarization in SN\,2004dt was high in \ion{Si}{2} lines, weak in 
\ion{Mg}{2} without showing \ion{O}{1} polarization 
\citep{Hoeflich_etal_2006, Wang_etal_2006_04dt}. 
The observation was reproduced by an off-center DDT model seen from 
$30^{\circ}$ relative the axis of symmetry. At this phase the photosphere 
passes the interface between explosive carbon and incomplete silicon 
burning. It could be understood in terms of covering the underlying 
scattering photosphere by optically thick lines in combination with the 
chemical gradients. Namely, all strong \ion{Si}{2} lines show similar 
polarization because both are optically thick at this phase, the 
magnesium polarization is formed in thin shell produced by explosive 
carbon burning. Oxygen shows little polarization because, in a C/O WD, 
in explosive carbon burning increases O to about 70\% but the constrast 
is too small with the unburned C/O layers at about 50\%. 

Similarly, SN\,2018gv shows no \ion{O}{1} polarization because there 
is neither an abundance jump in Oxygen in the line forming region at  
$-$13.6 and $-$0.5 day. The \ion{Si}{2} $\lambda$6355 line shows little 
polarization at $-$13.6 day because the line is partially optically 
thin as discussed above as indicated by the linear blue line wings 
(see above) and, thus, there is little asymmetric covering of the 
underlying photosphere. By $-$0.5 day, the underlying has entered the 
Si/S-rich region resulting $p$ of about 0.8\%. Compared to SN\,2004dt 
with a $p(\rm Si II)=$1.8\% \citep{Wang_etal_2006_04dt, Hoeflich_etal_2006}.
The line polarization may lower because, for the higher luminosity of 
SN\,2018gv, the Si region is shifted towards lower density and lower 
optical depth, at a different phase or we may see the object from a 
slightly larger inclination. The first effect likely is important 
because the flux in the blue wing of the \ion{Si}{2} line is hardly 
polarized. Future dense series of earlier-phase spectroplarimetry 
will allow for a separation of the effects. 

%
Now, we want to shift towards the continuum polarization which, early 
on, can be understood in terms of an asymmetric density/electron 
distribution. The low continuum polarization observed gives important 
limits on the overall this asymmetry. During the phase considered, 
the photosphere is formed in the C/O/Mg/Si/S layers, and the 
ionization fraction does hardly change because similar ionization 
potentials of the main electron donors. As a result, the electron 
distribution at the photosphere can be approximated by the density 
distribution. The polarization depends on the density structure and 
decreases with steeper density slopes because the incoming radiation 
at the photosphere becomes more isotropic. 

\begin{figure}[!h]
\epsscale{0.8}
\plotone{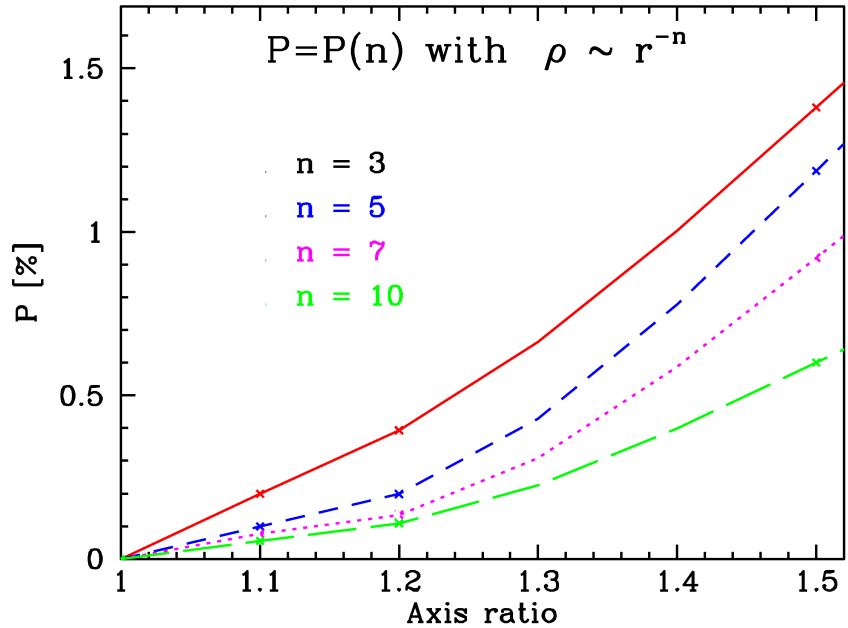}
\caption{Continuum polarization as a function of asymmetry in a scattering 
dominated frequency range formed in the region of intermediate mass elements. 
We assume a oblate ellipsoide with an axis ratio of A/B and depolarization 
at a Thomson optical depth of 5.} 
\label{pol_struc}
\end{figure}

In Fig.~\ref{pol_struc}, we give the the polarization seen equator-on for 
oblate ellipsoids with an axis ratio of A/B for various density slopes $n$ 
for structures realized in models. For SN\,2018gv, the continuum polarization 
is 0.20\%$\pm$0.13\% and 0.15\%$\pm$0.16\% at about $-$13.6 day and maximum 
light, respectively. The outer $(1-2)\times 10^{-2} M_{\odot}$ are aspherical 
with a well defined axis because the constant polarization angle. The size of 
asymmetry is about 10\% to 35\% when seen equator on. We see significant 
asymmetry early on. By maximum light, $p$ declined and it is consistent with 
no polarization. The size of asymmetry is between 0\% and 15\%. Note that $p$ 
goes roughly as $\propto {\rm sin}^2(\Theta)$ but, due to multiple scattering 
effects, the change of $p$ becomes less steep \citet{Hoeflich_1991}. 
 
It is beyond the scope of this paper for a full analysis, but we would like 
to put SN\,2018gv into context of different mechanism for producing asymmetry. 
a) Dynamical or head on collisions of WD show larger asymmetries and can be 
ruled out because they predict larger asymmetries in the inner layers or 
off-center energy sources \citep{Benz_etal_1990,  Pakmor_etal_2011, 
Sato_etal_2016, Katz_etal_2016, Garcia_Berro_etal_2017, GB_2017}; b) Rapidly 
rotating WDs close to $M_{\rm Ch}$ produce asymmetric initial configuration 
and are consistent with the polarization seen in this object 
\citep{Eriguchi_Mueller_1993, Yoshida_2018}; He-triggered detonations do produce 
strong asymmetries in the outer layers and, consistent with SN\,2018gv, show 
almost spherical inner layers \citep{Shen_Moore_2014}. One problem, though, may 
be that the early observations place the photosphere into He-rich layers and 
its burning products for all but WDs at the upper end of the masses possible. 
Both the flux and polarization spectra indicate no 'double-chemical'
structure of Si/S starting within the outer $3 \times 10^{-3} M_{\odot}$. 
The lack of \ion{O}{1} polarization puts an even stronger limit on a He layer 
atop $10^{-2} M_{\odot}$.

Though the picture of a rotating WD within the off-center delayed-detonation 
scenario seems to be consistent, we need earlier observations and denser 
series of spectroplarimetry obtained than obtained for SN\,2018gv.

\section{Summary~\label{section_summary}}
We present extensive UV and optical photometry, optical 
spectroscopy as well as optical linear spectropolarimetry of 
SN\,2018gv. We are able to draw the following conclusions: \\
\ \\
1) The rising light curve is consistent with a power-law exponent 
[$f(t) \propto (t-t_0)^{n}$]. Our fitting to the early $g'-$band 
light curve yields a rise time of $\sim$18 days and an index of $n\sim$2; \\
\ \\
2) We adopt the `CMAGIC' method and estimate the host-galaxy reddening 
towards SN\,2018gv to be negligible [i.e. 
$E(B-V)_{\rm host} = 0.028\pm0.027$ mag]; \\
\ \\
3) The light curves show that SN\,2018gv is a normal Type Ia SN with a 
typical $B-$band peak magnitude of $-19.06\pm0.05$ and a luminosity 
decline rate $\Delta m_{15}(B) = 0.96\pm0.05$ in the rest-frame; \\
\ \\
4) The comprehensive photometry allows us to construct the UV-optical 
pseudo-bolometric light curve of SN\,2018gv in the 1660$-$8180 \AA\ range. 
The bolometric luminosities are estimated after further correcting for the 
fraction of the NIR flux. The maximum bolometric luminosity gives 
$L = (1.186\pm0.064\pm0.022) \times 10^{43}$ erg s$^{-1}$, yielding a 
synthesized nickel mass of 0.56$\pm$0.08$M_{\odot}$; \\
\ \\
5) The early-time spectra of SN\,2018gv exhibit strong similarity to those of 
the normal Type Ia SN\,2011fe in most respects. No significant HV component 
is detected in the \ion{Si}{2} $\lambda$6355 absorption feature as early as 
$-$15.2 days relative to the $B-$band maximum. Strong HV features are 
unambiguously detected in \ion{Ca}{2} NIR3 features. At velocities from 
$-$27,000 to $-$20,000 km s$^{-1}$, they are detached from the photosphere; \\
\ \\
6) The earliest spectropolarimetry to date of a Type Ia SN at day $-$13.6 
has been obtained for SN\,2018gv. The degree of continuum polarization is 
as low as $\sim$0.2\%. The observation of low continuum polarization 
overlaid by significant line polarization is inconsistent with the 
double-degenerate violent merger case but consistent with the single-degenerate 
delayed-detonation and double-detonation models; \\
\ \\
7) An asymmetric polarization profile of the \ion{Si}{2} $\lambda$6355 
feature around the light curve peak has been observed. 
The very strong polarization at line center implies a sharp increase 
of the opacity gradient in the direction transverse to the line of sight 
as the photosphere recedes to the interface layer of Fe and IME. The 
sharp drop of polarization towards lower velocities suggests the presence 
of a distinct chemical `boundary layer', while the slow decline of 
polarization towards higher velocities indicates the formation of a 
rather diffuse and homogeneous ejecta profile above the photosphere; \\
\ \\
8) The flux and polarization spectra are consistent with classical 
off-center delayed detonation transitions with $M_{\rm Ch}$ mass WDs but 
originating from a rapidly rotating WD. However, our observations are 
insufficient to address a possible alignment of the symmetry axes in 
density and abundances. Double-detonation models would require masses 
close to $1.2 M_\odot$. \\
\ \\
9) Such an asymmetric \ion{Si}{2} $\lambda$6355 polarization profile 
around the SN peak luminosity may indicate that in opposite directions 
from the stellar center IMEs were produced over different distances. 
This is consistent with an off-center delayed detonation. 

Taken together, our observations suggest that SN\,2018gv resembles 
other normal Type Ia SNe in many respects. Polarimetry of infant SNe 
(SNe discovered within 1$-$2 days after their explosion) provide an 
important probe of the kinematics and chemical structures of SNe and 
their circumstellar environment during the final stages of the 
progenitor evolution. Wide-field, high-cadence transient surveys, 
such as the Zwicky Transient Facility \citep{Bellm_etal_2019}, ATLAS 
\citep{Tonry_etal_2011}, ASAS-SN \citep{Shappee_etal_2014}, and DLT40 
\citep{Tartaglia_etal_2018} will discover and monitor nearby infant 
SNe over the next few years. A large polarimetry sample of such objects 
will enable stringent constraints on the explosion mechanisms and the 
circumstellar environment of Type Ia SNe. 

\acknowledgments 
We are grateful to the European Organisation for Astronomical Research in 
the Southern Hemisphere for the generous allocation of observing time. 
We would like to especially thank the staff of the Paranal Observatory for 
their proficient and diligent support of this project in service mode. 
We acknowledge Gwen Eadie, Dino Bekte{\v s}evi{\'c}, and Bryce Bolin, 
co-observers on the night the DIS spectrum was obtained. 
The polarimetry studies in this work are based on observations made with 
ESO VLT at the Paranal Observatory under programme ID 0102.D-0528. 
PyRAF, PyFITS, STSCI$\_$PYTHON are products of the Space Telescope Science 
Institute, which is operated by AURA for NASA.
STScI is operated by the Association of Universities for Research in 
Astronomy, Inc., under NASA contract NAS5-26555. 
This research has made use of NASA's Astrophysics Data System Bibliographic Services. 
This research has made use of the SIMBAD database, operated at CDS, Strasbourg, France. 
This research has made use of the NASA/IPAC Extragalactic Database (NED) which 
is operated by the Jet Propulsion Laboratory, California Institute of Technology, 
under contract with the National Aeronautics and Space Administration. 
The research of Y. Yang is supported through a Benoziyo Prize Postdoctoral Fellowship. 
P. Hoeflich is supported by the NSF project ``Signatures of Type Ia Supernovae, 
New Physics and Cosmology'', grant AST-1715133. 
The research of J. Maund is supported through a Royal Society University Research Fellowship. 
The supernova research by L. Wang is supported by NSF award AST-1817099 and HST-GO-14139.001-A.
P. J. Brown was partially supported by a Mitchell Postdoctoral Fellowship. 
H.F. Stevance is supported by a PhD scholarship fron the University of Sheffield. 
This work at Rutgers University (S. W. Jha) is supported by NSF award AST-1615455. 
M. L. Graham acknowledges support from the DIRAC Institute in the 
Department of Astronomy at the University of Washington. 
The DIRAC Institute is supported through generous gifts from the Charles and 
Lisa Simonyi Fund for Arts and Sciences, and the Washington Research Foundation. 
This work makes used of data from the Las Cumbres Observatory Network. 
D. Hiramatsu, C. McCully, and G. Hosseinzadeh were supported by NSF grant AST-1313484. 
Research by D. J. Sand is supported by NSF grants AST-1821987 and 1821967. 
X. Wang is supported by the National Natural Science Foundation 
of China (NSFC grants 11178003 and 11325313). 
J. C. Wheeler is supported by NSF grant AST-1813825. 
J. Zhang is supported by the NSFC (grants 11773067, 11403096), the Youth 
Innovation Promotion Association of the CAS (grants 2018081) and the Western 
Light Youth Project of the CAS. 
\clearpage

\begin{table}[!h]
\caption{Photometric standards in the SN\,2018gv field}
\begin{scriptsize}
\begin{center}
\begin{tabular}{ccccccccc}
\hline
ID  &  $\alpha$(J2000)  & $\delta$(J2000)   &  $U$ (mag)    & $B$ (mag)     &  $V$ (mag)    &  $g$ (mag)    &  $r$ (mag)   &  $i$ (mag)    \\
\hline
1  &  8:05:39.190 & -11:23:57.77  &         &  13.199(028)  &  12.707(015)  &  12.897(027)  &  12.590(044)  &  12.533(044)  \\
2  &  8:05:34.667 & -11:23:32.26  &         &  14.029(027)  &  12.973(023)  &  13.439(040)  &  12.640(047)  &  12.402(029)  \\
3  &  8:05:38.418 & -11:30:10.37  &  11.80  &  11.856(035)  &  11.494(010)  &  11.612(041)  &  11.417(038)  &  11.432(028)  \\
4  &  8:05:34.874 & -11:22:33.51  &         &  13.008(033)  &  12.655(022)  &  12.754(023)  &  12.598(043)  &  12.638(035)  \\
5  &  8:05:44.471 & -11:32:12.54  &  11.07  &  10.888(022)  &  10.714(017)  &  10.729(030)  &  10.772(074)  &  10.870(039)  \\
6  &  8:06:03.681 & -11:28:56.15  &         &  13.276(041)  &  11.996(022)  &  12.597(047)  &  11.549(041)  &  11.166(025)  \\
7  &  8:05:15.921 & -11:34:21.17  &         &  13.545(050)  &  11.890(051)  &  12.631(063)  &  11.243(090)  &  10.072(000)  \\
8  &  8:05:08.379 & -11:31:24.17  &         &  12.948(021)  &  12.450(020)  &  12.644(024)  &  12.308(048)  &  12.239(027)  \\
9  &  8:06:06.937 & -11:26:14.69  &         &  13.309(036)  &  11.904(026)  &  12.577(048)  &  11.366(045)  &  10.914(038)  \\
10 &  8:06:07.016 & -11:31:35.82  &         &  12.222(020)  &  11.694(026)  &  11.904(033)  &  11.551(051)  &  11.443(041)  \\
11 &  8:06:04.411 & -11:24:29.13  &         &  13.381(032)  &  12.293(019)  &  12.786(045)  &  11.944(044)  &  11.680(026)  \\
12 &  8:04:46.849 & -11:23:58.66  &         &  13.279(015)  &  11.684(033)  &  12.448(044)  &  11.076(045)  &  10.496(090)  \\
13 &  8:04:59.549 & -11:19:29.13  &  13.73  &  12.901(009)  &  11.726(017)  &  12.247(043)  &  11.312(037)  &  10.968(022)  \\
14 &  8:05:09.374 & -11:16:13.25  &  10.51  &  10.836(020)  &  10.839(024)  &  10.740(036)  &  10.959(055)  &  11.205(029)  \\
15 &  8:05:52.649 & -11:23:37.18  &         &  13.838(022)  &  13.274(029)  &  13.501(032)  &  13.112(041)  &  13.038(027)  \\
16 &  8:05:41.910 & -11:27:49.40  &         &  14.812(026)  &  13.737(033)  &  14.240(067)  &  13.356(043)  &  13.071(015)  \\
17 &  8:05:30.515 & -11:24:07.41  &         &  14.503(041)  &  13.956(018)  &  14.183(034)  &  13.792(048)  &  13.717(025)  \\
18 &  8:05:48.271 & -11:28:03.66  &         &  14.380(025)  &  13.840(016)  &  14.068(032)  &  13.699(044)  &  13.603(044)  \\
19 &  8:06:05.047 & -11:21:19.65  &         &  13.645(022)  &  13.253(030)  &  13.378(013)  &  13.200(029)  &  13.153(042)  \\
20 &  8:05:15.276 & -11:25:40.51  &         &  14.811(022)  &  13.774(036)  &  14.219(022)  &  13.460(033)  &  13.141(028)  \\
21 &  8:05:19.620 & -11:20:45.77  &         &  13.686(022)  &  13.208(016)  &  13.382(034)  &  13.089(053)  &  13.019(012)  \\
22 &  8:05:05.024 & -11:20:18.45  &         &  13.324(020)  &  12.814(011)  &  13.009(029)  &  12.678(056)  &  12.584(022)  \\
23 &  8:05:00.250 & -11:25:08.76  &         &  13.549(011)  &  12.696(031)  &  13.056(030)  &  12.391(039)  &  12.180(034)  \\
24 &  8:05:10.005 & -11:16:43.26  &         &  14.011(023)  &  12.622(035)  &  13.342(030)  &  12.007(047)  &  11.407(013)  \\
25 &  8:05:46.770 & -11:31:36.23  &         &  13.927(031)  &  12.901(030)  &  13.367(044)  &  12.530(046)  &  12.228(023)  \\
26 &  8:05:18.257 & -11:26:31.45  &         &  15.140(049)  &  14.023(020)  &  14.521(035)  &  13.636(041)  &  13.288(021)  \\
27 &  8:05:05.733 & -11:17:44.99  &  13.74  &  12.718(022)  &  11.826(013)  &  12.219(032)  &  11.544(050)  &  11.298(022)  \\
\hline
\end{tabular}
\end{center}
\label{Table_phot_standard}
\end{scriptsize}
\end{table}
\clearpage

\begin{table}[!h]
\caption{Log of spectroscopic observations of SN\,2018gv \label{Table_log_spec}}
\begin{scriptsize}
\begin{tabular}{cccccc}
\hline
  UT Date      &      MJD      &  Phase$^a$ &        Range       &   Exposure    &  Instrument/Telescope  \\
   (2018)      &               &   (days)   &        (\AA\ )      &      (s)      &                        \\
\hline
Jan 16 12:08   &    58134.50   &   -15.18   &      3200$-$9900   &     2700      &  FLOYDS/LCO 2.0 m FTN  \\
Jan 17 08:13   &    58135.34   &   -14.34   &      3200$-$9900   &     2700      &  FLOYDS/LCO 2.0 m FTN  \\
Jan 17 11:53   &    58135.50   &   -14.19   &      3500$-$9900   &      300      &  GMOS-N/Gemini-N 8.1 m \\
Jan 18 02:20   &    58136.10   &   -13.59   &      4400$-$9200   & 4$\times$900  &  FORS2/VLT 8.2 m  \\
Jan 18 11:05   &    58136.46   &   -13.23   &      3200$-$9900   &     2700      &  FLOYDS/LCO 2.0 m FTN  \\
Jan 19 06:55   &    58137.29   &   -12.40   &      3600$-$10200  &      767      &  LRS2/HET 10 m  \\
Jan 20 10:04   &    58138.42   &   -11.27   &      3200$-$9900   &     2700      &  FLOYDS/LCO 2.0 m FTN  \\
Jan 23 09:42   &    58141.41   &    -8.28   &      3200$-$9900   &     1800      &  FLOYDS/LCO 2.0 m FTN  \\
Jan 25 07:57   &    58143.33   &    -6.36   &      3200$-$9900   &     1800      &  FLOYDS/LCO 2.0 m FTN  \\
Jan 25 20:35   &    58143.75   &    -5.94   &      3500$-$8900   &     2401      &  RSS/SALT 11 m  \\
Jan 28 06:33   &    58146.27   &    -3.41   &      3600$-$10200  &      507      &  LRS2/HET 10 m  \\
Jan 29 08:50   &    58147.37   &    -2.32   &      3200$-$9900   &     1800      &  FLOYDS/LCO 2.0 m FTN  \\
Jan 31 05:08   &    58149.21   &    -0.47   &      4400$-$9200   & 4$\times$120  &  FORS2/VLT 8.2 m  \\
Feb 04 10:13   &    58153.43   &     3.74   &      3200$-$9900   &     1800      &  FLOYDS/LCO 2.0 m FTS  \\
Feb 08 10:45   &    58157.45   &     7.76   &      3200$-$9900   &     1800      &  FLOYDS/LCO 2.0 m FTS  \\
Feb 13 06:38   &    58162.28   &    12.59   &      3200$-$9900   &     1800      &  FLOYDS/LCO 2.0 m FTN  \\
Mar 01 12:06   &    58178.50   &    28.82   &      4800$-$9300   &     1800      &  FLOYDS/LCO 2.0 m FTS  \\
Mar 09 13:45   &    58186.57   &    36.89   &      3300$-$9900   &     1800      &  FLOYDS/LCO 2.0 m FTS  \\
Mar 17 11:28   &    58194.48   &    44.79   &      3400$-$9900   &     1800      &  FLOYDS/LCO 2.0 m FTS  \\
Mar 22 03:49   &    58199.66   &    49.97   &      3800$-$9200   &      900      &  DIS/ARC 3.5 m  \\
Mar 23 11:28   &    58200.48   &    50.79   &      3300$-$9900   &     1800      &  FLOYDS/LCO 2.0 m FTS  \\
Mar 29 10:26   &    58206.43   &    56.75   &      4800$-$9300   &     1800      &  FLOYDS/LCO 2.0 m FTS  \\
Apr 05 10:43   &    58213.45   &    63.76   &      3200$-$9900   &     1800      &  FLOYDS/LCO 2.0 m FTS  \\
Apr 11 11:36   &    58219.48   &    69.80   &      3600$-$9300   &     2700      &  FLOYDS/LCO 2.0 m FTS  \\
Apr 17 11:04   &    58225.46   &    75.77   &      3700$-$9200   &     2700      &  FLOYDS/LCO 2.0 m FTS  \\
Apr 25 06:52   &    58233.29   &    83.60   &      3200$-$9900   &     2700      &  FLOYDS/LCO 2.0 m FTN  \\
\hline
\end{tabular}\\
{$^a$}{Relative to the $B-$band maximum light, MJD$_{Bmax}$ = 58,149.698$\pm$0.510 / 2018 Jan 31 16:45} \\
\end{scriptsize}
\end{table}

\begin{table}[!h]
\caption{LCO $UBg'Vr'i'$ photometry of SN\,2018gv. The table is published in its entirety in the electronic edition of the journal. 
A portion is displayed for guidance regarding its form and content.\label{Table_phot}}
\begin{tiny}
\begin{tabular}{cccccccccccc}
\hline
MJD$^a$ & $U$ (mag)   & MJD$^a$ & $B$ (mag)   & MJD$^a$ & $V$ (mag)   & MJD$^a$ & $g'$ (mag)  & MJD$^a$ & $r'$ (mag)  & MJD$^a$ & $i'$ (mag)  \\
\hline
...     & ...         & 134.078 & 16.763(008) & 134.086 & 16.426(008) & 134.090 & 16.562(006) & 134.094 & 16.417(008) & 134.098 & 16.777(014) \\
...     & ...         & 134.082 & 16.772(008) & 134.086 & 16.415(008) & 134.094 & 16.552(006) & 134.098 & 16.425(008) & 134.102 & 16.769(014) \\
134.863 & 16.330(013) & 134.871 & 16.123(005) & 134.879 & 15.896(006) & 134.883 & 15.969(004) & 134.887 & 15.914(006) & 134.891 & 16.254(011) \\
134.867 & 16.312(014) & 134.875 & 15.111(005) & 134.879 & 15.888(006) & 134.887 & 15.964(004) & 134.891 & 15.911(006) & 134.895 & 16.269(012) \\
 ...    &  ...        &  ....   &  ...        &  ...    &  ...        &  ...    &  ...        &  ...    &  ...        &  ...    &  ...        \\
150.820 & 12.557(002) & 150.828 & 12.813(002) & 150.828 & 12.861(003) & 150.832 & 12.818(002) & 150.836 & 12.941(003) & 150.836 & 13.593(006) \\
150.824 & 12.556(002) & 150.828 & 12.817(002) & 150.832 & 12.862(003) & 150.832 & 12.817(002) & 150.836 & 12.941(003) & 150.836 & 13.582(006) \\
 ...    &  ...        &  ....   &  ...        &  ...    &  ...        &  ...    &  ...        &  ...    &  ...        &  ...    &  ...        \\
254.980 & 17.218(044) & 254.984 & 16.920(013) & 254.988 & 16.472(013) & 254.992 & 16.542(009) & 254.996 & 16.743(016) & 254.996 & 17.164(034) \\
254.980 & 17.354(112) & 254.988 & 16.899(013) & 254.988 & 16.497(013) & 254.992 & 16.536(008) & 254.996 & 16.727(016) & 254.996 & 17.072(031) \\
264.723 & 17.427(084) & 264.727 & 17.061(030) & 264.730 & 16.717(029) & 264.734 & 16.653(017) & 264.738 & 17.021(033) & 264.738 & 17.393(064) \\
264.727 & 17.461(088) & 264.730 & 17.094(031) & 264.734 & 16.657(028) & 264.734 & 16.691(018) & 264.738 & 17.020(032) & 264.742 & 17.390(065) \\
...     & ...         & 286.961 & 17.404(026) & 286.965 & 17.136(023) & 286.965 & 17.038(013) & 286.969 & 17.620(031) & 286.973 & 18.056(070) \\
...     & ...         & 286.961 & 17.400(026) & 286.965 & 17.189(024) & 286.969 & 17.047(013) & 286.969 & 17.660(032) & 286.973 & 17.972(064) \\
\hline
\end{tabular}\\
{$^a$}{MJD - 58,000, MJD of the $B-$band maximum light;} 
\end{tiny}
\end{table}

\begin{table}[!h]
\caption{Photometric parameters of SN\,2018gv \label{Table_phot_para}}
\begin{scriptsize}
\begin{tabular}{cccccccc}
\hline
  Band  &  $\lambda_{pivot}$  & $A_{\lambda}^{\rm MW}$ & $A_{\lambda}^{\rm Host}$ & $t_{max}^a$ & $\Delta m_{15}$ & $m_{peak}$   &   $M_{peak}$  \\
        &  (\AA )        & (mag)                  & (mag$^b$)                    &  (day)      &  (mag$^c$)      &  (mag$^c$)        &   (mag$^c$)   \\
\hline
   $U$  &  3600               & 0.251                  & 0.131$\pm$0.125          &  47.518(430)  &  1.027(050)     &  12.230(046)      &  -19.442(048)  \\
   $B$  &  4365               & 0.210                  & 0.109$\pm$0.110          &  49.698(510)  &  0.963(054)     &  12.609(051)      &  -19.063(053)  \\
   $V$  &  5362               & 0.159                  & 0.087$\pm$0.086          &  50.598(520)  &  0.626(028)     &  12.696(025)      &  -18.976(027)  \\
  $g'$  &  4736               & 0.191                  & 0.102$\pm$0.097          &  49.928(545)  &  0.726(043)     &  12.641(044)      &  -19.031(045)  \\
  $r'$  &  6232               & 0.132                  & 0.068$\pm$0.068          &  50.078(450)  &  0.703(018)     &  12.778(043)      &  -18.894(044)  \\
  $i'$  &  7529               & 0.098                  & 0.053$\pm$0.052          &  47.688(405)  &  0.878(016)     &  13.342(055)      &  -18.330(066)  \\
\hline
\end{tabular}\\
{$^a$}{Uncertainties of maximum-light dates in unit of 0.01 days. The date is MJD$-$58100.} \\
{$^b$}{Uncertainties were not added when deducing the photometric parameters.} \\
{$^c$}{Uncertainties are indicated by the number in parentheses and in unit of 0.01 magnitude.}
\end{scriptsize}
\end{table}

\begin{table}[!h]
\caption{Fit Results of the different velocity components, velocities shown in absolute value. \label{Table_fitv}}
\begin{scriptsize}
\begin{tabular}{c|cc|cc|cc|cc|cc}
\hline
       &  \multicolumn{2}{c|}{\ion{Si}{2} $\lambda$5972} &  \multicolumn{2}{c|}{\ion{Si}{2} $\lambda$6355} &  \multicolumn{2}{c|}{\ion{C}{2} $\lambda$6580}  &  \multicolumn{2}{c|}{High Velocity Ca NIR3}      &  \multicolumn{2}{c}{Normal Velocity Ca NIR3}    \\
 Phase & $v$                 & $pEW$       & $v$                 & $pEW$       & $v$                 & $pEW$       & $v$                     & $pEW$       & $v$ & $pEW$                        \\
 (day) & ($\rm km \ s^{-1}$) & ( \AA )    & ($\rm km \ s^{-1}$) & (\AA)      & ($\rm km \ s^{-1}$) & (\AA)      & ($\rm km \ s^{-1}$)     & (\AA) & $\rm km \ s^{-1}$)   & (\AA) \\
\hline
-15.18 & 12894(277) & 11.2(1.4) & 16321(245) & 146.3(2.3) & 15038(254) & 10.4(1.2) & 27297(369) &  94.9(23.8) & 19233(610) & 238.2(33.6) \\
-14.34 & 11996(277) & 13.8(1.7) & 15322(245) & 138.2(2.4) & 14635(261) &  7.5(1.3) & 25636(373) &  78.2(37.6) & 18260(930) & 236.9(50.6) \\
-14.19 & 12172(138) & 19.6(1.9) & 15354(92)  & 136.0(2.5) & 14322(134) &  5.0(1.3) & 24942(203) & 150.9(11.5) & 16379(278) & 175.0(14.0) \\
-13.59 & 11999(375) & 15.2(2.8) & 14769(343) & 123.8(4.0) & 14209(386) &  4.1(2.1) & 24187(456) & 118.4(19.5) & 15804(582) & 172.5(27.5) \\
-13.23 & 11559(289) & 16.0(2.3) & 14390(246) & 124.0(2.9) & 14112(285) &  3.2(1.4) & 23941(424) & 101.5(29.1) & 16290(988) & 165.7(40.3) \\
-12.40 & 11569(115) & 24.2(1.5) & 14004(86)  & 113.5(1.9) & 13755(125) &  4.6(1.1) & 23528(330) & 102.2(14.2) & 16546(504) &  75.7(16.0) \\
-11.27 & 10619(316) & 12.0(2.3) & 13029(247) &  99.8(2.8) & 13309(342) &  2.8(1.6) & 22686(336) &  63.6(12.2) & 14142(589) &  94.7(19.3) \\
 -8.28 &  9225(507) &  8.5(3.0) & 11873(248) &  82.6(3.1) &            &           & 22200(302) &   35.5(6.3) & 12539(341) &  62.4(9.6)  \\
 -6.36 &  9667(422) &  5.7(2.5) & 11468(248) &  83.6(3.3) &            &           & 21579(316) &   32.0(6.5) & 11466(319) &  55.4(8.7)  \\
 -5.94 &  9117(522) & 14.6(3.6) & 11499(154) &  73.4(2.7) &            &           & 22148(241) &   26.3(5.5) & 12846(216) &  87.4(8.8)  \\
 -3.41 & 10490(275) & 10.1(2.0) & 11199(87)  &  81.4(2.1) &            &           & 21217(199) &   10.4(3.4) & 10905(541) &  24.8(7.2)  \\
 -2.32 &  9244(441) & 10.5(3.3) & 11022(248) &  89.2(3.6) &            &           & 20467(130) &   17.3(5.2) & 11639(283) &  75.5(8.1)  \\
 -0.47 & 10533(727) &  8.1(4.6) & 10984(345) &  84.2(5.3) &            &           & 20227(410) &    7.4(3.6) & 12551(350) & 107.4(6.8)  \\
  3.74 &            &           & 10673(251) &  92.7(5.7) &            &           &            &             & 11673(458) & 141.3(40.5) \\
  7.76 &            &           & 10650(249) &  92.1(5.0) &            &           & 18483$^a$  &             & 11626(311) & 275.5(25.3) \\
 \hline
\end{tabular}\\
{$^a$}{Multi-Gaussian component fitting failed. Velocity assigned by the local minimum and no error estimatation applied. } 
\end{scriptsize}
\end{table}

\begin{center}
\begin{scriptsize}
\begin{longtable}{cccc|cccc}
\caption{The pseudo-bolometric (UVO) and the estimated bolometric (UVOIR) luminosity of SN\,2018gv (left) and SN\,2011fe (right) \label{Table_bolo}}\\
\hline
Phase$^a$ & log\,$L$ (UVO) & Error$^b$      & log\,$L$ (UVOIR) & Phase$^c$ & log\,$L$ (UVO) &  Error$^b$      &  log\,$L$ (UVOIR) \\
Day       & (erg s$^{-1}$) & (erg s$^{-1}$) & (erg s$^{-1}$)   & Day       & (erg s$^{-1}$) &  (erg s$^{-1}$) &  (erg s$^{-1}$)   \\
\hline
\endfirsthead
\multicolumn{8}{c}%
{\tablename\ \thetable\ -- \textit{Continued from previous page}} \\
\hline
Phase$^a$ & log\,$L$ (UVO) & Error$^b$      & log\,$L$ (UVOIR) & Phase$^c$ & log\,$L$ (UVO) &  Error$^b$      &  log\,$L$ (UVOIR) \\
Day       & (erg s$^{-1}$) & (erg s$^{-1}$) & (erg s$^{-1}$)   & Day       & (erg s$^{-1}$) &  (erg s$^{-1}$) &  (erg s$^{-1}$)   \\
\hline
\endhead
\hline \multicolumn{8}{r}{\textit{Continued on next page}} \\
\endfoot
\hline
\endlastfoot

-15.61 & 41.463 & 0.022 & 41.532 & -16.31 & 40.962 & 0.016 & 41.030 \\
-14.81 & 41.693 & 0.022 & 41.761 & -15.95 & 41.151 & 0.016 & 41.219 \\
-14.57 & 41.747 & 0.022 & 41.815 & -15.32 & 41.409 & 0.016 & 41.478 \\
-14.38 & 41.805 & 0.022 & 41.874 & -14.95 & 41.535 & 0.016 & 41.603 \\
-12.72 & 42.192 & 0.023 & 42.261 & -14.31 & 41.713 & 0.016 & 41.782 \\
-12.34 & 42.266 & 0.023 & 42.335 & -13.95 & 41.813 & 0.016 & 41.882 \\
-11.62 & 42.382 & 0.023 & 42.452 & -12.96 & 42.039 & 0.016 & 42.108 \\
-11.42 & 42.434 & 0.023 & 42.503 & -12.33 & 42.156 & 0.016 & 42.226 \\
-10.55 & 42.562 & 0.024 & 42.631 & -11.96 & 42.238 & 0.016 & 42.308 \\
-10.45 & 42.574 & 0.024 & 42.643 & -11.33 & 42.345 & 0.016 & 42.414 \\
 -9.78 & 42.651 & 0.024 & 42.720 & -10.96 & 42.420 & 0.016 & 42.489 \\
 -9.60 & 42.670 & 0.024 & 42.739 & -10.33 & 42.501 & 0.016 & 42.571 \\
 -9.47 & 42.685 & 0.024 & 42.753 &  -9.95 & 42.562 & 0.016 & 42.632 \\
 -8.87 & 42.740 & 0.024 & 42.809 &  -9.33 & 42.630 & 0.016 & 42.699 \\
 -8.54 & 42.770 & 0.024 & 42.838 &  -8.96 & 42.674 & 0.016 & 42.743 \\
 -8.41 & 42.760 & 0.023 & 42.827 &  -8.34 & 42.721 & 0.016 & 42.789 \\
 -7.78 & 42.824 & 0.024 & 42.891 &  -7.96 & 42.767 & 0.016 & 42.835 \\
 -7.55 & 42.823 & 0.023 & 42.890 &  -7.34 & 42.801 & 0.016 & 42.869 \\
 -6.83 & 42.883 & 0.023 & 42.950 &  -6.94 & 42.840 & 0.016 & 42.908 \\
 -6.57 & 42.879 & 0.023 & 42.946 &  -6.34 & 42.874 & 0.016 & 42.940 \\
 -6.03 & 42.924 & 0.023 & 42.991 &  -4.96 & 42.942 & 0.016 & 43.007 \\
 -5.88 & 42.931 & 0.023 & 42.997 &  -4.34 & 42.964 & 0.017 & 43.028 \\
 -2.89 & 43.005 & 0.023 & 43.069 &  -3.34 & 42.989 & 0.016 & 43.053 \\
 -1.83 & 43.012 & 0.023 & 43.076 &  -2.34 & 43.003 & 0.016 & 43.067 \\
 -1.40 & 43.010 & 0.022 & 43.073 &  -1.34 & 43.011 & 0.016 & 43.074 \\
  1.14 & 42.998 & 0.022 & 43.059 &   0.65 & 43.006 & 0.016 & 43.067 \\
  2.13 & 42.983 & 0.022 & 43.043 &   1.65 & 42.992 & 0.016 & 43.052 \\
  3.23 & 42.966 & 0.022 & 43.024 &   2.65 & 42.984 & 0.016 & 43.043 \\
  4.09 & 42.948 & 0.021 & 43.005 &   3.03 & 42.957 & 0.017 & 43.015 \\
  5.15 & 42.928 & 0.021 & 42.985 &   3.65 & 42.963 & 0.017 & 43.020 \\
  6.97 & 42.879 & 0.021 & 42.935 &   4.65 & 42.954 & 0.017 & 43.010 \\
  8.33 & 42.847 & 0.024 & 42.903 &   6.64 & 42.902 & 0.017 & 42.957 \\
 10.35 & 42.789 & 0.021 & 42.848 &   9.04 & 42.807 & 0.018 & 42.864 \\
 13.57 & 42.676 & 0.022 & 42.750 &  10.01 & 42.734 & 0.018 & 42.793 \\
 16.38 & 42.580 & 0.022 & 42.668 &  11.01 & 42.693 & 0.019 & 42.755 \\
 19.39 & 42.494 & 0.022 & 42.606 &  12.01 & 42.668 & 0.019 & 42.734 \\
 22.52 & 42.419 & 0.022 & 42.562 &  12.99 & 42.632 & 0.019 & 42.701 \\
 25.16 & 42.364 & 0.021 & 42.524 &  14.00 & 42.599 & 0.020 & 42.673 \\
 28.33 & 42.302 & 0.022 & 42.481 &  14.98 & 42.557 & 0.020 & 42.637 \\
 31.50 & 42.233 & 0.022 & 42.423 &  15.99 & 42.524 & 0.020 & 42.612 \\
 34.54 & 42.179 & 0.022 & 42.379 &  20.97 & 42.423 & 0.018 & 42.551 \\
 39.39 & 42.066 & 0.022 & 42.260 &  21.97 & 42.370 & 0.017 & 42.505 \\
 42.41 & 42.006 & 0.022 & 42.189 &  22.97 & 42.364 & 0.017 & 42.507 \\
 43.50 & 41.993 & 0.022 & 42.174 &  24.97 & 42.311 & 0.016 & 42.471 \\
 46.47 & 41.953 & 0.023 & 42.130 &  25.97 & 42.300 & 0.016 & 42.468 \\
 46.84 & 41.954 & 0.022 & 42.131 &  33.97 & 42.112 & 0.016 & 42.311 \\
 53.47 & 41.874 & 0.022 & 42.029 &  35.96 & 42.062 & 0.016 & 42.262 \\
 57.84 & 41.830 & 0.022 & 41.974 &  36.96 & 42.044 & 0.016 & 42.243 \\
 61.77 & 41.795 & 0.022 & 41.929 &  37.97 & 42.023 & 0.016 & 42.219 \\
 65.43 & 41.744 & 0.022 & 41.870 &  43.95 & 41.936 & 0.016 & 42.116 \\
 73.33 & 41.670 & 0.022 & 41.777 &  80.39 & 41.538 & 0.016 & 41.628 \\
 77.38 & 41.633 & 0.022 & 41.731 &  92.41 & 41.418 & 0.017 & 41.498 \\
 78.39 & 41.629 & 0.022 & 41.724 &  94.43 & 41.409 & 0.017 & 41.489 \\
 87.72 & 41.549 & 0.022 & 41.629 &  96.45 & 41.384 & 0.017 & 41.464 \\
 97.73 & 41.448 & 0.022 & 41.528 &  97.38 & 41.365 & 0.017 & 41.445 \\
 --    & --     & --    & --     &  98.45 & 41.353 & 0.017 & 41.433 \\
 --    & --     & --    & --     &  99.38 & 41.339 & 0.017 & 41.419 \\
\end{longtable}
{$^a$}{Relative to the epoch of $B-$band maximum of SN\,2018gv (MJD = 58,149.698$\pm$0.510);} \\
{$^b$}{Uncertainty in the distance not included;} \\
{$^c$}{Relative to the epoch of $B-$band maximum of SN\,2011fe (MJD = 55,814.48$\pm$0.03));} \\
\end{scriptsize}
\end{center}

\appendix
\section{Appendix: VLT Spectropolarimetry Reduction~\label{App_VLT}}
All VLT spectropolarimetric observations were conducted with the 300V grism 
coupled to a 1$\arcsec$ slit. 
The high-pass was chosen to stop second order light, from the flux 
at $\lambda \lesssim4400$ \AA, contaminating the spectrum at wavelength 
$\lambda \> 6500$ \AA. This configuration provides a wavelength range of 
$\sim4400-9200$ \AA\, with a resolution of 11.0 \AA\ ($FWHM$) at a central 
wavelength of 5849 \AA, and a dispersion of $\sim$2.6 \AA\ pixel$^{-1}$. At 
both epochs, the spectropolarimetric observations consisted of four 
exposures each with the half-wave retarder plate positioned at angles 
of $0^{\circ}$, $45^{\circ}$, $22^{\circ}.5$, and $67^{\circ}.5$, respectively. 
Exposure times at each plate angle were chosen to be 15 minutes and 2 minutes 
for epoch 1 and epoch 2, respectively. The spectra were flux calibrated using 
one 60 s integration of the photometric standard star LTT3218, with the 
polarimetry optics in place and the retarder plate at 0 degrees. 

Spectra obtained at each retarder plate angle were bias subtracted, 
flat-field corrected and wavelength calibrated using standard tasks within 
{\sc IRAF}. The ordinary and extraordinary beams were processed separately, 
and the typical RMS error on the wavelength calibration gives $\sim$0.25 \AA. 

The Stokes parameters describe the polarization state of the 
electromagnetic radiation. $I$ gives the total intensity of the beam, $Q$ 
and $U$ can be recognized by the projection of the radiation E-vectors to 
different directions on the plane of the sky ($+Q\leftrightarrow$, 
$-Q\updownarrow$, $+U\neswarrow$, $-U\nwsearrow$). 
Briefly speaking, the Stokes parameters $Q$ and $U$ can be derived via 
Fourier transformation, as described in the VLT FORS2 User Manual 
\citep{Anderson_etal_2018}: 
\begin{equation}
\begin{aligned}
Q_{0} = \frac{2}{N} \sum\limits_{i=0}^{N-1} F(\theta_i) {\rm cos} (4\theta_i) \\
U_{0} = \frac{2}{N} \sum\limits_{i=0}^{N-1} F(\theta_i) {\rm sin} (4\theta_i),
\end{aligned}
\label{Eqn_stokes1}
\end{equation}
where $F(\theta_i)$ give the normalized flux difference between the 
ordinary ($f^{o}$) and extra-ordinary ($f^{e}$) beams: 
\begin{equation}
F(\theta_i) = \frac{f^{o}(\theta_i) - f^{e}(\theta_i)}{f^{o}(\theta_i) + f^{e}(\theta_i)}. 
\label{Eqn_stokes2}
\end{equation}
In the case of this study, four observations 
were carried out at retarder angles of 0, 22.5, 45, and 67.5 degrees ($N=4$). 
Therefore, $Q_{0} = [F(0^{\circ}) - F(45^{\circ})]/2$, and 
$U_{0} = [F(22^{\circ}.5) - F(67^{\circ}.5)]/2$. 
The measured Stokes parameters were also corrected for the offsets to 
the zero angle of the retarder plate ($-\Delta \chi(\lambda)$). Typical 
chromatic dependence of the zero angle was less than 4 degrees, since a 
super-achromatic half wave plate is used with FORS2 (see Figure 4.1 of 
the VLT FORS2 User Manual). The tabulated values are available from the 
FORS instrument description 
page\footnote{http://www.eso.org/sci/facilities/paranal/instruments/fors/inst/pola.html
}.

A small amount of wavelength-dependent instrumental polarization in FORS2 
($\lesssim0.1\%$) has been investigated by \citet{Fossati_etal_2007} and 
\citet{Siebenmorgen_etal_2014}. Further analytical quantification by 
\citet{Cikota_etal_2017_fors2} gives: 
\begin{equation}
\begin{aligned}
Q^{\rm Instr.}(\lambda) = 9.66 \times 10^{-8} \lambda + 3.29 \times 10^{-5} \\ 
U^{\rm Instr.}(\lambda) = 7.28 \times 10^{-8} \lambda - 4.54 \times 10^{-4},
\end{aligned}
\label{Eqn_pol_instr}
\end{equation}
where $\lambda$ is the observed wavelength in \AA. In many cases, in 
order to reduce slit losses, the 
instrument position angle, $\chi$, is aligned to the parallactic angle and 
not necessarily aligned to the north celestial meridian, i.e. $\chi \neq$0. 
Therefore, the correct measurement of the polarization position angle 
requires transforming the Stokes parameters from the instrumental reference 
frame to the sky reference frame. The instrumental polarization tends to be 
constant among different instrument position angles (see, i.e. Fig.~8 of 
\citealp{Siebenmorgen_etal_2014}) and needs to be corrected before the 
transformation between the instrument and sky reference frames (see, i.e. 
\citealp{Bagnulo_etal_2017}). Linear polarization measurement follows the 
transformation given by Equation (10) in \citet{Bagnulo_etal_2009}. 
Therefore, we write the expression of the Stokes parameters as the 
following: 
\begin{equation}
\begin{aligned}
Q =  (Q_0 - Q^{\rm Instr.}) \cos (2\chi) + (U_0 - U^{\rm Instr.}) \sin (2\chi), \\
U = -(Q_0 - Q^{\rm Instr.}) \sin (2\chi) + (U_0 - U^{\rm Instr.}) \cos (2\chi). 
\end{aligned}
\label{Eqn_stokes3}
\end{equation}
We set $\chi$ to 0 deg in our observations and subtract the 
instrumental polarization calculated from Equation~\ref{Eqn_pol_instr} 
to correct the instrumental effect. The wavelength scale of the Stokes 
parameters and calculated polarization were also corrected to the 
rest-frame by adopting the host galaxy recessional velocity (1582 km 
s$^{-1}$). 

\section{Appendix: `CMAGIC' Extinction Estimation~\label{App_CMAGIC}}
Following \citet{Wang_etal_2003}, we fit the linear region of the 
magnitude$-$color diagram from 5 to 27 days after the $B-$band 
maximum light with the CMAG relation: 
\begin{equation}
B = B_{BV} + \beta_{BV} (B-V)
\label{Eqn_cmag1}
\end{equation}
Here $\beta_{BV}$ and $B_{BV}$ denote the slope and the value for the
intercept at $(B-V) = 0$ of the linear region in the CMAG diagram, 
respectively. According to \citet{Wang_etal_2003}, the term 
$\mathcal{E}(B-V)$, namely `CMAGIC' color excess, is given by: 
\begin{equation}
\mathcal{E}(B-V) = \frac{(B_{\rm max} - B_{BV})}{\beta_{BV}}. 
\label{Eqn_cmag2}
\end{equation}
The `CMAGIC' color excess for a reddening-free SN, or $\mathcal{E}_0$, 
also shows a dependence on the $\Delta m_{B}^{15}$, which characterizes 
the width of the light curve. 
By fitting a sample of SNe with little or no color excess 
($B_{\rm max} - V_{\rm max} \textless 0.05$ mag) and also correcting 
that using $E(B-V)$ of \citet{Phillips_etal_1999}, the linear dependence 
of $\mathcal{E}_0$ on $\Delta m_{15}^{B}$ derived from this 
low-extinction sample of Type Ia SNe gives: 
\begin{equation}
\mathcal{E}_0 = (-0.118 \pm 0.013) + (0.249 \pm 0.043) (\Delta m_{15} - 1.1), 
\label{Eqn_cmag3}
\end{equation}
Finally, the color excesses of Type Ia SNe can be estimated using the formula: 
\begin{equation}
E_{BV}(B-V) = \mathcal{E}(B-V) - \mathcal{E}_0; 
\label{Eqn_cmag4}
\end{equation}

We fit the absolute $B$ magnitude and $B-V$ color of SN\,2018gv between $+$5 and 
$+$27 days to the CMAG relation described by Equation~\ref{Eqn_cmag2}. Both 
quantities have been corrected for the Galactic extinction. The result gives 
$\mathcal{E}^{\rm 18gv}(B-V) = -0.127 \pm 0.018$ mag, 
$\mathcal{E}^{\rm 18gv}_0 = -0.155 \pm 0.020$ mag, and 
$E^{\rm 18gv}_{BV} = 0.028 \pm 0.027$ mag. 

A sanity test was carried out by applying the same procedure to the $B$ and 
$V-$band photometry of SN\,2011fe published in \citet{Munari_etal_2013}. 
In order to be consistent with our extinction estimation of SN\,2018gv, we 
applied only the Galactic extinction, $E(B-V) = 0.01$ mag to the photometry 
and color of SN\,2011fe based on the estimated host Na {\sc ID} absorption 
features from high-resolution spectroscopy \citep{Patat_etal_2013}. 
A Cepheid distance modulus $\mu_0 = 29.04$ was used for SN\,2011fe 
\citep{Shappee_etal_2011}. Finally, we conducted the CMAGIC fitting to 
estimate the host extinction: 
$\mathcal{E}^{\rm 11fe}(B-V) = -0.125 \pm 0.018$ mag, 
$\mathcal{E}^{\rm 11fe}_0 = -0.125 \pm 0.020$ mag, and the color excess of 
SN\,2011fe from the host gives $E^{\rm 11fe}_{BV} = 0.000 \pm 0.027$ mag. 
This is consistent with the evidence for very little dust extinction of 
SN\,2011fe in its host galaxy 
\citep{Tammann_etal_2013, Patat_etal_2013, Pereira_etal_2013, Zhang_etal_2016}. 

\section{Appendix: Steps Of The Pseudo-Bolometric Light Curve Construction~\label{App_SED}}
(1) {\bf Extinction correction and magnitude conversion.} 
We correct for the Galactic extinction and the host extinction listed in 
Table~\ref{Table_phot_para} to the $UBg'Vr'i'$-band photometry of SN\,2018gv. 
The $UBV$-band photometry was converted to the AB system adopting the 
linear offsets given in Table~1 of \citet{Blanton_etal_2007}. Specifically, 
$U_{\rm AB} = U_{\rm Vega} + 0.79$, $B_{\rm AB} = B_{\rm Vega} - 0.09$, 
and $V_{\rm AB} = V_{\rm Vega} + 0.02$; 

(2) {\bf Synthetic photometry on template spectra.} 
We adopt the spectral template of Type Ia SNe built by \citet{Hsiao_etal_2007} 
and register the spectral templates to the nearest photometric phase. For each 
photometric epoch, we then perform synthetic photometry on the appropriate 
spectrum for the $UBg'Vr'i'$ bandpasses; 

In practice, our synthetic photometry was carried out by first calculating the 
mean photon flux density for different bandpasses: 
\begin{equation}
\langle \lambda F_{b}(\lambda) \rangle = \frac{\int \lambda F{_\lambda} 
T_{b}(\lambda) d\lambda}{\int \lambda T_{b}(\lambda) d\lambda}, 
\label{Eqn_syn1}
\end{equation}
where $F_{\lambda}$ is the flux spectrum in unit of 
erg cm$^{-2}$ s$^{-1}$ \AA$^{-1}$, and $T_{b}(\lambda)$ gives the 
unitless throughput for a certain bandpass, which is denoted as $b$. 
Then we synthesize the magnitude in the AB system to compare with our 
photometry of the SN. According to the {\it Synphot} User's Guide for the 
{\it HST} {\it Synphot 
\footnote{http://www.stsci.edu/institute/software\_hardware/stsdas/synphot}
} software, we first obtain the ST magnitude, which is similarly defined as 
the AB system but gives constant flux per unit wavelength interval rather 
than unit frequency: 
\begin{equation}
STMAG_{b} = -2.5 \times {\rm log}_{10} \langle \lambda F_{b}(\lambda) \rangle - 21.1.  
\label{Eqn_syn2}
\end{equation}
The conversion between the ST and AB systems is given by: 
\begin{equation}
ABMAG_{b} = STMAG_{b} - 5{\rm log}_{10} \lambda_{b}^{\rm pivot} + 18.692 - ZP_{b}. 
\end{equation}
The source-independent pivot wavelength is defined by: 
\begin{equation}
\lambda_{b}^{\rm pivot} = \sqrt{\frac{\int \lambda T_{b}(\lambda) 
d\lambda}{\int T_{b}(\lambda) d\lambda / \lambda}}, 
\label{Eqn_syn3}
\end{equation}
and the zero points for $UBgVri$ bandpasses can be found in Table~3 of 
\citet{Pickles_etal_2010}; 

(3) {\bf Scaling factors between template spectra and photometry.} 
We calculate the differences between the photometry of SN\,2018gv after 
applying the corrections in step (1) and the synthesized AB magnitude of 
each template spectrum for each band obtained in step (2). The difference 
magnitudes were converted into flux space to obtain the scale factors 
between the observations of SN\,2018gv and template spectra, shown in the 
middle panel of Fig.~\ref{Fig_construct_swift_bolo}. Therefore, 
multiplying the scale factor for each bandpass to 
$\langle \lambda F_{b}(\lambda) \rangle$ calculated from the corresponding 
template spectrum gives the mean photon flux density of SN\,2018gv at a 
certain bandpass. This is shown by the filled symbols in the upper panel 
of Fig.~\ref{Fig_construct_swift_bolo}. Abscissae are given by the 
pivot wavelength of each bandpass calculated by Equation~\ref{Eqn_syn3}. 
Three different colors and symbols illustrate the construction of the 
spectral energy distribution (SED) of SN\,2018gv at three epochs. For each 
epoch, the abscissa of the leftmost and the rightmost points are determined 
by $\lambda_{U}^{\rm pivot} - FWHM_{U}$ and $\lambda_{i}^{\rm pivot} + FWHM_{i}$, 
respectively, while the two ordinates are assigned to 
$\langle \lambda F_{U}(\lambda) \rangle / 2$ and 
$\langle \lambda F_{i}(\lambda) \rangle / 2$, respectively; 

(4) {\bf The supernova SED.} 
We connect the mean photon flux density at $UBg'Vr'i'$ bandpasses and the 
leftmost and the rightmost boundaries described in step (3) to construct 
the optical SED of SN\,2018gv (SED-dots), shown by the dotted lines in the 
upper panel of Fig.~\ref{Fig_construct_swift_bolo}. We also warp the 
template spectra using linear interpolation to the scale factors (SED-warp). 
The warped spectra at three epochs are also presented by the 
solid lines in the upper panel of Fig.~\ref{Fig_construct_swift_bolo}; 

(5) {\bf Integrate the SED to obtain the pseudo-bolometric luminosity.} 
For each epoch, we integrate the SED over the wavelength $\sim1660-8180$ \AA\ 
for both SED-dots and SED-warp to calculate the pseudo-bolometric luminosity. 
The pseudo-bolometric luminosities of SN\,2018gv before $t = 100$ days 
calculated using SED-warp are on average (4.0$\pm1.1$)\% higher than those 
using SED-dots. This discrepancy results from the construction of SEDs, 
see \citet{Brown_etal_2016}. 

The errors due to photometric uncertainties of each bandpass were computed 
through a Monte Carlo re-sampling approach using the photometric errors. 
The typical error before $t = 100$ days amounts to 0.4\% of the luminosity. 
The total error of the pseudo-bolometric luminosity is dominated by the 
systematical difference between choices of the SED. We characterize this by 
calculating the variance between the SED-dots and SED-warp integrations. 
Finally, for each epoch, we add this systematical uncertainty and the error 
due to photometry in quadrature to obtain the final uncertainties in the 
pseudo-bolometric light curve. The median value of the final uncertainties 
is 3.0\%$\pm0.8$\%. The result of the host-galaxy extinction estimation is 
consistent with no extinction within the associated uncertainties (see 
Section~\ref{Section_extinction} and Table~\ref{Table_phot_para}). 
Therefore, we did not include the uncertainties in the estimation of the 
host galaxy extinction. 

\section{Appendix: ISP Estimation~\label{App_ISP}}
We identify spectral regions that are likely to be intrinsically depolarized 
due to the overlap of many Fe absorption wings. This ``line blanketing'' has 
a significant effect on opacities, and the spectral regions are significantly 
depolarized since the line blanketing opacity dominates over electron 
scattering opacity. As suggested in \citet{Howell_etal_2001} and 
\citet{Maund_etal_2013}, we start by considering the wavelength region of 
$4800-5600$ \AA\ (region A) and $5100-5300$ \AA\ (region B) as the 
intrinsically depolarized regions of SN\,2018gv. We notice that there are 
suspicious line polarization patterns, at epoch 1, in the flux-normalized 
Stokes $Q$ and $U$ spectra (see panels b-c of Fig.~\ref{Fig_iqu_ep1} and 
Section~\ref{section_other_lines}), possibly associated with the 
\ion{S}{2} $\lambda$5454 and $\lambda$5640 lines at a velocity of 
$v\sim-$13,800 km s$^{-1}$. The expected effects of line-blanketing, instead, 
appear particularly prominent at wavelengths below $\sim$5000 \AA\ 
\citep{Leonard_etal_2005}. We also consider the additional wavelength 
region C, defined as $4800-5100$ \AA. 

We measure the Stokes parameters by taking the error-weighted mean value 
across these wavelength ranges. The measured values at epoch 1 are: 
\begin{equation}
\begin{aligned}
Q^{A}_{\rm ISP1} =  0.19\% \pm0.10\%, U^{A}_{\rm ISP1} = -0.41\% \pm0.07\%,\\
Q^{B}_{\rm ISP1} =  0.26\% \pm0.06\%, U^{B}_{\rm ISP1} = -0.39\% \pm0.08\%，\\
Q^{C}_{\rm ISP1} =  0.08\% \pm0.07\%, U^{C}_{\rm ISP1} = -0.37\% \pm0.06\%;
\end{aligned}
\label{Eqn_stokes4}
\end{equation}
At epoch 2, the Stokes parameters over the same wavelength ranges are: 
\begin{equation}
\begin{aligned}
Q^{A}_{\rm ISP2} =  0.08\% \pm0.12\%, U^{A}_{\rm ISP2} = -0.49\% \pm0.08\%, \\
Q^{B}_{\rm ISP2} =  0.07\% \pm0.15\%, U^{B}_{\rm ISP2} = -0.50\% \pm0.08\%, \\
Q^{C}_{\rm ISP2} =  0.06\% \pm0.13\%, U^{C}_{\rm ISP2} = -0.48\% \pm0.07\%. \\
\end{aligned}
\label{Eqn_stokes5}
\end{equation}
At epoch 1, the possible polarization associated with \ion{S}{2} 
$\lambda$5454 and $\lambda$5640 makes the Stokes $Q$ and $U$ measured in 
regions A and B systematically $0.11\%$ and $0.08\%$ higher than those 
measured at epoch 2, respectively. It can be seen that $Q^{C}_{\rm ISP1}$ 
is consistent with $Q^{C}_{\rm ISP2}$. For each spectral region, the 
measurements at both epochs are consistent within the associated 
uncertainties. Considering that the observation at epoch 2 achieved a 
higher S/N, which provides a more accurate estimate of the ISP, 
we adopt the error-weighted mean Stokes parameters measured over the 
three regions at epoch 2 to represent the Stokes parameters for the ISP: 
$Q_{\rm ISP} =  0.07\% \pm0.16$\%, $U_{\rm ISP} = -0.49\% \pm0.09$\%. 
The uncertainties were estimated by adding the average errors of the 
Stokes parameters over the three wavelength ranges, the standard deviation 
of the Stokes parameters over the three wavelength ranges, and the 
uncertainties of the weighted mean Stokes parameters in quadrature. 

\end{document}